\documentclass[11pt,a4paper]{article}

\pdfoutput=1

\usepackage{epsfig,amsfonts,amssymb}
\usepackage[vcentermath]{youngtab}
\usepackage{amsmath,graphics}
\usepackage{slashed}
\usepackage[utf8]{inputenc}
\usepackage{scalerel}
\usepackage{graphicx}
\usepackage{stackengine}
\usepackage{hyperref}
\usepackage{cleveref}
\usepackage{amsmath,mleftright}
\usepackage{xparse}
\usepackage{slashed}
\usepackage{amssymb}
\usepackage{cancel}
\usepackage{multirow}
\usepackage{color}                                                      
\usepackage{xcolor}
\usepackage{bbm}
\usepackage{url}
\usepackage{longtable}
\usepackage{array}
\usepackage{bbold}
\numberwithin{equation}{section}

\newcommand{\ads}{\mathrm{AdS_4}}

\newcommand{\Action}{\mathcal{S}}
\newcommand{\RR}{\mathbbm{R}}

\newcommand{\Tr}{\mathop{\mathrm{tr}}}

\newcommand{\sign}{\mathop{\mathrm{sign}}}
\newcommand{\g}{\dagger}
\newcommand{\NN}{\mathcal{N}}
\newcommand{\WW}{\mathcal{W}}

\newcommand{\ZZ}{\mathcal{Z}}
\newcommand{\VV}{\mathcal{V}}
\newcommand{\Ep}{(\varepsilon_1+i\varepsilon_2)}
\newcommand{\Em}{(\varepsilon_1-i\varepsilon_2)}
\newcommand{\E}{\varepsilon_3}
\newcommand{\Ev}{\varepsilon}

\newcommand{\m}{{{(j)}}}

\newcommand{\n}{{{(j-1)}}}
\newcommand{\p}{{{(j+1)}}}
\newcommand{\1}{{{(1)}}}
\newcommand{\2}{{{(2)}}}
\newcommand{\3}{{{(3)}}}
\newcommand{\5}{{{(5)}}}
\newcommand{\6}{{{(6)}}}
\newcommand{\4}{{(4)}}
\newcommand{\7}{{(7)}}
\newcommand{\nn}{\nonumber}

\newcommand{\ben}{\begin{eqnarray*}}
\newcommand{\en}{\end{eqnarray*}}
\newcommand{\mm}{\mathfrak{m}}
\newcommand{\N}{\mathfrak{n}}
\newcommand{\s}{{{(s)}}}

\newcommand{\superN}{\mathcal{N}}

\newcommand{\Bigsbrk}[1]{\Bigl[#1\Bigr]}

\ifx\href\asklfhas\newcommand{\href}[2]{#2}\fi
\ifx\arxivref\asklfhas\newcommand{\arxivref}[1]{\href{http://arxiv.org/abs/#1}%
{#1}}\fi
\ifx\doiref\asklfhas\newcommand{\doiref}[2]{\href{http://dx.doi.org/#1}{#2}}\fi

\newcommand{\be}{\begin{eqnarray}}
\newcommand{\ee}{\end{eqnarray}}

\usepackage{url}
\usepackage{longtable}

 \usepackage{hyperref}
\def\one{{\hbox{ 1\kern-.8mm l}}}
\def\zero{{\hbox{ 0\kern-1.5mm 0}}}

\begin{document}

\baselineskip 24pt

\begin{center}
{\Large \bf Superconformal Index for $\mathcal{N}=3$ $\widehat{ADE}$ Chern-Simons Quiver Gauge Theories}

\end{center}

\vskip .6cm
\medskip

\vspace*{4.0ex}

\baselineskip=18pt

\centerline{\large \rm   Moumita Patra}

\vspace*{4.0ex}

\centerline{\large \it National Institute of Science Education and Research Bhubaneshwar,}

\centerline{\large \it  P.O. Jatni, Khurda, 752050, Odisha, INDIA}
\vspace*{1.0ex}

\centerline{\large \it Homi Bhabha National Institute, Training School Complex,}

\centerline{\large \it  Anushakti Nagar, Mumbai, India 400085}

\vspace*{4.0ex}
\centerline{E-mail:  mpatra91@niser.ac.in }

\vspace*{5.0ex}

\centerline{\bf Abstract} \bigskip
\thispagestyle{empty}
We compute superconformal indices for   $\mathcal{N} = 3$ $\widehat{ADE}$ Chern-Simons quiver  gauge theories with a product gauge group $\prod_i U(N)_i$, using the method of supersymmetric localization. 
We also perform a large $N$ analysis of the index. This index includes contribution from non zero magnetic flux sector.
The fact that these theories have a  weakly coupled UV completion in terms of $\mathcal{N}= 3$ supersymmetric Chern-Simons Yang-Mills theories enables us to apply the localization technique. Such theories have dual M-theory description on $\mathrm{AdS}_4\times M_7$, where $M_7$ is a tri-Sasaki Einstein manifold.

\vfill \eject

\baselineskip=18pt

\newpage
\setcounter{page}{1}
\renewcommand{\thefootnote}{\arabic{footnote}}
\setcounter{footnote}{0}

\hrule
\tableofcontents
\vspace{8mm}
\hrule
\vspace{4mm}



\section{\label{1}Introduction}
Superconformal(SCF) indices in various dimensions have been studied extensively for the past many years. 
The  index of a $D$ dimensional SCF theory is defined as,
\be 
I(\beta_j)=\Tr_\mathcal{H}\Big[(-1)^F\, e^{-\gamma\{Q,Q^\g \}}e^{-\sum_j \beta_jt_j}\Big]
\ee 
where, the trace is over the Hilbert space of radially quantized theory, 
$(-1)^F$ is fermion number operator, $Q,Q^\g$ are supercharges and 
$t_j$'s  are the generators of	the Cartan subalgebra of the superconformal and flavor symmetry algebra commuting with the $Q$'s.  
\smallskip

One  finds $I$ is independent of $\gamma$, i.e  it receives contributions only from those BPS states that cannot combine into long representations.  Therefore  the index does not change under variation of any continuous parameter of the  theory and hence are guaranteed to be protected. This also implies that the index of strongly coupled  theories can be computed by setting the  coupling to zero.
SCF index in $D=4$ dimension was first constructed in \cite{Kinney:2005ej} and then extended to $D=3,5,6$ in \cite{Bhattacharya:2008zy}. In  \cite{Kinney:2005ej} it was shown that the index of the field theory   perfectly agrees with the multiparticle index  of graviton  on $AdS_5\times S^5$. This agreement provides a check on the AdS/CFT conjecture \cite{Maldacena:1997re}.
Moreover  the study on the index is useful in understanding general super-conformal field theories and their classifications. 
\smallskip 

SCF indices in $D=3$ have been of great interest in establishing $\ads/\mathrm{CFT}_3$ correspondence and studying low energy limit of multiple  $M2$ branes. The field theory side of this correspondence is  $\NN=6$ superconformal Chern-Simons(CS) matter theory. Such a theory, widely known as ABJM theory,  was constructed in \cite{abjm}  whose gravity dual is M-theory on $\ads\times S^7/\mathbbm{Z}_k$, where $k$ is the CS level. In particular, the ABJM
theory  describes $N M2$ branes at a  $\mathbbm{C}^4/\mathbbm{Z}_k$ orbifold singularity.
\smallskip 

As a natural extension of ABJM theory is  to construct theories which can arise as world volume theories of multiple M2 branes probing other hyperK\"{a}hler singularities. A large class of such theories are conjectured to be holographically dual to M-theory on $\ads\times Y$,  where $Y$ is Sasaki-Einstein seven-manifolds. The low energy theories are CS quiver gauge theories, whose field content can be summarised by   $\widehat{ADE}$ quiver diagrams  \cite{jaff}\cite{gulotta}\cite{Szabo:2014zua}. In this note we compute superconformal indices in this class of theories using method of supersymmetric localization following \cite{Kim:2009wb}\footnote{The deformation we  use to carry out supersymmetric localization is same as \cite{bkk} which is slightly different from \cite{Kim:2009wb}. This will be explained in section \ref{sec_localization}.}. The idea behind supersymmetric localization is  described below.
\smallskip

The localization technique involves deforming the theory by a $Q$-exact term by adding $t\,Q(V)$ to the action, where $t$ is a continuous parameter and  V is some expression involving the fields of the theory. The only thing one has to keep in mind that this deformation should preserve some supersymmetry of the theory. 
Consider a theory with action  $\mathcal{S}[\phi]$ and  $Q$ is a symmetry of the action such that
\be 
\label{properties_supercharge}
Q \mathcal{S}=0\quad and\quad Q^2=0\,.
\ee 
We deform the action by a $Q$ exact term, i.e $tQV[\phi]$. 
Now, the path integral of the deformed theory looks as follows,
\be 
\int [\mathcal{D}\phi]\,	e^{-\mathcal{S}}\to \int [\mathcal{D}\phi]\, e^{-\mathcal{S}+tQV[\phi]}\;.
\ee 
It can be shown using eqn.\ref{properties_supercharge}  that,
\be 
\frac{d	}{dt} \int [\mathcal{D}\phi]\, e^{-\mathcal{S}+tQV[\phi]}=0
\ee 
when the integration measure is $Q$ invariant.  Hence we can adjust $t$ as per our convenience for computation. Usually we have a free theory in the $t\to \infty$ and in this limit
 the dominant contribution of $e^{tQV[\phi]}$ to the path integral comes from those field configurations for  which,
 \be 
  QV[\phi]=0\;.
  \ee 
  The field configurations which satisfy the above is called the saddle point equation.
At the end of the computation  we set $t=0$ which brings us back to the original theory. 
\smallskip 

One also needs the manifold on which we deform the theory to be compact to apply method of localization. Therefore for three dimensions we do the computations on $S^2\times S^1$ and relate it to the flat space theory by a conformal transformation. Moreover,  we need the theory to be superconformal at the quantum level since the supersymmetry transformation $Q$ that is used in localization generates the superconformal algebra.  But if we add a superpotential, it breaks the quantum SCF symmetry as it has anomalous dimension.  In localization we will see that the choice of superpotential will not matter as the localization demands that all matter fields should vanish. 

\smallskip
Now, we write our final result below which will be explained in later sections. The SCF index of $\widehat{A}$-type quiver with gauge group $\prod_{\m=1}^n U(N)_{k_\m}$ and constraint $\sum_{\m=1}^n k_\m=0$, is
\begin{align}\label{final_index_an}
&			I(x)=
		x^{\epsilon_0}\int\frac{1}{\rm (symmetry)}
		\prod_{\m=1}^n	\left[\frac{d\alpha_\m^{\mathfrak{m}}}{2\pi}\right]\prod_{\mathfrak{{m}}<\mathfrak{{n}}}
			\left[2\sin\left(\frac{\alpha_\m^\mathfrak{m}\!-\!\alpha_\m^\mathfrak{n}}{2}\right)\right]^2\times e^{i\sum_{\m=1}^n k_\m\sum_{\mathfrak{m}=1}^N q_\m^\mathfrak{m}\alpha_\m^\mathfrak{{m}}\!}&\nn\\
&\prod_{\hat{\mathfrak{m}},\mathfrak{n}=1}^N\exp\Bigg[\sum_{\m=1}^n\sum_{r=1}^\infty \frac{1}{r}\Big[ e^{-ir(\alpha_\m^\mathfrak{n}-\alpha_\p^{\mathfrak{\hat{m}}})}f^{+}_{\mathfrak{n}\hat{\mathfrak{m}}\m}(x^r) +e^{ir(\alpha_\m^\mathfrak{n}-\alpha_\p^{\mathfrak{\hat{m}}})}f^{-}_{\mathfrak{n}\hat{\mathfrak{m}}\m}(x^r)\Bigg]&\nn \\
&			\times\!\prod_{\mathfrak{{m}},\mathfrak{{n}}=1}^N\!\exp\left[\sum_{r=1}^\infty\frac{1}{r}
\sum_{\m=1}^n			f^{{\rm adj}}_{\mathfrak{{m}}\mathfrak{{n}}\m}(x^r)e^{-ir(\alpha^\mathfrak{{m}}_\m\!-\!\alpha^\mathfrak{n}_\m)}
			\right]&
\end{align}
where, $\epsilon_0$ is the zero point energy of the vacuum,
\begin{equation}
	\epsilon_0=\sum_{\m=1}^n\Bigg(\frac{1}{2}\sum_{\mathfrak{m},\mathfrak{n}=1}^N|q^\mathfrak{m}_\m\!-\!q^\mathfrak{n}_\p|-\sum_{\mathfrak{m}<\mathfrak{n}}|q^\mathfrak{m}_\m\!-\!q^\mathfrak{n}_\m|\Bigg)
\end{equation}
$\rm (symmetry)$ is the order of the Weyl group of unbroken gauge symmetry, $ \hat{\mathfrak{m}},\mathfrak{n}=1,2, \cdots, N$ are gauge indices, $\alpha_\m^\mathfrak{{m}}$'s are holonomy variables along $S^1$, $q_\m^\mathfrak{{m}}$'s are magnetic fluxes. and,
\begin{align}
	\label{eqn:matter_contribution}
&f^{+}_{\mathfrak{n}\hat{\mathfrak{m}}\m}(x)=\,x^{\frac{1}{2}}\sum_{j=\frac{|q_\m^{\mathfrak{n}}-q_\p^{\hat{\mathfrak{m}}}|}{2}}^\infty x^{2j}-\,x^{\frac{3}{2}}\sum_{j=\frac{|q_\m^{\mathfrak{n}}-q_\p^{\hat{\mathfrak{m}}}|}{2}}^\infty x^{2j}=f^{-}_{\mathfrak{n}\hat{\mathfrak{m}}\m}(x)&\nn\\
&f^{\mathrm{adj}}_{\mathfrak{n}{\mathfrak{m}}\m }(x)=(-1+\delta_{q_{\mathfrak{m}}q_{\mathfrak{n}}})\,x^{|q^\mathfrak{m}_\m-q^\mathfrak{n}_\m|}\;.&
\end{align}

The SCF index of $\widehat{D}$-type quiver with gauge group $U(2N)^{n-3}\times U(N)^4, (n>3)$ and whose CS levels satisfy $\sum_{\m=5}^n 2k_\m+\sum_{\m=1}^4 k_\m=0$, is
\begin{align}\label{final_index_dn}
	&			I(x)=
x^{\epsilon_0}\int\frac{1}{\rm (symmetry)}
	\prod_{\m=1}^{n+1}\left[\frac{d\alpha_\m^{\mathfrak{m}}}{2\pi}\right]	\prod_{\mathfrak{{m}}<\mathfrak{{n}}}
	\left[2\sin\left(\frac{\alpha_\m^\mathfrak{m}\!-\!\alpha_\m^\mathfrak{n}}{2}\right)\right]^2\times e^{i\sum_{\m=5}^{n+1} k_\m\sum_{\mathfrak{m}=1}^{2N} q_\m^\mathfrak{m}\alpha_\m^\mathfrak{{m}}\!}&\nn\\
	&\times e^{i\sum_{\m=1}^4 k_\m\sum_{\mathfrak{m}=1}^N t_\m^\mathfrak{m}\alpha_\m^\mathfrak{{m}}\!}\times \prod_{\hat{\mathfrak{m}},\mathfrak{n}=1}^{2N}\exp\Bigg[\sum_{\m=5}^n\sum_{r=1}^\infty \frac{1}{r}\Big[ e^{-ir(\alpha_\m^\mathfrak{n}-\alpha_\p^{\mathfrak{\hat{m}}})}f^{+}_{\mathfrak{n}\hat{\mathfrak{m}}\m}(x^r) +e^{ir(\alpha_\m^\mathfrak{n}-\alpha_\p^{\mathfrak{\hat{m}}})}f^{-}_{\mathfrak{n}\hat{\mathfrak{m}}\m}(x^r)\Bigg]&\nn \\
	&\prod_{\hat{\mathfrak{m}},\mathfrak{n}=1}^N\exp\Bigg[\sum_{\m=1}^2\sum_{r=1}^\infty \frac{1}{r}\Big[ e^{-ir(\alpha_\m^\mathfrak{n}-\alpha_\5^{\mathfrak{\hat{m}}})}f^{+}_{\mathfrak{n}\hat{\mathfrak{m}}\m}(x^r) +e^{ir(\alpha_\m^\mathfrak{n}-\alpha_\5^{\mathfrak{\hat{m}}})}f^{-}_{\mathfrak{n}\hat{\mathfrak{m}}\m}(x^r)\Bigg]&\nn\\
	&\times\prod_{\hat{\mathfrak{m}},\mathfrak{n}=1}^N\exp\Bigg[\sum_{\m=3}^4\sum_{r=1}^\infty \frac{1}{r}\Big[ e^{-ir(\alpha_\m^\mathfrak{n}-\alpha_{(n+1)}^{\mathfrak{\hat{m}}})}f^{+}_{\mathfrak{n}\hat{\mathfrak{m}}\m}(x^r) +e^{ir(\alpha_\m^\mathfrak{n}-\alpha_{(n+1)}^{\mathfrak{\hat{m}}})}f^{-}_{\mathfrak{n}\hat{\mathfrak{m}}\m}(x^r)\Bigg]&\nn\\
	&			\times\!\prod_{\mathfrak{{m}},\mathfrak{{n}}=1}^{2N}\!\exp\left[\sum_{r=1}^\infty\frac{1}{r}
	\sum_{\m=5}^{n+1} f^{{\rm adj}}_{\mathfrak{{m}}\mathfrak{{n}}\m}(x^r)e^{-ir(\alpha^\mathfrak{{m}}_\m\!-\!\alpha^\mathfrak{n}_\m)}
	\right]\times\!\prod_{\mathfrak{{m}},\mathfrak{{n}}=1}^{N}\!\exp\left[\sum_{r=1}^\infty\frac{1}{r}
	\sum_{\m=1}^4 f^{{\rm adj}}_{\mathfrak{{m}}\mathfrak{{n}}\m}(x^r)e^{-ir(\alpha^\mathfrak{{m}}_\m\!-\!\alpha^\mathfrak{n}_\m)}
	\right]&
\end{align}
 and,
 \be
 \epsilon_0 	&=&\frac{1}{2}\sum_{\m=5}^n\sum_{\mathfrak{m},\mathfrak{n}=1}^{2N}|q^\mathfrak{m}_\m\!-\!q^\mathfrak{n}_\p| +\frac{1}{2}\sum_{\m=1}^2\sum_{\mathfrak{m},\mathfrak{n}=1}^N|q^\mathfrak{m}_\m\!-\!q^\mathfrak{n}_\5|+\frac{1}{2}\sum_{\m=3}^4\sum_{\mathfrak{m},\mathfrak{n}=1}^N|q^\mathfrak{m}_\m\!-\!q^\mathfrak{n}_{n+1}|\nn\\
 &-&\sum_{\m=5}^{n+1}\sum_{\mathfrak{m}<\mathfrak{n}=1}^{2N}|q^\mathfrak{m}_\m\!-\!q^\mathfrak{n}_\m|-\sum_{\m=1}^4\sum_{\mathfrak{m}<\mathfrak{n}=1}^N|q^\mathfrak{m}_\m\!-\!q^\mathfrak{n}_\m|
 \ee
\begin{align}
\textit{for $\m=5,\cdots, n$}\quad	&f^{+}_{\mathfrak{n}\hat{\mathfrak{m}}\m}(x)=\,x^{\frac{1}{2}}\sum_{j=\frac{|q_\m^{\mathfrak{n}}-q_\p^{\hat{\mathfrak{m}}}|}{2}}^\infty x^{2j}
-\,x^{\frac{3}{2}}\sum_{j=\frac{|q_\m^{\mathfrak{n}}-q_\p^{\hat{\mathfrak{m}}}|}{2}}^\infty x^{2j}=	f^{-}_{\mathfrak{n}\hat{\mathfrak{m}}\m}(x)&\nn\\
	&f^{\mathrm{adj}}_{\mathfrak{n}{\mathfrak{m}}\m }(x)=(-1+\delta_{q_{\mathfrak{m}}q_{\mathfrak{n}}})\,x^{|q^\mathfrak{m}_\m-q^\mathfrak{n}_\m|}&\nn\\ \nn\\
\textit{for $\m=1,2$}\quad	&f^{+}_{\mathfrak{n}\hat{\mathfrak{m}}\m}(x)=\,x^{\frac{1}{2}}\sum_{j=\frac{|t_\m^{\mathfrak{n}}-q_\5^{\hat{\mathfrak{m}}}|}{2}}^\infty x^{2j}
-\,x^{\frac{3}{2}}\sum_{j=\frac{|t_\m^{\mathfrak{n}}-q_\5^{\hat{\mathfrak{m}}}|}{2}}^\infty x^{2j}=	f^{-}_{\mathfrak{n}\hat{\mathfrak{m}}\m}(x)&\nn\\
\textit{for $\m=3,4$}\quad	&f^{+}_{\mathfrak{n}\hat{\mathfrak{m}}\m}(x)=\,x^{\frac{1}{2}}\sum_{j=\frac{|t_\m^{\mathfrak{n}}-q_{(n+1)}^{\hat{\mathfrak{m}}}|}{2}}^\infty x^{2j}
-\,x^{\frac{3}{2}}\sum_{j=\frac{|t_\m^{\mathfrak{n}}-q_{n+1}^{\hat{\mathfrak{m}}}|}{2}}^\infty x^{2j}=	f^{-}_{\mathfrak{n}\hat{\mathfrak{m}}\m}(x)&\nn\\
\textit{for $\m=1,\cdots ,4$}\quad	&f^{\mathrm{adj}}_{\mathfrak{n}{\mathfrak{m}}\m }(x)=(-1+\delta_{t_{\mathfrak{m}}t_{\mathfrak{n}}})\,x^{|t^\mathfrak{m}_\m-t^\mathfrak{n}_\m|}\;.&
\end{align}
where $t_\m$'s are magnetic charges for $\m=1,\cdots, 4$. We write the final expression of index for $\hat{E}_6$ in section \ref{sec_e6_index}.
\smallskip

The rest of the paper is organized as follows. In section \ref{action quivers} we describe the general structure of $\NN=3$ $\widehat{ADE}$ quiver theories. In section \ref{superconformal_index} we define the SCF index in these class of theories followed by $Q$ exact deformation in section \ref{sec_localization}. Then in section \ref{sec_index_computation} we explicitly compute the index for $\widehat{A}$-type quiver. The indices for $\widehat{D}$ and $\widehat{E}_6$ quivers are computed in section \ref{sec_dn_index} and section \ref{sec_e6_index} respectively following which in section \ref{sec:large_N} we perform the large $N$ analysis of the indices computed. We present a consistency check of our results in appendix \ref{app:consistency_check}. 
Appendix \ref{app:a3} and \ref{app:d4} contains explicit computation of the index integral in the large $N$ limit for $\widehat{A}_3$ and $\widehat{D}_4$ quiver.

\section{Action of $\NN=3$ Chern-Simons quiver theories}\label{action quivers}
We start by briefly describing the $\NN=3$ SCF CS theory \cite{charge_ade} with the following superspace  action,
\be
\label{eqn:totalaction}
\Action=\Action_{\mathrm{CS}}+\Action_{\mathrm{mat}}+\Action_{\mathrm{pot}}\;.
\ee
The $\NN=3$ multiplet contains a vector superfield $\VV(\sigma, A_\mu,\chi_\sigma)$ and a chiral superfield $\Phi(\phi,\chi_\phi)$ in the adjoint representation of the gauge group and in the matter part one has two chiral superfields $\ZZ(Z,\zeta)$  and $ \WW(W,\omega)$ in the bi-fundamental representation under the gauge groups. Here $\sigma, A_\mu, \chi_\sigma, \chi_\phi, \phi, Z, W, \zeta, \omega$ are component fields. $\NN=1$ superfields are used  to compute the action. 
\begin{table}
	\centering
	\begin{tabular}{ | m{2cm} | m{2cm} |  m{2cm}|  m{2cm}|   } 
		\hline
		Fields & $h$ & $j_3$   & $\epsilon=R$     \\ 
		\hline 
		$\phi_\m$         & $1$ & $0$ &$1$  \\ \vspace{.2cm}
		$\phi^\g_\m$         & $-1$ & $0$ &$1$    \\ \vspace{.2cm}
		$\chi_{\phi\m}$         & $1$ & $\pm \frac{1}{2}$ &$\frac{3}{2}$   \\ \vspace{.2cm}
		$\chi_{\phi\m}^\g$         & $-1$ & $\pm \frac{1}{2}$ &$\frac{3}{2}$  \\ 
		\hline\vspace{.3cm}
		$A_{\mu\m}$         & $0$ & $(-1,0,1)$ &$1$  \\ \vspace{.2cm}
		$\sigma_\m$         & $0$ & $0$ &$1$ \\ \vspace{.2cm}
		$\chi_{\sigma\m}$         & $0$ & $\pm \frac{1}{2}$ &$\frac{3}{2}$   \\ \vspace{.2cm}
		$\chi_{\sigma\m}^\g $         & $0$ & $\pm \frac{1}{2}$ &$\frac{3}{2}$    \\  \hline \vspace{.2cm}
		$Z_\m$ & $\frac{1}{2}$  & $0$  &$\frac{1}{2}$ \\ \vspace{.2cm}
		$Z^{\g}_\m$ & $-\frac{1}{2}$  & $0$  &$\frac{1}{2}$ \\  \vspace{.2cm}
		$\zeta_\m$ & $\frac{1}{2}$  & $\pm \frac{1}{2}$  &$1$ \\  \vspace{.2cm}
		$\zeta^\g_\m$ & $-\frac{1}{2}$  & $\pm \frac{1}{2}$  &$1$ \\ \hline \vspace{.2cm}
		$W_\m$ & $\frac{1}{2}$  & $0$  &$\frac{1}{2}$ \\ \vspace{.2cm}
		$W^{\g }_\m$ & $-\frac{1}{2}$  & $0$ &$\frac{1}{2}$ \\  \vspace{.2cm}
		$\omega_\m$ & $\frac{1}{2}$  & $\pm \frac{1}{2}$ &$1$ \\ 
		\vspace{.2cm}
		$\omega^\g_\m$ & $-\frac{1}{2}$  & $\pm \frac{1}{2}$ &$1$ \\ \hline
		\vspace{.2cm}
		$(Q_1+iQ_2)_\pm$ & $1$  & $\pm \frac{1}{2}$ &$\frac{1}{2}$ \\ \vspace{.2cm}
		$S_\pm$ & $-1$  & $\mp \frac{1}{2}$ &$-\frac{1}{2}$ \\ 
		\hline
	\end{tabular}
	\caption{Charges of component fields and supercharges}
	\label{tab:charges}
\end{table}
We write the component action of $\widehat{A} $-type quiver or necklace quiver with gauge group $\prod_{\m=1}^r U(N)_\m$ below, details of the computation can be found in  \cite{charge_ade}. The other quivers are explained in latter sections.
The CS term  on $\RR^{1,2}$ is,
\begin{eqnarray}
	\label{eqn:cs}
	\mathcal{S}_\mathrm{CS}&=&\int d^3x\sum_{\m=1}^n \Tr\Bigg(-2\kappa_\m \sigma_\m D_\m+\kappa_\m\epsilon^{\mu\nu\lambda}({A_\mu}_\m\partial_\nu {A_\lambda}_\m+\frac{2i}{3}{A_\mu}_\m {A_\nu}_\m {A_\lambda}_\m)\nonumber\\&+&\frac{i{\kappa}_\m}{2}{\chi_\sigma}_\m{\chi^\dagger_\sigma}_\m+\frac{i\kappa_\m}{2}{\chi^\dagger_\sigma}_\m{\chi_\sigma}_\m\Bigg)
\end{eqnarray}
where, $\kappa_\m=\frac{k_\m}{4\pi}$.
The matter part of the action on $\RR^{1,2}$ is,
\be
\mathcal{S}_{\mathrm{mat}}&=&\int d^3x \sum_{\m=1}^{n}\Tr\Bigg(-(\mathcal{D}_m Z_\m)(\mathcal{D}^m Z^\dagger_\m)+i\zeta^\dagger_\m\slashed{\mathcal{D}} \zeta_\m -Z^\dagger_\m Z_\m{D}_\p +Z^\dagger_\m D_\m Z_\m\nn\\
& -& iZ^\dagger_\m(\zeta_\m {\chi_\sigma}_\p-{\chi_\sigma}_\m \zeta_\m)- i\zeta^\dagger_\m(Z_\m{\chi^\dagger_\sigma}_\p -{\chi^\dagger_\sigma}_\m Z_\m) - i \zeta^\dagger_\m(\sigma_\m\zeta_\m\nn\\&-&\zeta_\m{\sigma}_\p)- Z^\dagger_\m Z_\m{\sigma}_\p^2 +  2 Z^\dagger_\m\sigma_\m Z_\m{\sigma}_\p -Z^\dagger_\m \sigma^2_\m Z_\m+F_\m^\dagger F_\m\nn\\
& - &(\mathcal{D}_m W_\m)(\mathcal{D}^m W^\dagger_\m)+i\omega_\m\slashed{\mathcal{D}} \omega^\g_\m -W^\g_\m W_\m D_\m +W^\g_\m {D}_{{(j+1)}} W_\m \nonumber\\&-&iW^\g_\m(\omega_\m {\chi_\sigma}_\m-{\chi_\sigma}_\p \omega_\m) - i\omega^\g_\m (W_\m{\chi^\g_\sigma}_\m -{\chi^\g_\sigma}_\p W_\m)+ i \omega^\g_\m\Big(\omega_\m\sigma_\m\nn\\
&-&\sigma_\p\omega_\m\Big)- W^\g_\m W_\m\sigma^2_\m +  2 W^\g_\m{\sigma}_\p W_\m{\sigma}_\m -W^\g_\m {\sigma}^2_\p W_\m +G^\g_\m G_\m\Bigg)\nn\\
\end{eqnarray}
and the superpotential is,
\begin{eqnarray}
	\label{eqn:pot}
	\mathcal{S}_\mathrm{pot}&=&\int d^3x\sum_{\m=1}^n\Tr\Bigg( \phi_\m Z_\m G_\m-\phi_\m\zeta_\m\omega_\m+\phi_\m F_\m W_\m-{\chi_\phi}_\m Z_\m\omega_\m\nn\\
	&-&{\chi_\phi}_\m\zeta_\m W_\m + {F_\phi}_\m Z_\m W_\m
	-{\phi}_\m W_{\n} F_\n+\phi_\m\omega_\n\zeta_\n
	\nn\\
	&-&\phi_\m G_\n Z_\n+{\chi_\phi}_\m \omega_\n Z_\n+{\chi_\phi}_\m W_\n\zeta_\n - {F_\phi}_\m W_\n Z_\n \nn \\
	&+&\phi^\g_\m W^\g_\m F^\g_\m+ \phi^\g_\m\omega^\g_\m \zeta^\g_\m+ \phi^\g_\m G^\g_\m Z^\g_\m+{\chi^\g_\phi}_\m  W^\g_\m \zeta^\g_\m+{\chi^\g_\phi}_\m\omega^\g_\m Z^\g_\m\nn\\
	&+&{F^\g_\phi}_\m W^\g_\m Z^\g_\m-{\phi}^\g_\m Z^\g_\n G^\g_\n 
	-{\phi}^\g_\m\omega^\g_\n\zeta^\g_\n -{\phi}^\g_\m F^\g_\n W^\g_\n\nn\\
	&-&{\chi_\phi}^\g_\m Z^\g_\n \omega^\g_\m
	-{\chi_\phi}^\g_\m \zeta^\g_\n W^\g_\n
	-{F^\g_\phi}_\m Z^\g_\n W^\g_\n \nonumber\\
	&+&\frac{\kappa_\m}{2} \Big(2\phi_\m {F_\phi}_\m-{\chi_\phi}_\m{\chi_\phi}_\m+2 \phi^\g_\m {F^\g_\phi}_\m +{\chi^\g_\phi}_\m {\chi^\g_\phi}_\m  \Big)\Bigg)\;.
\end{eqnarray}
We observe that $\sigma, \phi, \chi_\sigma, \chi_\phi$ are all auxiliary fields. For our computation we translate the above action on $S^2\times S^1$ whose details are presented in appendix A of \cite{Kim:2009wb}. This R-symmetry group of this theory is $SU(2)_R$ and the supercharges are,
\be
Q_{a\alpha}^{~b}= Q_i(\sigma_i)_a^{~b}
\ee
where, $i=1,2,3$ is $SO(3)$ vector index, $a,b=1,2$ are $SU(2)_R$ indices, $\alpha=\pm $ is  spinor index and $(\sigma_i)_a^{~b}$ are Pauli matrices.
In this paper we compute the superconformal index associated with $(Q_1+iQ_2)_-$ and its conjugate.
Charges of component fields under various global symmetries are summarized in table \ref{tab:charges}. 

\section{The superconformal index}\label{superconformal_index}
The superconformal algebra of $\NN=3$ theory is $Osp(3|4)$ whose bosonic subalgebra is $SO(3)\times SO(3,2)$. Therefore we have one Cartan $h$ from $SO(3)$ R-symmetry group. The Cartans of $SO(2)\times SO(3)\subset SO(3,2)$ are energy $\epsilon$ and angular momentum $j_3$ respectively.
We pick the charge $(Q_1+iQ_2)_-\equiv Q$ and its conjugate $S$ to define the index as following,
\be
\label{eqn:definition_index}
I=\Tr\Bigg( (-1)^F e^{-\beta^\prime\{Q,S  \}}e^{-\beta(\epsilon+j_3)}\Bigg)\;.
\ee
 The algebra of $Q$ and $S$ is as follows,
 \be
 \{Q,S\}=\epsilon-j_3-h,\quad Q^2=S^2=0\;.
 \ee
From the table of charges \eqref{tab:charges} we observe that only one combination of the generators $\epsilon+j_3$ commutes with $Q$ and $S$ since they have non zero $h$ value. If we compare our superconfromal index to that of ABJM\cite{Kim:2009wb} we find that they agree by setting $h_1=h_2=0$ in the later. This is obvious since we have a smaller R symmetry group $SO(3)$ which is a subgroup of $SO(6)$. The index $I$ does not depend on the parameter $\beta^\prime$ as only those sates for which $\{Q,S\}=0$ contributes to it.
\smallskip 

Since the index is related to the partition function of the theory on $S^2\times S^1$, it is evaluated by doing a path integral on $S^2\times S^1$ with the fields obeying the following  boundary condition along $S^1$, 
\be
\label{eqn:boundary condition}
\mathcal{X}_{\mathbbm{R}^3}(r=\beta+\beta^\prime) = e^{-\mathrm{dim}{(\mathcal{X})}(\beta+\beta^\prime)}\mathcal{X}_{\mathbbm{R}^3}(r=1)
\ee
where $\mathcal{X}$ is a generic field on $\RR^3$ and $\beta+\beta^\prime$ is the radius of $S^1$.  The insertion in the path integral
\be 
e^{-\beta^\prime(\epsilon-j_3-h) }e^{-\beta(\epsilon+j_3)}=\exp\Big[-\epsilon(\beta+\beta^\prime)-j_3(\beta-\beta^\prime)+h\beta^\prime\Big]
\ee 
twists the boundary condition which is undone by treating them as background gauge fields and replacing the derivative in the action as follows,
\be
\partial_\tau\rightarrow\partial_\tau-\frac{\beta-\beta^\prime}{\beta+\beta^\prime}\, j_3+ \frac{\beta^\prime}{\beta+\beta^\prime}\, h
\ee
where $\tau=$ is the Euclidean time in radial quantization.
Now we treat all the fields to be periodic in $\tau\sim \tau+(\beta+\beta^\prime)$.

\section{Localization}\label{sec_localization}
In this section we set up  the localization problem. The first step is to add a Q-exact term to the action which preserves some supersymmetry. The Q-exact terms in this case are,
the Maxwell kinetic term,
\be
\Action_{\mathrm{YM}} = \frac{1}{4g^2} \int d^3x\, d^2\theta\,  \sum_{\m=1}^n \Tr\Bigsbrk{ \mathcal{U}^{\alpha}_\m \mathcal{U}_{\alpha\m}}  \; 
\ee
where, $g$ is a coupling of mass dimension $\frac{1}{2}$   and ${\mathcal{U}_{\alpha}}_\m = \frac{1}{4} \bar{D}^2 e^{\VV_\m} D_\alpha e^{-\VV_\m}$ is the super field strength and kinetic terms of the adjoint chiral multiplet,
\be
\Action_{\mathrm{adj}} = \frac{1}{g^2} \int d^3x\,d^4\theta\: \sum_{\m=1}^n\Tr \Bigsbrk{
	- \bar{\Phi}_\m e^{-\VV_\m} \Phi_\m e^{\VV_\m}
} \,\,.
\ee
which is needed in order to preserve $\NN=3$ supersymmetry. We write the component action of the above Q-exact deformation directly in $\RR^3$ (details of radial quantization can be found in \cite{charge_ade}) below,
\be\label{eqn:ym}
	\mathcal{S}_{YM}&=&\sum_{\m=1}^n \int d^3x \Tr\Big(+\frac{1}{2g^2}{F_{\mu\nu}}_\m F^{\mu\nu}_\m - \frac{i}{g^2}{\chi_\sigma}_\m\slashed{\mathcal{D}}{\chi_\sigma^\g}_\m + \frac{1}{g^2}\mathcal{D}_m\sigma_\m\, \mathcal{D}^m\sigma_\m -\frac{1}{g^2}D_\m^2\nn\\
&-&\frac{i}{g^2}{\chi_\sigma}_\m[\sigma_\m,{\chi^{\dagger}_{\sigma}}_\m]\Big)\nn\\
	&=&\sum_{\m=1}^n\frac{1}{g^2}\int d^3x \Tr\Big( \big((\star F_{\m})_\mu-\mathcal{D}_\mu\sigma_\m\big)^2  -{\chi_\sigma}_\m\slashed{\mathcal{D}}{\chi_\sigma^\g}_\m  -D_\m^2-{\chi_\sigma}_\m[\sigma_\m,{\chi^{\dagger}_{\sigma}}_\m]\Big)\nn\\
\ee
\begin{eqnarray}
	\label{eqn:adj}
\hspace*{-1.5cm}	\mathcal{S}_{\mathrm{adj}}&=&\sum_{\m=1}^n\frac{1}{g^2}\int d^3x \Tr\Bigg (\mathcal{D}_m \phi^\g_\m\,\mathcal{D}^m \phi_\m)-{\chi_\phi}_\m \slashed{\mathcal{D}}{\chi_\phi^\g}_\m +\phi^\g_\m[{\chi_\sigma}_\m,{\chi_\phi}_\m] -{\chi_\phi^\g}_\m[\sigma_\m,{\chi_\phi}_\m]\nonumber
	\\&-&i{\chi_\phi^\g}_\m[{\chi_\sigma^\g}_\m,\phi_\m] -[\sigma_\m, \phi_\m][\sigma_\m,\phi^\g_\m] +\phi^\g_\m \phi_\m D_\m - \phi^\g_\m  D_\m\phi_\m-{F_\phi^\g}_\m {F_\phi}_\m\Bigg)\;.\nn\\
\end{eqnarray}

This is a good point to explain the difference between the deformations used in \cite{Kim:2009wb} and in \cite{bkk}(which is what we use). The deformation  in eqn.(2.17) of \cite{Kim:2009wb} is a Q-exact deformation as required to carry out supersymmetric localization. In \cite{Kim:2009wb} only one dynamical scalar  $\sigma$ is turned on through $\mathcal{S}_{\rm YM}$, while have turned on three dynamical scalars $\sigma,\phi, \phi^\g$  through the deformation term $\mathcal{S}_{\rm adj}$. This is simply because we have not eliminated the IR auxiliary fields $\phi, \phi^\g$ in the Lagrangian.
It is necessary to add $\mathcal{S}_{\rm adj}$  to preserve $\NN=3$ supersymmetry since the $Q$-exact deformation has to preserve some amount of supersymmetry in order to carry out the localization procedure. However, if one eliminates the auxiliary adjoint chiral multiplet $\Phi$ in the IR Lagrangian then we do not need to add $\mathcal{S}_{\mathrm{adj}}$ and consequently one would obtain same results as in \eqref{final_index_an} etc of our paper by using the same deformation used in \cite{Kim:2009wb}, because the saddle point equations \eqref{monopole_solution} (2.21, 2.22 in \cite{Kim:2009wb}) will remain unchanged. The saddle point equations are the ones that affects the superconformal index.

Also the deformation we use is a time dependent one since we rescale the coupling parameter $g$ by \cite{charge_ade},
\be 
g_{\RR^3}\longrightarrow e^{-\tau/2}g_{\RR\times S^2}
\ee 
while going from $\RR^3$ to $\RR\times S^2$ via a conformal transformation. This transformation relates the strong and weak coupling regime of the theory by $\tau\to +\infty$ and $\tau\to -\infty$ respectively. By this deformation one breaks the dilatational symmetry while in \cite{Kim:2009wb} the translational symmetry is broken by adding the factor of $r$ in the deformation $\Action_{\mathrm{YM}} = \frac{1}{4g^2} \int d^3x\, d^2\theta\,\, r\Tr\Bigsbrk{ \mathcal{U}^{\alpha} \mathcal{U}_{\alpha}} $. 
The deformation we have  used is very similar to \cite{Kim:2009wb} as they add  Maxwell term to the action which can be seen by comparing \eqref{eqn:ym}  and eqn.(2.18) of \cite{Kim:2009wb}. The only difference is the presence of $r$ in the deformation of \cite{Kim:2009wb}. The factor of $r$ is absorbed in the coupling $g^2$ in our case while we perform radial quantization. Hence this does not affect the saddle point equation therefore the final result of the index. 

 The simplest way to understand the deformation we used in this paper is as follows:\\
(i) We start with an $\NN=3$ superconformal CS matter theory constructed in \cite{jaff}. \\
(ii) We deform the theory by preserving $\NN=3$ supersymmetry following \cite{bkk}, which is done by the deformations in eqn.\eqref{eqn:ym} and eqn.\eqref{eqn:adj}. 

A special case of the undeformed theory constructed in \cite{jaff} preserves $\NN=6$ supersymmetry for two gauge groups. Hence the deformation we use does not preserve $\NN=6$ supersymmetry.

Coming back to the localization method,  the full action of the theory is,
\be 
e^{\mathcal{S}}\to e^{\mathcal{S}+t(\mathcal{S}_{YM}+\mathcal{S}_{\mathrm{adj}})}
\ee
where $t=\frac{1}{g^2}$ is a continuous parameter. The next step of localization method is to take $t\to \infty$ which sets,
\be 
\label{eqn:localization_manifold}
\mathcal{S}_{YM}+\mathcal{S}_{\mathrm{adj}}=0\;.
\ee 
The field configurations satisfying the above equation is the localization manifold and we perform the path integral only on those fields configuration instead of the infinite dimensional manifold as in the original theory. We wish to find a bosonic solution to  \eqref{eqn:localization_manifold} therefore we set all the fermions to zero clasically.
Eliminating the auxiliary fields $D_\m$ and $F_{\phi\m}$  by using,
\be
D_\m=\frac{1}{2}\Big( \phi^\g_\m \phi_\m - \phi_\m \phi^\g_\m\Big),\qquad\qquad F_{\phi\m}=0
\ee
we find,
\begin{eqnarray}
	\mathcal{S}_{YM}+
	\mathcal{S}_{\mathrm{adj}}&=& \frac{1}{g^2}\int d^3x \Tr\Bigg[ \frac{1}{2} \Big((*F_\m)_{\mu}  - \mathcal{D}_\mu\sigma_\m\Big)^2 +\frac{1}{4}\Big( \phi^\g_\m \phi_\m - \phi_\m \phi^\g_\m\Big)^2\nn\\[2mm]
	&+& \lvert \mathcal{D}^m \phi_\m\rvert ^2+ \lvert [\sigma_\m, \phi_\m]\rvert^2 \Bigg]\;.
\end{eqnarray}
Saddle point equations are,
\be
\label{eqn:saddlepoints}
\Big(\star F_\m\Big)_{\mu}=\mathcal{D}_\mu\sigma_\m, \quad D_\m=0,  \quad \phi_\m=0,\quad F_{\phi\m}=0\;.
\ee
We can choose Dirac monopole solution for the gauge fields and turn on adjoint scalar $\sigma_\m$ 
\begin{eqnarray}
	\label{monopole_solution}
	A_{(j)} = \frac{H_\m}{2r} (\pm 1 - \cos \theta)\,d\varphi, \quad \sigma_{(j)} = - \frac{H_\m}{2r}
\end{eqnarray}
which satisfies the above equation.   And $H_\m=\mathrm{diag}(q^{\scaleto{1}{4pt}}_\m,q^{\scaleto{2}{4pt}}_\m, \cdots, q^{\scaleto{N}{4pt}}_\m)$ and $q_i\in \mathbbm{Z}$ are the magnetic charges.
The  supersymmetry variations of the adjoint fermions,
\begin{align}
\delta\chi_{\phi\m}    &=\frac{(i-1)}{2\sqrt{2}}\Ev_3\epsilon^{\mu\nu\lambda}F_{\mu\nu\m}\gamma_\lambda-\frac{(i+1)}{\sqrt{2}}\E\slashed{\mathcal{D}}\sigma_\m  + \frac{(i+1)}{\sqrt{2}}\Ep\slashed{\mathcal{D}}\phi_\m &\nn\\&-\frac{(i+1)}{\sqrt{2}}\Ep[\sigma_\m,\phi_\m]-\frac{(i+1)}{\sqrt{2}}\E D_\m+\frac{(i+1)}{\sqrt{2}} F_{\phi\m}&\nonumber\\ \nn\\
\delta\chi_{\sigma\m}    &=\frac{(1+i)}{2\sqrt{2}}\Em\epsilon^{\mu\nu\lambda}F_{\mu\nu\m}\gamma_\lambda +\frac{(i-1)}{\sqrt{2}}\Em\slashed{\mathcal{D}}\sigma  + \frac{(i-1)}{\sqrt{2}}\E\slashed{\mathcal{D}}\phi +\frac{(i-1)}{\sqrt{2}}\E[\sigma,\phi]&\nonumber\\-&\frac{(i-1)}{\sqrt{2}}\Em   D-\frac{(i-1)}{\sqrt{2}} \E F_{\phi\m} &
\end{align}
vanish for the solution obtained in eqn.\eqref{monopole_solution}. To make the supersymmetry variations of the  matter fermions zero,
\begin{align}
	\delta\zeta_\m &=\frac{(1+i)}{\sqrt{2}}\E(\phi^\dagger_\m Z_\m + Z_\m {\phi}^\dagger_\p)+\frac{(1+i)}{\sqrt{2}}\Ep(\slashed{\mathcal{D}} Z_\m+\sigma Z_\m -Z\hat{\sigma})\nonumber\\&+\frac{(1+i)}{\sqrt{2}}\Em(\phi^\dagger W^\dagger + W^\dagger\hat{\phi}^\dagger)-\frac{(1+i)}{\sqrt{2}}\E(\slashed{\mathcal{D}} W+\sigma W^\dagger- W^\dagger\hat{\sigma})\nn\\ \nn\\
	\delta\omega_\m &=\frac{(1+i)}{\sqrt{2}}\E(\slashed{\mathcal{D}} Z^\dagger+\hat{\sigma} Z^\dagger -Z^\dagger\sigma)+\frac{(1+i)}{\sqrt{2}}\Ep(\slashed{\mathcal{D}} W+\hat{\sigma} W- W{\sigma})\nonumber\\&+\frac{(1+i)}{\sqrt{2}}\Em(\hat{\phi}^\dagger Z^\dagger + Z^\dagger{\phi}^\dagger)-\frac{(1+i)}{\sqrt{2}}\E(\hat{\phi}^\dagger W + W{\phi}^\dagger)
\end{align}
 we choose $Z_\m=W_\m=0$ as this is the only solution which satisfies the boundary condition  eqn.\eqref{eqn:boundary condition} on $S^2\times S^1$. 
In eqn.\eqref{monopole_solution} we have $A_{r\m}=0$. But one can turn on these modes in the following way,
\be 
A_{r\m}=\frac{1}{(\beta+\beta^\prime)r}\, \rm diag(\alpha_\m^1, \alpha_\m^2, \cdots, \alpha_\m^N)
\ee 
which is still compatible with the saddle point solutions. The factor of $\beta+\beta^\prime$ in the denominator is kept for  convenience in the computation. The entries $\alpha_\m$'s satisfies the periodicity condition,
\be 
\alpha_\m^{\mathfrak{{m}}}\sim \alpha_\m^{\mathfrak{{m}}}+2\pi,\quad (\mathfrak{{m}}=1,2,\cdots, N)
\ee 
along $S^1$ as needed for the gauge invariance under large gauge transformations.

\section{Index computation}\label{sec_index_computation}
Schematically the index looks like the following,
\be
\label{eqn:schematic_index}
\textit{ classical contribution}.\;\;\textit{Faddeev-Popov measure}.\;\;  \frac{\det_{ \chi_\sigma} \det_{ \chi_\phi} \det_\zeta \det_\omega }{\det_{A_\mu}\det_{\sigma,\phi}\det_{Z,W} }.\; \textit{zero point energy}\nn\\
\ee
The classical contribution comes from only CS part of the action in eqn.\eqref{eqn:cs}, which is,
\be 
e^{i\sum_{\m=1}^n k_\m\sum_{\mathfrak{m}=1}^N q_\m^\mathfrak{m}\alpha_\m^\mathfrak{{m}}\!}\;.
\ee 
The Faddeev-Popov measure is computed while we handle the gauge fields.
Now let us move on to the one loop determinant calculations in the $g\to 0$ limit. 
\subsection{One loop determinant of matter  scalars}
\par We first determine the one loop determinants of the matter scalar fields $Z_\m, W_\m, Z^\g_\m, W^\g_\m$. The quadratic part of the action for these fields on $S^2 \times S^1$ is,
\begin{eqnarray}
\mathcal{S}^E(Z,W, Z^\g, W^\g) &=&\int d\tau\, d\Omega\, \sum_{\m=1}^n \mathrm{tr} \Big(
-  Z^\g_\m \mathcal{D}_m\mathcal{D}^m Z_\m-   W^\g_\m \mathcal{D}^m \mathcal{D}_m W_\m+\frac{1}{4}Z^\g_\m Z_\m+\frac{1}{4}W_\m W^\g_\m \nn\\
&+&  Z^\g_\m \sigma^2_\m Z_\m+W^\g_\m W_\m\sigma^2_\m + Z^\g_\m Z_\m \sigma^2_\p+W_\m W^\g_\m \sigma^2_\p\nn\\
&-&2Z^\g_\m \sigma_\m Z_\m \sigma_\p-2W_\m \sigma_\m W^\g_\m \sigma_\p \Big)\;.
\end{eqnarray}
Plugging in the background fields from eqn.\eqref{monopole_solution} in the  above, we obtain,
\begin{align}
&\mathcal{S}^E(Z,W, Z^\g, W^\g) =\int d\tau\, d\Omega\,\sum_{\hat{\mathfrak{m}},\mathfrak{n}=1}^N \sum_{\m=1}^n  \Big(
- Z^{\g }_{\m \hat{\mathfrak{m}}\mathfrak{n}} \mathcal{D}_m\mathcal{D}^m Z_{\m \mathfrak{n}\hat{\mathfrak{m}}}-   W^{\g}_{\m \mathfrak{n}\hat{\mathfrak{m}}} \mathcal{D}^m \mathcal{D}_m W_{\m \hat{\mathfrak{m}}\mathfrak{n}}&\nn\\&+\frac{1}{4}Z^{\g }_{\m \hat{\mathfrak{m}}\mathfrak{n}} Z_{\m \mathfrak{n}\hat{\mathfrak{m}}}
+\frac{1}{4}W_{\m \hat{\mathfrak{m}}\mathfrak{n}} W^{\g}_{\m \mathfrak{n}\hat{\mathfrak{m}}} +\frac{1}{4}  Z^{\g jm}_{\m \hat{\mathfrak{m}}\mathfrak{n}} (q_\m^{\mathfrak{n}} - q_\p^{\hat{\mathfrak{m}}})^2   Z_{\m \mathfrak{n}\hat{\mathfrak{m}}}+\frac{1}{4} W^{\g}_{\m \mathfrak{n}\hat{\mathfrak{m}}} (q_\m^{\mathfrak{n}}-q_\p^{\hat{\mathfrak{m}}})^2\, W_{\m \hat{\mathfrak{m}}\mathfrak{n}}\Big)&\nn\\
\end{align}
where we have  written the trace in terms of matrix components. Each of these fields are function of $\tau, \Omega$, where $\Omega$ denotes $S^2$ co-ordinates. We expand all the fields as,
\be
\label{mode_expansion_scalar}
\Phi_{\mathfrak{m}\mathfrak{\hat{n}}}(\tau,\Omega)=\sum_{jm}\Phi_{\mathfrak{m}\mathfrak{\hat{n}}}(\tau)\, Y_{qjm}(\Omega)\;.
\ee
Now this reduces to  a problem of a free particle on $S^2$ in the presence of a magnetic monopole of charge $(q_\m^{\mathfrak{n}}-q_\p^{\hat{\mathfrak{m}}})\equiv q$ at the center of the sphere. The monopole spherecial harmonics $Y_{qjm}$ satisfies,
\be
- \mathcal{D}_m\mathcal{D}^m\, Y_{qjm}=\Big(j(j+1)-\frac{1}{4}q^2\Big)\, Y_{qjm}
\ee 
where the quantum numbers are as follows,
\be
j=\frac{q_\m^{\mathfrak{n}}- q_\p^{\mathfrak{\hat{m}}}}{2}, \frac{(q_\m^{\mathfrak{n}}- q_\p^{\mathfrak{\hat{m}}})}{2}
+1,\cdots,\qquad m=-j,-j+1,\cdots, +j
\ee
Substituting this back into the action we obtain the following,
\begin{eqnarray}
\mathcal{S}^E(Z,W, Z^\g, W^\g)&=&\int d\tau\, d\Omega\,\sum_{\hat{\mathfrak{m}},\mathfrak{n}=1}^N \sum_{\m=1}^n  \Bigg(
Z^{\g jm}_{\m \hat{\mathfrak{m}}\mathfrak{n}}\Bigg( -\mathcal{D}_\tau^2 +( j+\frac{1}{2})^2\Bigg)\, Z^{jm}_{\m \mathfrak{n}\hat{\mathfrak{m}}}\nn\\
& +&W^{\g jm}_{\m \mathfrak{n}\hat{\mathfrak{m}}} \Bigg( -\mathcal{D}_\tau^2 +( j+\frac{1}{2})^2\Bigg)\, W^{jm}_{\m \hat{\mathfrak{m}}\mathfrak{m}}
\end{eqnarray}
with,
\be
\label{eqn:derivative_on_sphere}
\hspace*{-1cm} \mathcal{D}_\tau=\partial_\tau+ i\eta\frac{\alpha_\m^\mathfrak{n}-\alpha_\p^{\mathfrak{\hat{m}}}}{\beta+\beta^\prime} -\frac{\beta-\beta^\prime}{\beta+\beta^\prime}\, j_3+ \frac{\beta^\prime}{\beta+\beta^\prime}\, h,\qquad \textit{ $\eta=+1$  for $Z, W^\g$ and -1 for  $W, Z^\g$}\;.\nn\\
\ee
The one loop effective action of the complex scalars $\Gamma(Z,W, Z^\g, W^\g)$ is,
\be
e^{\Gamma(Z,W, Z^\g, W^\g)}&=&1/
\det  \Big( -\mathcal{D}_\tau^2++( j+\frac{1}{2})^2\Big)\;.
\ee 
Now, following \cite{Aharony:2005bq} we choose an eigenbasis of the operator $\partial_\tau$ whose time dependence is given by $ \prod_{r=-\infty}^\infty e^{2\pi i r \tau/\beta+\beta^\prime}$ and using the identity
	\be 
{\prod_{r=-\infty}^\infty \Bigg(  (2\pi r+\alpha)^2+\Delta^2  \Bigg)=\sin\Big(\frac{(\alpha+i\Delta)}{2}\Big)\, \sin\Big(\frac{(\alpha-i\Delta)}{2}\Big)}
\ee 
we find,
\be 
&&\det  \Big( -\mathcal{D}_\tau^2++( j+\frac{1}{2})^2\Big)\nn\\
&=&\sum_{\hat{\mathfrak{m}},\mathfrak{n}=1}^N\sum_{\m=1}^n \prod_{j,j_3} \sin\Bigg[\frac{1}{2}\Big( \eta(\alpha_\m^\mathfrak{n}-\alpha_\p^{\mathfrak{\hat{m}}})+i\beta ( \epsilon_j+ j_3)+i\beta^\prime(\epsilon_j- j_3-h)\Big) \Bigg]\nn\\
&&\sin\Bigg[\frac{1}{2}\Big( -\eta(\alpha_\m^\mathfrak{n}-\alpha_\p^{\mathfrak{\hat{m}}})+i\beta(\epsilon_j- j_3)+i\beta^\prime (j_3 + \epsilon_j +h)\Big) \Bigg]
\ee
where $\epsilon_j=j+\frac{1}{2}$. Here we have set an overall factor, which is a function of $n, \beta+\beta^\prime$ to unity(See \cite{Aharony:2005bq} for details).    Observe that in the sine factors above all charges except energy $\epsilon_j$ has opposite signs. This is interpreted as contribution from particle and anti-particle. Therefore,
\be
&&\Gamma(Z,W, Z^\g, W^\g)\nn\\
&=&-\sum_{\hat{\mathfrak{m}},\mathfrak{n}=1}^N\sum_{\m=1}^n\log\Bigg(
\prod_{j,j_3}\sin\Bigg[\frac{1}{2}\Big( \eta(\alpha_\m^\mathfrak{n}-\alpha_\p^{\mathfrak{\hat{m}}})+i\beta ( \epsilon_j+ j_3)+i\beta^\prime(\epsilon_j- j_3- h) \Big) \Bigg]\Bigg)\nn\\
&=&-\sum_{\hat{\mathfrak{m}},\mathfrak{n}=1}^N\sum_{\m=1}^n\log
\prod_{j,j_3}\Bigg[ e^{\frac{i}{2}\Big( \eta(\alpha_\m^\mathfrak{n}-\alpha_\p^{\mathfrak{\hat{m}}})+i\beta ( \epsilon_j+ j_3)+i\beta^\prime(\epsilon_j- j_3- h) \Big)} - e^{-\frac{i}{2}\Big( \eta(\alpha_i-\tilde{\alpha}_j)+i\beta ( \epsilon_j+ j_3)+i\beta^\prime(\epsilon_j- j_3- h) \Big)}\Bigg] \nn\\
&=&-
\sum_{\hat{\mathfrak{m}},\mathfrak{n}=1}^N\sum_{\m=1}^n\sum_{j,j_3} \Bigg[\frac{i\eta}{2}(\alpha_\m^\mathfrak{n}-\alpha_\p^{\mathfrak{\hat{m}}})+\frac{\beta}{2} ( \epsilon_j+ j_3)+\frac{1}{2}\beta^\prime(\epsilon_j- j_3-h) \Bigg]\nn\\
&+&\sum_{\m=1}^n\sum_{j}\sum_{\rm matter\; scalars}\sum_{r=1}^\infty \frac{1}{r}\Big[ e^{ -in(\alpha_\m^\mathfrak{n}-\alpha_\p^{\mathfrak{\hat{m}}})} x^{n ( \epsilon_j+ j_3)}(x^\prime)^{n(\epsilon_j- j_3-h)}\Big],\qquad\quad x=e^{-\beta}, x^\prime=e^{-\beta^\prime}\;.\nn\\
\ee
Substituting values of various charges in the last term we find,\\
for, $Z_\m, h=\frac{1}{2}$
\be 
&&x^{ j+\frac{1}{2}+ j_3}(x^\prime)^{j+\frac{1}{2}- j_3-h}
=x^{ \frac{1}{2}}(x^\prime)^{2j}+x^{ \frac{3}{2}}(x^\prime)^{2j- 1}+...+x^{ 2j+\frac{1}{2}}
\ee 
for, $W_\m, h=\frac{1}{2}$
\be 
&&x^{ j+\frac{1}{2}+ j_3}(x^\prime)^{j+\frac{1}{2}- j_3-h}
=x^{ \frac{1}{2}}(x^\prime)^{2j}+x^{ \frac{3}{2}}(x^\prime)^{2j- 1}+...+x^{ 2j+\frac{1}{2}}
\ee 
for, $Z^\g_\m, h=-\frac{1}{2}$
\be 
&&x^{ j+\frac{1}{2}+ j_3}(x^\prime)^{j+\frac{1}{2}- j_3-h} 
=x^{ \frac{1}{2}}(x^\prime)^{2j+1}+x^{ \frac{3}{2}}(x^\prime)^{2j}+...+x^{ 2j+\frac{1}{2}}(x^\prime)
\ee 
for, $W^\g_\m, h=-\frac{1}{2}$
\be 
&&x^{ j+\frac{1}{2}+ j_3}(x^\prime)^{j+\frac{1}{2}- j_3-h} 
=x^{ \frac{1}{2}}(x^\prime)^{2j+1}+x^{ \frac{3}{2}}(x^\prime)^{2j}+...+x^{ 2j+\frac{1}{2}}(x^\prime)\;.
\ee 
We write the final expression as following,
\be
\hspace*{-2.5cm}\Gamma(Z,W, Z^\g, W^\g)
&=&-
\sum_{\hat{\mathfrak{m}},\mathfrak{n}=1}^N\sum_{\m=1}^n\sum_{j,j_3} \Bigg[\frac{i\eta}{2}(\alpha_\m^\mathfrak{n}-\alpha_\p^{\mathfrak{\hat{m}}})+\frac{\beta}{2} ( \epsilon_j+ j_3)+\frac{1}{2}\beta^\prime(\epsilon_j- j_3-h) \Bigg]\nn\\
&+&\sum_{\hat{\mathfrak{m}},\mathfrak{n}=1}^N\sum_{\m=1}^n\sum_{r=1}^\infty \frac{1}{r}\Big[ e^{-ir(\alpha_\m^\mathfrak{n}-\alpha_\p^{\mathfrak{\hat{m}}})}f^{+B}_{\mathfrak{n}\hat{\mathfrak{m}}}(x^r, (x^\prime)^r) +e^{ir(\alpha_\m^\mathfrak{n}-\alpha_\p^{\mathfrak{\hat{m}}})}f^{-B}_{\mathfrak{n}\hat{\mathfrak{m}}}(x^r, (x^\prime)^r)\Big]\nn\\
\ee
where,
\be 
f^{+B}_{\mathfrak{n}\hat{\mathfrak{m}}}(x^r, (x^\prime)^r)&=&\sum_{j=\frac{|q_\m^{\mathfrak{n}}-q_\p^{\hat{\mathfrak{m}}}|}{2}}^\infty\Big(x^{ \frac{1}{2}}(x^\prime)^{2j+1}+x^{ \frac{3}{2}}(x^\prime)^{2j}+...+x^{ 2j+\frac{1}{2}}(x^\prime)\Big)\nn\\
&+&\sum_{j=\frac{|q_\m^{\mathfrak{n}}-q_\p^{\hat{\mathfrak{m}}}|}{2}}^\infty\Big(x^{ \frac{1}{2}}(x^\prime)^{2j}+x^{ \frac{3}{2}}(x^\prime)^{2j- 1}+...+x^{ 2j+\frac{1}{2}}\Big)
\ee 
which is the contribution from $Z_\m$ and $W^\g_\m$ and 
\be 
f_{\mathfrak{n}\hat{\mathfrak{m}}}^{-B}(x^r, (x^\prime)^r)&=&
\sum_{j=\frac{|q_\m^{\mathfrak{n}}-q_\p^{\hat{\mathfrak{m}}}|}{2}}^\infty\Big(x^{ \frac{1}{2}}(x^\prime)^{2j}+x^{ \frac{3}{2}}(x^\prime)^{2j- 1}+...+x^{ 2j+\frac{1}{2}}\Big)\nn\\
&+&\sum_{j=\frac{|q_\m^{\mathfrak{n}}-q_\p^{\hat{\mathfrak{m}}}|}{2}}^\infty\Big(x^{ \frac{1}{2}}(x^\prime)^{2j+1}+x^{ \frac{3}{2}}(x^\prime)^{2j}+...+x^{ 2j+\frac{1}{2}}(x^\prime)\Big)\;,
\ee 
which comes from $Z^\g_\m$ and $W_\m$.

\subsection{One loop determinant of matter fermions}
We introduce the following notation, 
\begin{equation}
	\label{eqn:xi-SU2R}
	\xi^{a}_\m =\left( \begin{array}{cc}
		\omega^\dagger_\m  \, e^{i\pi/4} \\[2mm] \zeta_\m \, e^{-i\pi/4} 
	\end{array}\right) ,\hspace{1cm}
	{\xi^\dagger_{a}}_\m =\left(\begin{array}{cc}  \omega_\m \, e^{-i\pi/4} \\[2mm] \zeta^\dagger_\m \, e^{i\pi/4} 
	\end{array}\right).
\end{equation}
to write down the quadratic terms of the matter fermions in the action in a compact form. After substituting the saddle point solutions the quadratic term of the relevant part of the action is,
\begin{eqnarray}
\mathcal{L}^E_\mathrm{matter} (\xi,\xi^\g)
&=&\sum_{\m=1}^n \sum_{\mathfrak{n},\mathfrak{\hat{m}}=1}^N \Bigg(
-i\xi^\g_{a\m \mathfrak{\hat{m}n}}\slashed{\mathcal{D}}\xi^a_{\m \mathfrak{n\hat{m}}}
- \frac{i}{2} \,\xi^\g_{a\m \mathfrak{\hat{m}n}}q_{\mathfrak{n} \mathfrak{\hat{m}}} \,(\sigma_3)^a_{\,\,b}\,\xi^b_{\m \mathfrak{n\hat{m}}}\Bigg)\,\,.
\end{eqnarray}
To perform the path integral we expand the spinor in monopole spinor harmonics basis, which are basically the eigenvectors of Dirac operator $\slashed{\mathcal{D}}_S$ on $S^2$ in presence of a magnetic monopole, as follows,
\be
\label{monopole-harmonics-basis}
\psi(\tau,\Omega)=\sum_m \psi_m(\tau)\Upsilon^0_{qm}(\Omega)+\sum_{jm\varepsilon}\psi^\varepsilon_{jm}(\tau)\Upsilon^\varepsilon_{qjm}(\Omega)
\ee
where, $\varepsilon=\pm 1 $. The machinery of monopole spinor harmonics and its orthogonality properties  can be found in Appendix C of \cite{bkk} and we briefly summarize the spectrum of the Dirac operator  $\slashed{\mathcal{D}}_S$ below.
\be
\textit{ for $q\neq 0:$}\quad\;
j=\frac{|q|-1}{2}, \frac{|q|+1}{2}, \frac{|q|+3}{2},...\qquad m=-j,-j+1,...,j\,,\nn\\
\textit{for $q= 0:$}\qquad\qquad
j= \frac{|q|+1}{2}, \frac{|q|+3}{2},...\qquad m=-j,-j+1,...,j\,\,.
\ee
Eigenvalue equations of $\slashed{\mathcal{D}}_S $ are,
\be
&&\slashed{\mathcal{D}}_S \Upsilon^0_{qm} =0 \qquad\qquad\qquad\,\, \textit{for}\,\, j=\frac{|q|-1}{2},\, q\neq 0\,,\\
&&\slashed{\mathcal{D}}_S \Upsilon^\pm_{qjm} =i\Delta^\pm_{jq}\Upsilon^\pm_{qjm} \qquad\,\, \textit{for}\,\, j=\frac{|q|+1}{2},\frac{|q|+3}{2},\cdots
\ee
where, $\Delta^{\pm}_{jq}=\pm\frac{1}{2}\sqrt{(2j+1)^2-q^2}$. $\Upsilon^0_{qm}$ is called the zero mode since it has zero eigenvalue.
Performing the path integral we find,
\be 
 e^{-\Gamma(\xi,\xi^\g )}=\int [d\xi][d\xi^\g]\,e^{-\mathcal{S}}
= \det \Bigg(-i \mathbbm{1}\,\mathcal{D}_\tau 
-iq\,\sign(q)\,  \, \sigma_3\Bigg)\, \det \begin{pmatrix} -i\mathcal{D}_\tau&  \Delta^- -iq\sigma_3\\ \Delta^+-iq\sigma_3 & -i\mathcal{D}_\tau \end{pmatrix}\;.\nn\\
\ee 
For the zero modes we have,
\be
\det \Bigg(-i \mathbbm{1}\,\mathcal{D}_\tau 
-iq\,\sign(q)\,  \, \sigma_3\Bigg)=-\mathcal{D}_\tau ^2+q^2
\ee
and for the non-zero modes we get,
\be 
\det \begin{pmatrix} -i\mathcal{D}_\tau & \Delta^--iq\sigma_3\\ \Delta^+-iq\sigma_3 & -i\mathcal{D}_\tau \end{pmatrix} =\Big((j+\frac{1}{2})^2-\mathcal{D}_\tau^2\Big)\;.
\ee 
Therefore the one loop determinant of fermions is,
\be 
\Gamma(\xi,\xi^\g)
&=& \sum_{\hat{\mathfrak{m}},\mathfrak{n}=1}^N\sum_{\m=1}^n\sum_{j,j_3}\frac{1}{2}\Big( i\eta(\alpha_\m^\mathfrak{n}-\alpha_\p^{\mathfrak{\hat{m}}}) +\beta(\epsilon_j+j_3) +\beta^\prime(\epsilon_j- j_3-h) \Big)\nn\\
&+&\sum_{\m=1}^n\sum_{j}\sum_{\rm matter\; fermions}\sum_{r=1}^\infty \frac{1}{r}\Big[ e^{ in\eta(\alpha_\m^\mathfrak{n}-\alpha_\p^{\mathfrak{\hat{m}}})} x^{n ( \epsilon_j+ j_3)}(x^\prime)^{n(\epsilon_j- j_3-h)}\Big]\;.\nn\\
\ee 
Substituting values of various charges in the last term we find,\\
for $\zeta_\m, h=\frac{1}{2}$
\be 
&&x^{j+\frac{1}{2}+j_3} {(x^\prime)}^{j+\frac{1}{2}- j_3-h}=x^{j+\frac{1}{2}+j_3} {(x^\prime)}^{j- j_3}\nn\\
&&=x^{ \frac{1}{2}}(x^\prime)^{2j}+x^{ \frac{3}{2}}(x^\prime)^{2j- 1}+...+x^{ 2j+\frac{1}{2}}
\ee 
for $\omega_\m, h=\frac{1}{2}$
\be 
&&x^{j+\frac{1}{2}+j_3} {(x^\prime)}^{j+\frac{1}{2}- j_3-h}\nn\\
&&=x^{ \frac{1}{2}}(x^\prime)^{2j}+x^{ \frac{3}{2}}(x^\prime)^{2j- 1}+...+x^{ 2j+\frac{1}{2}}
\ee 
for $\zeta^\g_\m, h=-\frac{1}{2}$
\be 
&&x^{j+\frac{1}{2}+j_3} {(x^\prime)}^{j+\frac{1}{2}- j_3-h}=x^{j+\frac{1}{2}+j_3} {(x^\prime)}^{j+1- j_3}\nn\\
&&=x^{ \frac{1}{2}}(x^\prime)^{2j+1}+x^{ \frac{3}{2}}(x^\prime)^{2j}+...+x^{ 2j+\frac{1}{2}}(x^\prime)
\ee 
for $\omega^\g_\m, h=-\frac{1}{2}$
\be 
&&x^{j+\frac{1}{2}+j_3} {(x^\prime)}^{j+\frac{1}{2}- j_3-h}\nn\\
&&=x^{ \frac{1}{2}}(x^\prime)^{2j+1}+x^{ \frac{3}{2}}(x^\prime)^{2j}+...+x^{ 2j+\frac{1}{2}}(x^\prime)\;.
\ee 
The above is the contribution from non zero modes. Now, for the zero modes, i.e $j=\frac{|q|-1}{2}$ we have $m=2j+1=|q|$ unpaired states\footnote{This is explainced nicely in \cite{bkk} with plot of energy spectrum.} with energy $j+\frac{1}{2}$. By unpaired states we mean that there is no states with energy $-(j+\frac{1}{2})$. This can be seen by solving the eigenvalue problem for the zero mode eigen vectors which are,
\be 
(-i\slashed{\mathcal{D}}_S-q\sigma_3)\Upsilon^0_{qm}=-q\sigma_3 \Upsilon^0_{qm}\;.
\ee 
Now using the explicit eigen vectors 
\be 
\Upsilon^0_{qm}=\begin{pmatrix}
	Y_{qm}\\0
\end{pmatrix}\quad q>0,\qquad\qquad \Upsilon^0_{qm}=\begin{pmatrix}
	0\\Y_{qm}
\end{pmatrix}\quad\qquad q<0
\ee 
we find,
\be 
-q\sigma_3 \Upsilon^0_{qm}=q\begin{pmatrix}
	-1&0\\0&1
\end{pmatrix}= -q\Upsilon^0_{qm},\quad q>0\nn\\
= q\Upsilon^0_{qm},\quad q<0\;.
\ee 
We get negative eigenvalue for both the cases.
Therefore we have contribution only from the fields $\zeta^\g,\omega^\g$ which is,
\be 
&&\sum_{j_3=-j}^j\sum_{\zeta^\g,\omega^\g} x^{j+\frac{1}{2}+j_3} {(x^\prime)}^{j+1- j_3}\nn\\ [3mm]
&=&2\Big(x^{ \frac{1}{2}}(x^\prime)^{2j+1}+x^{ \frac{3}{2}}(x^\prime)^{2j}+...+x^{ 2j+\frac{1}{2}}(x^\prime)\Big)|_{j=\frac{|q_\m^{\mathfrak{n}}-q_\p^{\hat{\mathfrak{m}}}|-1}{2}}\;.
\ee 
Combining the zero mode and the non zero modes we write the final answer as, 
\begin{align}
&\Gamma(\xi,\xi^\g)
= \sum_{\hat{\mathfrak{m}},\mathfrak{n}=1}^N\sum_{\m=1}^n\sum_{j,j_3}\frac{1}{2}\Big( i\eta(\alpha_\m^\mathfrak{n}-\alpha_\p^{\mathfrak{\hat{m}}}) +\beta(\epsilon_j+j_3) +\beta^\prime(\epsilon_j- j_3-h) \Big)&\nn\\
&+\sum_{\hat{\mathfrak{m}},\mathfrak{n}=1}^N\sum_{\m=1}^n\sum_{r=1}^\infty \frac{1}{r}\Big[ e^{-ir(\alpha_\m^\mathfrak{n}-\alpha_\p^{\mathfrak{\hat{m}}})}f^{+F}_{\mathfrak{n}\hat{\mathfrak{m}}}(x^r, (x^\prime)^r) +e^{ir(\alpha_\m^\mathfrak{n}-\alpha_\p^{\mathfrak{\hat{m}}})}f^{-F}_{\mathfrak{n}\hat{\mathfrak{m}}}(x^r, (x^\prime)^r)\Big]&
\end{align} 
where,
\be 
f^{+F}_{\mathfrak{n}\hat{\mathfrak{m}}}&=&\,\sum_{j=\frac{|q_\m^{\mathfrak{n}}-q_\p^{\hat{\mathfrak{m}}}|+1}{2}}^\infty\Big(x^{ \frac{1}{2}}(x^\prime)^{2j}+x^{ \frac{3}{2}}(x^\prime)^{2j- 1}+...+x^{ 2j+\frac{1}{2}}\Big)\nn\\
&+& \,\sum_{j=\frac{|q_\m^{\mathfrak{n}}-q_\p^{\hat{\mathfrak{m}}}|-1}{2}}^\infty\Big(x^{ \frac{1}{2}}(x^\prime)^{2j+1}+x^{ \frac{3}{2}}(x^\prime)^{2j}+...+x^{ 2j+\frac{1}{2}}(x^\prime)\Big)
\ee
which is the contribution from $\zeta_\m$ and $\omega^\g_\m$ and 
 \be 
f^{-F}_{\mathfrak{n}\hat{\mathfrak{m}}}&=&\,\sum_{j=\frac{|q_\m^{\mathfrak{n}}-q_\p^{\hat{\mathfrak{m}}}|+1}{2}}^\infty\Big(x^{ \frac{1}{2}}(x^\prime)^{2j}+x^{ \frac{3}{2}}(x^\prime)^{2j- 1}+...+x^{ 2j+\frac{1}{2}}\Big)\nn\\
&+& \,\sum_{j=\frac{|q_\m^{\mathfrak{n}}-q_\p^{\hat{\mathfrak{m}}}|-1}{2}}^\infty\Big(x^{ \frac{1}{2}}(x^\prime)^{2j+1}+x^{ \frac{3}{2}}(x^\prime)^{2j}+...+x^{ 2j+\frac{1}{2}}(x^\prime)\Big)\;.
\ee 
which is the contribution from $\zeta^\g_\m$ and $\omega_\m$.

\subsection{Final result of matter sector}

Now, adding contributions from matter scalars and fermions we find,
\begin{align}
	&\Gamma(\mathrm{ matter})
	=- \sum_{\hat{\mathfrak{m}},\mathfrak{n}=1}^N\sum_{\m=1}^n\sum_{j,j_3}\frac{1}{2}(-1)^F\Big( i\eta(\alpha_\m^\mathfrak{n}-\alpha_\p^{\mathfrak{\hat{m}}}) +\beta(\epsilon_j+j_3) +\beta^\prime(\epsilon_j- j_3-h) \Big)&\nn\\
	&+\sum_{\hat{\mathfrak{m}},\mathfrak{n}=1}^N\sum_{\m=1}^n\sum_{r=1}^\infty \frac{1}{r}\Big[ e^{-ir(\alpha_\m^\mathfrak{n}-\alpha_\p^{\mathfrak{\hat{m}}})}f^{+}_{\mathfrak{n}\hat{\mathfrak{m}}\m}(x^r) +e^{ir(\alpha_\m^\mathfrak{n}-\alpha_\p^{\mathfrak{\hat{m}}})}f^{-}_{\mathfrak{n}\hat{\mathfrak{m}}\m}(x^r)\Big]\;.&
\end{align} 
Adding the series one obtains,
\be 
\label{f_mn_an}
f^{+}_{\mathfrak{n}\hat{\mathfrak{m}}\m}(x)=f^{-}_{\mathfrak{n}\hat{\mathfrak{m}}\m}(x)&&=\,x^{\frac{1}{2}}\sum_{j=\frac{|q_\m^{\mathfrak{n}}-q_\p^{\hat{\mathfrak{m}}}|}{2}}^\infty x^{2j}-\,x^{\frac{3}{2}}\sum_{j=\frac{|q_\m^{\mathfrak{n}}-q_\p^{\hat{\mathfrak{m}}}|}{2}}^\infty x^{2j}\nn\\
&=&x^{|q_\m^{\mathfrak{n}}-q_\p^{\hat{\mathfrak{m}}}|}.\frac{x^{\frac{1}{2}}}{1-x^2}-x^{|q_\m^{\mathfrak{n}}-q_\p^{\hat{\mathfrak{m}}}|}.\frac{x^{\frac{3}{2}}}{1-x^2}\;.
\ee 
We find the contribution  $x^\prime$ dependent terms perfectly cancel between the bosons and the fermions as it should.
\subsection{One loop determinant of adjoint fields}
In this section we evaluate the one loop determinants of adjoint fields viz $A_\mu, \sigma,\phi, \chi_\phi, \chi_\sigma$.
We want to evaluate the following path integral,
\be 
e^{\Gamma(A_\mu, \phi, \phi^\g,\sigma)}=\int [dA_\mu][d\sigma][d\phi][d\phi^\g]\, e^{\mathcal{S}_{YM}+\mathcal{S}_{adj}}\;.
\ee 
For this computation we use the fields on $\RR^3$ subjected to the boundary condition \eqref{eqn:boundary condition}.
Now we pick the  quadratic fluctuation of $\mathcal{S}_{YM}+\mathcal{S}_{adj}$ since these terms will be $g$ independent after the following rescaling,
\be 
\delta A_\mu\to {g}\,\delta A_\mu,\quad \delta \sigma\to {g}\,\delta \sigma,\quad \delta\phi\to {g}\, \delta\phi,\quad \delta\phi^\g\to {g}\, \delta\phi^\g\;.
\ee 
The  quadratic fluctuation is,
\be
\delta(\mathcal{S}_{YM}+\mathcal{S}_{adj})&=&\sum_\m\int d^3x\, \Big(  \vec{D} \times \delta \vec{A}- \vec{D} \sigma+i[\sigma, \delta \vec{A} ] \Big)_{\m\mu} \Big( \vec{D} \times \delta \vec{A}- \vec{D} \sigma+i[\sigma, \delta \vec{A} ] \Big)^\mu_\m\nn\\
&+&|\vec{D}.  \delta\phi_\m+i[\delta A_{\mu},\phi_\m] |^2+ |[\sigma_\m, \delta\phi_\m]+[\delta\sigma_\m, \phi_\m]|^2\;.
\ee
Now substituting the  saddle point equations \eqref{monopole_solution} in the quadratic fluctuation  we find,
\be
\delta(\mathcal{S}_{YM}+\mathcal{S}_{adj})&=&\sum_\m\int d^3x\, \Big(  \vec{D} \times \delta \vec{A}- \vec{D} \sigma+iq_\m\, \delta \vec{A}  \Big)_{\m\mu} \Big( \vec{D} \times \delta \vec{A}- \vec{D} \sigma+iq_\m\,\delta \vec{A}  \Big)^\mu_\m\nn\\
&+&|\vec{D}.  \delta\phi_\m|^2+q^2 | \delta\phi_\m|^2,\quad\quad\qquad q=\frac{q^\mathfrak{m}_\m-q^\mathfrak{n}_\m }{2}
\ee
To perform the path integral we follow the method of \cite{Kim:2009wb} and expand the fields using monopole spherical harmonics,
\begin{equation}
	\label{eqn: monopole_harmonics_expansion}
	\delta\vec{A}=\sum_{n=-\infty}^\infty\sum_{j,m}\sum_{\lambda=0,\pm
		1}a^\lambda_{njm}r^{-i\frac{2\pi n}{\beta+\beta^\prime}}
	\vec{C}^\lambda_{jm}\ ,\ \
	\delta\sigma=\sum_{n,j,m}b_{njm}r^{-i\frac{2\pi n}{\beta+\beta^\prime}}
	\frac{Y_{jm}}{r},\quad \delta\phi= \sum_{njm}d_{njm}\,r^{-i\frac{2\pi n}{\beta+\beta^\prime}}  \, \frac{Y_{jm}}{r}
\end{equation}
where, $\vec{C}^\lambda_{jm}$ is monopole vector spherical harmonics. Construction of monopole vector spherical harmonics from monopole scalar harmonics $Y_{jm}$ and its properties can be found in \cite{Weinberg:1993sg} and \cite{Kim:2009wb}. 
Substituting this in the quadratic fluctuation and using the orthogonality property of monopole harmonics we obtain,
\be 
\label{eqn:pathint_gauge}
e^{\Gamma(A_\mu, \phi, \phi^\g,\sigma)}&=&\int [dA_\mu][d\sigma][d\phi][d\phi^\g]\, \prod_{n=-\infty}^\infty\prod_{\mathfrak{m},\mathfrak{n}=1}^N\prod_{\m=1}^r\prod_{j,m} e^{v^T_{-n,j,-m} {\tilde{M}}{M}\,v_{n,j,m}}\,\, e^{d^*_{n,j,m}\,\mathcal{N}\,\,d_{n,j,m}}\nn\\
&=& \prod_{n=-\infty}^\infty\prod_{\mathfrak{m},\mathfrak{n}=1}^N\prod_{\m=1}^r\prod_{j,m}\int [dA_\mu][d\sigma] e^{v^T_{-n,j,-m} {\tilde{M}}{M}\,v_{n,j,m}} \int [d\phi][d\phi^\g]\, \,\, e^{d^*_{n,j,m}\,{N}\,\,d_{n,j,m}}\nn\\
&=&\frac{1}{{\Big(\det({\tilde{M}}{M})\Big)^\frac{1}{2}} }.\frac{1}{\det({N})} 
\ee 
\footnote{The number of nodes $n$ should not be confused with the mode $n$ which runs from $-\infty$ to $+\infty$}where,
\be
\hspace*{-2cm}&&v=\begin{pmatrix}
	a_+\\a_-\\a_0\\b
\end{pmatrix},\quad
{M}^q_n=\begin{pmatrix}
	-\Lambda+iq&0&is_+&-s_+\\0&\Lambda+iq&-is_-&-s_-\\
	-is_+&is_-&iq&i\Lambda+1
\end{pmatrix},\quad {\tilde{M}}=( {M}^{-q}_{-n,j,-m})^T=\begin{pmatrix}
	\Lambda-iq&0&-is_-
	\\0&-\Lambda-iq&is_+
	\\is_-&-is_+&-iq
	\\-s_-&-s_+&-i\Lambda+1
\end{pmatrix}\nn\\
\hspace*{-2cm}&&{N}
=(i\Lambda^\prime+\frac{1}{2}).(-i\Lambda^\prime+\frac{1}{2})+  (\mathcal{J}+iq). (\mathcal{J}-iq),
\ee
with,
\be 
&&\Lambda=\frac{2\pi  n}{\beta+\beta^\prime}+\frac{\alpha^\mathfrak{m}_\m\!-\alpha^\mathfrak{n}_\m}
{\beta\!+\!\beta^\prime} -i\frac{\beta-\beta^\prime}{\beta+\beta^\prime}\,j_3,
\quad \mathcal{J}=j(j+1)-q^2,\quad s_\pm=\sqrt{\frac{\mathcal{J}^2\pm q}{2}}\nn\\ [3mm]
&&\Lambda^\prime=\frac{2\pi  n}{\beta+\beta^\prime}+\frac{\alpha^\mathfrak{m}_\m\!-\alpha^\mathfrak{n}_\m}
{\beta\!+\!\beta^\prime} -i\frac{\beta-\beta^\prime}{\beta+\beta^\prime}\,j_3+ \frac{\beta^\prime}{\beta+\beta^\prime}\, h+\frac{1}{2}\;.
\ee 

\subsubsection{Determinant of adjoint complex scalars }
From above we observe that after substituting the saddle point solutions the action of $\phi_\m $ gets decoupled from $A_{\mu\m}, \sigma_\m$. Therefore the effective action for the complex adjoint scalars $\phi_\m,\phi_\m ^\g$ is,
\be 
e^{\Gamma(\phi, \phi^\g)}&=&\int[d\phi][d\phi^\g]\, \prod_{n=-\infty}^\infty\prod_{\mathfrak{m},\mathfrak{n}=1}^N\prod_{\m=1}^r\prod_{j,m}  e^{d^*_{n,j,m}\,{N}\,\,d_{n,j,m}}\nn\\
&=&\frac{1}{\det({N})} \;.
\ee 
Following same procedure as before we find,
\begin{align}
&\det({N})
=\prod_{n=-\infty}^\infty \Big(\frac{2\pi n}{\beta+\beta^\prime}+\frac{\alpha^\mathfrak{m}_\m\!-\alpha^\mathfrak{n}_\m}
{\beta\!+\!\beta^\prime} -i\frac{\beta-\beta^\prime}{\beta+\beta^\prime}\,j_3+ \frac{\beta^\prime}{\beta+\beta^\prime}\, h+\frac{1}{2}\Big)^2+ (j+\frac{1}{2})^2&\nn\\
&= \prod_{j,j_3} (-2i)\sin\Bigg[\frac{1}{2}\Big( -(\alpha_\m^\mathfrak{n}-\alpha_\m^{\mathfrak{{m}}})+i\beta ( \epsilon_j+ j_3+\frac{1}{2})+i\beta^\prime(\epsilon_j- j_3-h+\frac{1}{2})\Big) \Bigg]&\nn\\
&(-2i)\sin\Bigg[\frac{1}{2}\Big( -(\alpha_\m^\mathfrak{n}-\alpha_\m^{\mathfrak{{m}}})+i\beta(\epsilon_j- j_3-\frac{1}{2})+i\beta^\prime (j_3 + \epsilon_j +h-\frac{1}{2})\Big) \Bigg],\quad \epsilon_j=j+\frac{1}{2}&
\end{align}
The one loop effective action is,
\be 
\Gamma(\phi_\m,\phi_\m^\g) &=&-
\sum_{\mathfrak{m},\mathfrak{n}=1}^N\sum_{\m=1}^n\sum_{j,j_3} \Bigg[\frac{i}{2}(\alpha_\m^\mathfrak{n}-\alpha_\m^{\mathfrak{{m}}})+\frac{\beta}{2} ( \epsilon_j+ j_3+\frac{1}{2})+\frac{\beta^\prime}{2}(\epsilon_j- j_3-h+\frac{1}{2})\nn\\ &+&\frac{\beta}{2} ( \epsilon_j+ j_3-\frac{1}{2})+\frac{\beta^\prime}{2}(\epsilon_j- j_3-h-\frac{1}{2})\Bigg]\nn\\
&+&\sum_{\mathfrak{m},\mathfrak{n}=1}^N\sum_{\phi_\m,\phi_\m^\g}\sum_{\m=1}^n\sum_{j,j_3} \frac{1}{r} e^{ -ir(\alpha_\m^\mathfrak{n}-\alpha_\m^{\mathfrak{{m}}})} x^{r ( \epsilon_j+ j_3+\frac{1}{2})}(x^\prime)^{r(\epsilon_j- j_3-h+\frac{1}{2})}\;.
\ee 
Putting the charges, for $\phi_\m , h=1$ we obtain 
\be 
&&x^{ ( \epsilon_j+ j_3+\frac{1}{2})}(x^\prime)^{\epsilon_j- j_3-h+\frac{1}{2}}
=\sum_{j={\frac{|q^\mathfrak{m}_\m-q^\mathfrak{n}_\m |}{2}}}^\infty x(x^\prime)^{2j}+x^{  2}(x^\prime)^{2j-1}+...+x^{  2j+1}\;.
\ee 
and for $\phi^\g, h=-1$ we get,
\be 
&&x^{ ( \epsilon_j+ j_3-\frac{1}{2})}(x^\prime)^{\epsilon_j- j_3-h-\frac{1}{2}}
=\sum_{j={\frac{|q^\mathfrak{m}_\m-q^\mathfrak{n}_\m| }{2}}}^\infty (x^\prime)^{2j+1}+x(x^\prime)^{2j}+...+x^{  2j}(x^\prime)\;.
\ee 

\subsubsection{Gauge fixing and  the determinant }
Now let us move on to the computation of the determinant $A_{\mu\m}, \sigma_\m$ for which we need to calculate $\det M$. Now the matrix $M$ depends on some results of monopole vector harmonics and changes after we do the gauge fixing. So let us first fix the gauge and calculate the determinant case by case. 
We work in  Coulumb gauge,
\be
\mathcal{D}_a A^a=0,\quad \textit{a=1,2 are  $S^2$ co-ordinates}
\ee
which implies the following upon substituting eqn.\eqref{eqn: monopole_harmonics_expansion},
\be
\label{Coulumb_gauge}
\boxed{s_+ a^+_{njm}+ s_- a^-_{njm} =0}\;.
\ee
We write some results of \cite{Weinberg:1993sg} here which will be important for gauge fixing.
\begin{itemize}
	\item (1) For $ j=q-1\geq 0$: one multiplet $\vec{C}^+$ is non zero.
	\item (2) For $ j=q> 0$: two multiplets $\vec{C}^+, \vec{C}^0$ are non zero.
	\item (3) For $ j=q= 0$: one multiplet $ \vec{C}^0$ is non zero.
	\item (4) For $ j\geq q$: three multiplets $\vec{C}^+, \vec{C}^-, \vec{C}^0$ are present.
\end{itemize}
Now we incorporate the gauge condition in the path integral by changing the measure, which finally gives,
\be
Z=\int [\mathcal{D}A_\mu]\,e^{-S}=\int d\alpha \int \Delta_{FP}\delta\big( f[A_\mu^\alpha]\big) [\mathcal{D}A_\mu]e^{-S}
\ee
where,
\be
f[A_\mu^\alpha]=0
\ee
is the gauge condition that  we impose. Now we compute the gauge condition for different values of $j$ we listed above and compute the Faddeev- Popov determinant $\Delta_{FP}$.\\
 \underline{ \textbf{  For $\boldsymbol{ j=q-1\geq 0}$}}\\
 In this case the gauge condition is automatically satisfied because of the following.
	\be 
	s_+=0,\quad s_-=\sqrt{-q},\quad{s_+ a^+_{njm} =0}
	\ee 
	Therefore we do not have to fix the gauge.
	The integrand for gauge fields \eqref{eqn:pathint_gauge} becomes(the scalar harmonics are also absent),
	\be
	v=\left(\begin{array}{c}a^+\end{array}\right),\quad
	{M}= \begin{pmatrix}
		-\Lambda+iq
	\end{pmatrix}
	\ee 
The path integral is,
	\be 
\prod_{\mathfrak{m},\mathfrak{n}=1}^N\prod_{\m=1}^n\prod_{j_3=-j}^{j}\sin\Bigg[\frac{1}{2}\Big(\beta(j+1+j_3) + i(\alpha^\mathfrak{m}_\m\!-\alpha^\mathfrak{n}_\m)
+\beta^\prime(j+1-j_3)\Big)\Bigg]
\ee 
where, we have put $q=j+1$ and used the following identity,
	\be
	\sin x=x\prod_{n=-\infty}^\infty (1+\frac{x}{n\pi})\;.
	\ee 
\underline{\textbf{  For $ \boldsymbol{j=q> 0}$}}\\
	The gauge condition in this case is,
	\be
	{a_+=0}
	\ee
	Since there are no $a^-$ modes, the gauge is already fixed and from eqn.\eqref{eqn:pathint_gauge}
	we have,
	\be
	v=\left(\begin{array}{c}a^0\\b\end{array}\right),\quad\quad
{M}= \begin{pmatrix}
		i\sqrt{q}&-\sqrt{q}\\
		iq&i\Lambda+1
	\end{pmatrix}
	\ee 
	The path integral is,
	\be  \frac{\sqrt{q}}{\beta+\beta^\prime}\sin\Bigg[\frac{1}{2}\Big(\beta(j+1+j_3)+\beta^\prime(j+1-j_3)+i(\alpha^\mathfrak{m}_\m\!-\alpha^\mathfrak{n}_\m) \Big)\Bigg] \;.
	\ee 
\underline{\textbf{  For $ \boldsymbol{j=q= 0}$}}\\
In this case we have, 
	\be 
q^\mathfrak{m}_\m=q^\mathfrak{n}_\m 
	\ee 
	The Coulomb gauge condition does not fix the the gauge redundancy since both $a^\pm$ vanishes in this case. Therefore we impose,
	\be
	\frac{d}{d\tau} \int_{S^2}\, A_\tau=0
	\ee
	The Coulomb gauge condition for infinitesimal gauge transformation $\epsilon$,
	\be
	\delta A_m\to \delta A_m +\partial_m \epsilon
	\ee
	is,
	\be
	\partial^a\partial_a\epsilon=0,\quad since \,\, q=0
	\ee
	Now we can expand $\epsilon$ as scalar harmonics and obtain following eigenvalue equation,
	\be 
	\partial^2 Y_{jm}=0
	\ee 
	where, 
	\be 
	j=0,1,2,\cdots
	\ee 
	Therefore for $j=0$ we can have a non-zero eigenvector. This implies that there exists non zero $\epsilon$ for which $\partial^a\partial_a\epsilon=0$.
	The corresponding Fadeev-Popov determinant is computed following \cite{Aharony:2003sx} which is,
	\be
\prod_{\mathfrak{m}<\mathfrak{n}}\Big(2 \sin \frac{\alpha^\mathfrak{m}_\m-\alpha^{\mathfrak{n}}_\m}{2}\Big)^2\;.
	\ee
 \underline{\textbf{  For $ j>q$: three multiplets $\vec{C}^+, \vec{C}^-, \vec{C}^0$}}\\
To satisfy the   Coulomb gauge condition in  eqn.\eqref{Coulumb_gauge}  we choose the following,
\be
a_+=s_- a,\qquad a_-=-s_+ a
\ee
Substituting this in eqn.\eqref{eqn:pathint_gauge} we have,
\be 
{M}= \begin{pmatrix}
	-s_-(\Lambda-iq)& is_+&-s_+\\
	-s_+(\Lambda+iq)&-is_-&-s_-\\
	-2is_+s_-&iq&i\Lambda+1
\end{pmatrix}\;.
\ee 
The path integral is,
	\be 
	&& \prod_{\mathfrak{m},\mathfrak{n}=1}^N\prod_{\m=1}^n\prod_{j_3=-j}^{j} \Big((j+\frac{1}{2})^2+(\Lambda -\frac{i}{2})^2\Big)\nn\\
&&=\prod_{\mathfrak{m},\mathfrak{n}=1}^N\prod_{\m=1}^n\prod_{j_3=-j}^{j}\sin\Bigg[\frac{1}{2}\Big(\beta(j+1+j_3) + i(\alpha^\mathfrak{m}_\m\!-\alpha^\mathfrak{n}_\m)
+\beta^\prime(j+1-j_3)\Big)\Bigg]\nn\\
&& \sin\Bigg[\frac{1}{2}\Big(\beta(j-j_3) + i(\alpha^\mathfrak{m}_\m\!-\alpha^\mathfrak{n}_\m)
+\beta^\prime(j+j_3)\Big)\Bigg]\;.
\ee 

Combining all the cases the final answer is\footnote{Since the final answer is  symmetric in $j_3\to -j_3$, while combining all the modes we have replaced $j_3\to -j_3$ in some terms},
\begin{align}
&\Gamma(A_{\mu\m},\sigma_\m)
=-\frac{1}{2}{\Tr}_B\Big(  \beta(j-j_3)+\beta^\prime(j+j_3) +\beta(j+1+j_3)+\beta^\prime(j+1-j_3)  \Big)& \nn\\
&+\sum_{\m=1}^n\sum_{\mathfrak{m},\mathfrak{n}=1}^N\sum_{r=1}^\infty\frac{1}{r}\Bigg[ e^{ -ir(\alpha_\m^\mathfrak{n}-\alpha_\m^{\mathfrak{{m}}})}  x^ {r(j-j_3)} (x^\prime)^{r(j+j_3)}
+ e^{ -ir(\alpha_\m^\mathfrak{n}-\alpha_\m^{\mathfrak{{m}}})}  x^{r(j+1+j_3)}(x^\prime)^{r(j+1-j_3)} \Bigg]&
\end{align}	
where, ${\Tr}_B$ denotes sum over the following  modes. We compute the letter indices for each case below.\\
\underline{for the modes with $j>q$}
\begin{align}
& x^ {j-j_3} (x^\prime)^{(j+j_3)}+x^{j+1+j_3}(x^\prime)^{j+1-j_3}&\nn\\
&=\sum_{j=\frac{|q^\mathfrak{m}_\m-q^\mathfrak{n}_\m| }{2}+1}^\infty \Big[ x^ {2j} +x^ {2j-1} (x^\prime)+\cdots+  (x^\prime)^{2j}\Big]+\sum_{j=\frac{|q^\mathfrak{m}_\m-q^\mathfrak{n}_\m |}{2}+1}^\infty  \Big[x(x^\prime)^{2j+1}+x^{2}(x^\prime)^{2j}+\cdots+ x^{2j+1}(x^\prime)\Big]&
\end{align} 
\underline{for the modes with $j=q=\frac{|q^\mathfrak{m}_\m-q^\mathfrak{n}_\m| }{2}$}
\be 
&&\hspace*{-1cm}x^{j+1+j_3}(x^\prime)^{j+1-j_3}
=x(x^\prime)^{|q^\mathfrak{m}_\m-q^\mathfrak{n}_\m |+1}+x^{2}(x^\prime)^{|q^\mathfrak{m}_\m-q^\mathfrak{n}_\m |}+\cdots+ x^{|q^\mathfrak{m}_\m-q^\mathfrak{n}_\m |+1}(x^\prime)
\ee 
\underline{For the modes with $j=q=\frac{|q^\mathfrak{m}_\m-q^\mathfrak{n}_\m |}{2}-1$}
\be 
&&\hspace*{-1cm}x^{j+1+j_3}(x^\prime)^{j+1-j_3}
=x(x^\prime)^{|q^\mathfrak{m}_\m-q^\mathfrak{n}_\m |-1}+x^{2}(x^\prime)^{|q^\mathfrak{m}_\m-q^\mathfrak{n}_\m |-2}+\cdots+ x^{|q^\mathfrak{m}_\m-q^\mathfrak{n}_\m |-1}(x^\prime)
\ee 
The final result is,
		\begin{align}
			&\Gamma(A_\mu,\sigma)
			=-\frac{1}{2}\sum_{\m=1}^n\sum_{\mathfrak{m},\mathfrak{n}=1}^N{\Tr}_B\Big(  \beta(j-j_3)+\beta^\prime(j+j_3) +\beta(j+1+j_3)+\beta^\prime(j+1-j_3)+i(\alpha^\mathfrak{m}_\m\!-\alpha^\mathfrak{n}_\m)  \Big)& \nn\\
			&+\sum_{\m=1}^n\sum_{\mathfrak{m},\mathfrak{n}=1}^N\sum_{r=1}^\infty\frac{1}{r}e^{-in
			(\alpha^\mathfrak{m}_\m\!-\alpha^\mathfrak{n}_\m)}\Bigg(\sum_{j=\frac{|q|}{2}+1}^\infty \Big[ x^ {2j} +x^ {2j-1} (x^\prime)+\cdots+  (x^\prime)^{2j}\Big]&\nn\\
		&+\sum_{j=\frac{|q|}{2}+1}^\infty  \Big[x(x^\prime)^{2j+1}+x^{2}(x^\prime)^{2j}+\cdots+ x^{2j+1}(x^\prime)\Big]&\nn\\
			&+\Big(x(x^\prime)^{|q^\mathfrak{m}_\m-q^\mathfrak{n}_\m|+1}+x^{2}(x^\prime)^{|q^\mathfrak{m}_\m-q^\mathfrak{n}_\m|}+\cdots+ x^{|q^\mathfrak{m}_\m-q^\mathfrak{n}_\m|+1}(x^\prime)\Big)\;& \nn\\ 
			&+\Big(x(x^\prime)^{|q^\mathfrak{m}_\m-q^\mathfrak{n}_\m|-1}+x^{2}(x^\prime)^{|q^\mathfrak{m}_\m-q^\mathfrak{n}_\m|-2}+\cdots+ x^{|q^\mathfrak{m}_\m-q^\mathfrak{n}_\m|-1}(x^\prime)\Big)|_{|q^\mathfrak{m}_\m-q^\mathfrak{n}_\m|\geq 1}\Bigg)&
		\end{align}

\subsection{One loop determinants of adjoint fermions}
We write the quadratic action of the adjoint fermions on $\RR\times S^2$ in the following form \cite{charge_ade},
\be
&&\hspace*{-1cm}\mathcal{S}= \int d\tau\, d\Omega\,\,\sum_{\m=1}^n \mathrm{tr} \Bigg(\,
-  i \lambda_{1a\m} \slashed{\mathcal{D}} \lambda^{1a}_\m   +\frac{i}{2}  {\lambda^\g_{1a}}_\m  {(\sigma_3)}^{a}_{\,\,\,b}(q_\m^{\mathfrak{m}}-q_\m^{\mathfrak{n}}+\frac{1}{2}) {\lambda^{1b}}_\m \Bigg)
\ee
where $a,b=1,2$,
\be
&&\lambda_{11}=  \chi^\g_{\sigma\m}e^{i\pi/4},\qquad {\lambda^\g_{11}}=-\chi_{\sigma\m}e^{i\pi/4},
\qquad \lambda^{11}=\chi_{\sigma\m}e^{-i\pi/4},\qquad \lambda^{12}=\chi^\g_{\phi\m}e^{-i\pi/4} \nn\\
&&\lambda_{12}=\chi_{\phi\m}e^{i\pi/4},\qquad \lambda^\g_{12}=-\chi_{\phi\m}e^{i\pi/4}\;.
\ee
The reason we introduced the above notation so that we can  relate the action to that of matter fermions. Observe that this action is same as the matter fermions after replacing 
\be 
(q_\p^{\mathfrak{\hat{m}}}-q_\m^{\mathfrak{n}})\to -(q_\m^{\mathfrak{{m}}}-q_\m^{\mathfrak{n}})+\frac{1}{2}\;.
\ee 
Since the final answer depends on $q^2$, the sign change in $q$ does not affect anything. But the factor of $\frac{1}{2}$, which arises due to radial quantization shifts the eigenvalue $\epsilon_j$ of the operator by $\pm \frac{1}{2}$. For the non zero modes we have both positive and negative energy states with energy $\pm \epsilon_j=\pm(j+\frac{1}{2})$ while for the zero modes we have $\epsilon_0=\frac{|q|-1}{2}$. Therefore the energies get shifted as follows,
\be 
+\epsilon_j \to \epsilon_j-\frac{1}{2},\quad \epsilon_j \to \epsilon_j+\frac{1}{2},\quad \epsilon_0\to \frac{|q|}{2}\;.
\ee 
The determinant of the operator,
\be 
&&\det\Bigg[D_\tau +\epsilon_j-\frac{1}{2}\Bigg]
\propto \Bigg[\beta (\epsilon_j\pm\frac{1}{2}+j_3)+\beta^\prime(\epsilon_j\pm\frac{1}{2}-j_3-h)\Bigg]
\ee 
for various fields is, $\chi_\phi, (h=-1)$
\be 
\Bigg[\beta (\epsilon_j-\frac{1}{2}+j_3)+\beta^\prime(\epsilon_j-\frac{1}{2}-j_3-h)\Bigg]
=\Bigg[\beta (j+j_3)+\beta^\prime(j+1-j_3)\Bigg]
\ee
 $\chi_\phi^\g, (h=1)$
\be 
\Bigg[\beta (\epsilon_j+\frac{1}{2}+j_3)+\beta^\prime(\epsilon_j+\frac{1}{2}-j_3-h_3)\Bigg]
=\Bigg[\beta (j+1+j_3)+\beta^\prime(j-j_3)\Bigg]\;.
\ee
The letter index of $\chi_{\phi\m}$ is,
\be 
x^{j+j_3}\, (x^\prime)^{j+1-j_3}=-\sum_{j=\frac{|q^\mathfrak{m}_\m-q^\mathfrak{n}_\m|+1}{2}}^\infty \Big( (x^\prime)^{2j+1}+x\, (x^\prime)^{2j}+...+x^{2j}\, (x^\prime)\Big)\;.
\ee 
The letter index of $\chi^\g_{\phi\m}$ is,
\be 
x^{j+1+j_3}\,(x^\prime)^{j-j_3}=-\sum_{j=\frac{|q^\mathfrak{m}_\m-q^\mathfrak{n}_\m|-1}{2}}^\infty\Big(x\,(x^\prime)^{2j}+x^{2}\,(x^\prime)^{2j-1}+...+x^{2j+1}\Big)\;.
\ee 
For $\chi_\sigma, (h=0)$ we have
\be 
\Bigg[\beta (\epsilon_j-\frac{1}{2}+j_3)+\beta^\prime(\epsilon_j-\frac{1}{2}-j_3-h_3)\Bigg]
=\Bigg[\beta (j+j_3)+\beta^\prime(j-j_3)\Bigg]
\ee
and the letter index is,
\be 
x^{ j+j_3}\,(x^\prime)^{j-j_3}=-\sum_{j=\frac{|q^\mathfrak{m}_\m-q^\mathfrak{n}_\m|+1}{2}}^\infty\Big(\,(x^\prime)^{2j}+x\,(x^\prime)^{2j-1}+...+x^{ 2j}\Big)\;.
\ee 
Similarly for $\chi_\sigma^\g, (h=0)$ we have,
\be 
\Bigg[\beta (\epsilon_j+\frac{1}{2}+j_3)+\beta^\prime(\epsilon_j+\frac{1}{2}-j_3-h_3)\Bigg]
=\Bigg[\beta (j+1+j_3)+\beta^\prime(j+1-j_3)\Bigg]
\ee
and the letter index is,
\be 
x^{j+1+j_3}\,(x^\prime)^{j+1-j_3}=-\sum_{j=\frac{|q^\mathfrak{m}_\m-q^\mathfrak{n}_\m|-1}{2}}^\infty\Big(x(x^\prime)^{2j+1}+x^{2}\,(x^\prime)^{2j}+...+x^{2j+1}\,(x^\prime)\Big)\;.
\ee 
Combining the above  final result for adjoint fermions is,
\begin{align}
&\Gamma(\chi_\phi,\chi_\phi^\g, \chi_\sigma,\chi_\sigma^\g)=\frac{1}{2}\sum_{\m=1}^n\sum_{\mathfrak{m},\mathfrak{n}=1}^N{\Tr}_F\Bigg[\beta (j+j_3)+\beta^\prime(j+1-j_3)+\beta (j+1+j_3)+\beta^\prime(j-j_3)
&\nn\\&		+\beta (j+j_3)+\beta^\prime(j-j_3)+i(\alpha^\mathfrak{m}_\m\!-\alpha^\mathfrak{n}_\m)\Bigg]&
\nn\\&+\frac{1}{2}\sum_{\m=1}^n\sum_{\mathfrak{m},\mathfrak{n}=1}^N\sum_{r=1}^\infty e^{-ir
	(\alpha^\mathfrak{m}_\m\!-\alpha^\mathfrak{n}_\m)}\frac{1}{r}\Bigg[-\sum_{j=\frac{|q^\mathfrak{m}_\m-q^\mathfrak{n}_\m|+1}{2}}^\infty \Big( (x^\prime)^{2j+1}+x\, (x^\prime)^{2j}+...+x^{2j}\, (x^\prime)\Big)&\nn\\&-\sum_{j=\frac{|q^\mathfrak{m}_\m-q^\mathfrak{n}_\m|-1}{2}}^\infty\Big(x\,(x^\prime)^{2j}+x^{2}\,(x^\prime)^{2j-1}+...+x^{2j+1}\Big)-\sum_{j=\frac{|q^\mathfrak{m}_\m-q^\mathfrak{n}_\m|+1}{2}}^\infty\Big(\,(x^\prime)^{2j}+x\,(x^\prime)^{2j-1}+...+x^{ 2j}\Big)&\nn\\
&-\sum_{j=\frac{|q^\mathfrak{m}_\m-q^\mathfrak{n}_\m|-1}{2}}^\infty\Big(x(x^\prime)^{2j+1}+x^{2}\,(x^\prime)^{2j}+...+x^{2j+1}\,(x^\prime)\Big)\Bigg]&
	\end{align}
\subsection{Final result of adjoint sector}
Now, let us combine the final results of fermions and bosons.\\
\underline{ Contribution from $A_\mu, \sigma,\chi_\phi,  \chi_\phi^\g$}
\be
&&+\sum_{j=\frac{|q|}{2}+1}^\infty \Big[ x^ {2j} +x^ {2j-1} (x^\prime)+\cdots+  (x^\prime)^{2j}\Big]+\sum_{j=\frac{|q|}{2}+1}^\infty  \Big[x(x^\prime)^{2j+1}+x^{2}(x^\prime)^{2j}+\cdots+ x^{2j+1}(x^\prime)\Big]\nn\\
&&\hspace*{-1cm}\Big(x(x^\prime)^{|q_{\mathfrak{{m}}\mathfrak{n}}|+1}+x^{2}(x^\prime)^{|q_{\mathfrak{{m}}\mathfrak{n}}|}+\cdots+ x^{|q_{\mathfrak{{m}}\mathfrak{n}}|+1}(x^\prime)\Big)+\Big(x(x^\prime)^{|q|-1}+x^{2}(x^\prime)^{|q_{\mathfrak{{m}}\mathfrak{n}}|-2}+\cdots+ x^{|q_{\mathfrak{{m}}\mathfrak{n}}|-1}(x^\prime)\Big)|_{|q_{\mathfrak{{m}}\mathfrak{n}}|\geq 1}\nn\\
&-&\sum_{j=\frac{|q|+1}{2}}^\infty \Big( (x^\prime)^{2j+1}+x\, (x^\prime)^{2j}+...+x^{2j}\, (x^\prime)\Big)-\sum_{j=\frac{|q|+1}{2}}^\infty\Big(x\,(x^\prime)^{2j}+x^{2}\,(x^\prime)^{2j-1}+...+x^{2j+1}\Big)\nn\\
&-&\Big(x\,(x^\prime)^{|q|-1}+x^{2}\,(x^\prime)^{|q|-2}+...+x^{|q|}\Big),\qquad\qquad q=q^\mathfrak{m}_\m-q^\mathfrak{n}_\m
\ee
We observe that the series for $|q|\geq 1$ combines with the last line and gives
\be 
-x^{|q|}
\ee 
The remaining terms completely cancel as follows,\\
\underline{$O(x^0)$}
\be 
&&\sum_{j=\frac{|q|}{2}+1}(x^\prime)^{2j}-\sum_{j=\frac{|q|+1}{2}}(x^\prime)^{2j+1}\nn\\
&&=\Big[(x^\prime)^{|q|+2}+(x^\prime)^{|q|+4}+...\Big]-\Big[(x^\prime)^{|q|+2}+(x^\prime)^{|q|+4}+...\Big]=0
\ee 
\underline{$O(x^1)$}
\be 
&&\sum_{j=\frac{|q|}{2}+1}(x^\prime)^{2j-1}+\sum_{j=\frac{|q|}{2}+1}(x^\prime)^{2j+1}+(x^\prime)^{|q|+1}
-\sum_{j=\frac{|q|+1}{2}}(x^\prime)^{2j}-\sum_{j=\frac{|q|+1}{2}}(x^\prime)^{2j}\nn\\ [3mm]
&&=\Big[(x^\prime)^{|q|+1}+(x^\prime)^{|q|+3}+...\Big]+\Big[(x^\prime)^{|q|+1}+(x^\prime)^{|q|+3}+(x^\prime)^{|q|+5}+...\Big]\nn\\
&&-\Big[(x^\prime)^{|q|+1}+(x^\prime)^{|q|+3}+...\Big]-\Big[+(x^\prime)^{|q|+1}+(x^\prime)^{|q|+3}+...\Big] =0
\ee 
and so on. Contribution from the  remaining fields of adjoint sector cancels completely as shown below.\\
\underline{ Contribution from $\chi_\sigma, \chi_\sigma^\g,\phi, \phi^\g$}
\begin{align}
&-\sum_{j=\frac{|q|+1}{2}}^\infty\Big(\,(x^\prime)^{2j}+x\,(x^\prime)^{2j-1}+...+x^{2j}\Big)-\sum_{j=\frac{|q|-1}{2}}^\infty\Big(x(x^\prime)^{2j+1}+x^{2}\,(x^\prime)^{2j}+...+x^{2j+1}\,(x^\prime)\Big)&\nn\\
&+\sum_{j=\frac{|q|}{2}}^\infty x(x^\prime)^{2j}+x^{  2}(x^\prime)^{2j-1}+...+x^{  2j+1}+ \sum_{j=\frac{|q|}{2}}^\infty (x^\prime)^{2j+1}+x(x^\prime)^{2j}+...+x^{  2j}(x^\prime)&
\end{align}
Changing limit of of summation of the first term from $\sum_{j=\frac{|q|+1}{2}}^\infty$ to $\sum_{j=\frac{|q|}{2}}^\infty$ by making $2j\to 2j+1$ and changing limit of of summation of the first term from $\sum_{j=\frac{|q|-1}{2}}^\infty$ to $\sum_{j=\frac{|q|}{2}}^\infty$ by making $2j\to 2j-1$ we get,
\begin{align}
&-\sum_{j=\frac{|q|}{2}}^\infty\Big(\,(x^\prime)^{2j+1}+x\,(x^\prime)^{2j}+...+x^{2j+1}\Big)-\sum_{j=\frac{|q|}{2}}^\infty\Big(x(x^\prime)^{2j}+x^{2}\,(x^\prime)^{2j}+...+x^{2j}\,(x^\prime)\Big)&\nn\\
&+\sum_{j=\frac{|q|}{2}}^\infty x(x^\prime)^{2j}+x^{  2}(x^\prime)^{2j-1}+...+x^{  2j+1}+ \sum_{j=\frac{|q|}{2}}^\infty (x^\prime)^{2j+1}+x(x^\prime)^{2j}+...+x^{  2j}(x^\prime)=0&
\end{align}
Therefore the final result of adjoint sector is,
\begin{align}
	&\Gamma(\mathrm{ adjoint})
=\sum_{{\mathfrak{m}},\mathfrak{n}=1}^N\sum_{\m=1}^n\sum_{r=1}^\infty \frac{1}{r}\Big[ e^{-ir(\alpha_\m^\mathfrak{n}-\alpha_\m^{\mathfrak{{m}}})}f^{\mathrm{adj}}_{\mathfrak{n}{\mathfrak{m}}\m}(x^r)\Big]&
\end{align} 
where,
\be 
f^{\mathrm{adj}}_{\mathfrak{n}{\mathfrak{m}}\m }(x)=(-1+\delta_{q_{\mathfrak{m}}q_{\mathfrak{n}}})\,x^{|q^\mathfrak{m}_\m-q^\mathfrak{n}_\m|}
\ee

\section{$\widehat{D}$-type quiver}\label{sec_dn_index}
The index computation for $\hat{D}$ quiver is very similar to the $\hat{A}$ case. We consider the gauge group of the theory to be $U(2N)^{n-3}\times U(N)^4$\cite{nishioka}\cite{gulotta}\cite{non-toric} as they are the correct candidate for  gravity duals of $\ads\times M_7$ vacua of M-theory, where $M_7$ is a tri-Sasakian manifold. Such  theories have been considered in the context of $\widehat{ADE}$ matrix models as they have a nice large $N$ limit. In this case $\m$ runs from 1 to r+1. We call $\m=1,\cdots, 4$ external nodes and $\m=5,\cdots, n+1$ internal nodes. The CS levels satisfy the following constraint,
\be 
k_\1+k_\2+k_\3+k_\4+2 \big(k_\5+ \cdots + k_{(n+1)}\big)=0\,\,.
\ee 
The matter action on $\RR\times S^2$  is,
\begin{align}
	\label{action_dn}
	&\mathcal{L}_{\mathrm{mat}}=
	\sum_{\m=1}^2 \Tr\Big[{-\mathcal{D}_m Z_{\m}\mathcal{D}^m Z^\dagger_{\m} + \frac{1}{4}Z^\g_\m Z_\m+i\zeta^\dagger_{\p}\slashed{\mathcal{D}} \zeta_{\m}}   - i {\zeta^\dagger_{\m}\big(\sigma_{\m}\zeta_{\m}-\zeta_{\m}{\sigma}_5}\big) - Z^\dagger_{\m} Z_{\m}{\sigma^2}_{\5} &\nn\\ &+  2 Z^\dagger_{\m}\sigma_{\m} Z_{\m}{\sigma}_5 -Z^\dagger_{\m} \sigma^2_{\m} Z_{\m} -(\mathcal{D}_m W_{\m})(\mathcal{D}^m W^\g_{\m})+i\omega_{\m}\slashed{\mathcal{D}} \omega^\dagger_{\m}  
	+ i \omega^\dagger_{\m}\big(\omega_{\m}\sigma_{\m}-{\sigma}_\5\omega_{\m}\big)& \nn\\ & - W^\dagger_{\m} W_{\m}\sigma^2_{\m} +  2 W^\dagger_{\m}{\sigma}_5 W_{\m}{\sigma}_{\m} -W^\dagger_{\m} {\sigma}^2_5 W_{\m}\Big]+ \sum_{\m=3}^4\Big[{-\mathcal{D}_m Z_{\m}\mathcal{D}^m Z^\dagger_{\m}+\frac{1}{4}Z^\g_\m Z_\m+i\zeta^\dagger_{\m}\slashed{\mathcal{D}} \zeta_{\m}}&\nn\\
	&- i \zeta^\dagger_{\m}\big(\sigma_{\m}\zeta_{\m}-\zeta_{\m}{\sigma}_{n+1}\big) - Z^\dagger_{\m} Z_{\m}{\sigma^2}_{n+1} +  2 Z^\dagger_{\m}\sigma_{\m} Z_{\m}{\sigma}_{n+1} -Z^\dagger_{\m} \sigma^2_{\m} Z_{\m} -(\mathcal{D}_m W_{\m})(\mathcal{D}^m W^\dagger_{\m})&\nn\\
	&+i\omega_{\m}\slashed{\mathcal{D}} \omega^\dagger_{\m} + i \omega^\dagger_{\m}\big(\omega_{\m}\sigma_{\m}-{\sigma}_{(n+1)}\omega_{\m}\big) - W^\dagger_{\m} W_{\m}\sigma^2_{\m} +  2 W^\dagger_{\m}{\sigma}_{(n+1)} W_{\m}{\sigma}_{\m} -W^\dagger_{\m} {\sigma}^2_{r+1} W_{\m} \Big]&\nn\\
& + \sum_{\m=5}^r \Big[{-\mathcal{D}_m Z_{\m}\mathcal{D}^m Z^\dagger_{\m}+\frac{1}{4}Z^\g_\m Z_\m+i\zeta^\dagger_{\m}\slashed{\mathcal{D}} \zeta_{\m}}  - i \zeta^\dagger_{\m}\big(\sigma_{\m}\zeta_{\m}-\zeta_{\m}{\sigma}_{\p}\big) - Z^\dagger_{\m} Z_{\m}{\sigma^2}_{\p} &\nn\\
&+  2 Z^\dagger_{\m}\sigma_{\m} Z_{\m}{\sigma}_{\p} -Z^\dagger_{\m} \sigma^2_{\m} Z_{\m}
 -(\mathcal{D}_m W_{\m})(\mathcal{D}^m W^\dagger_{\m})+i\omega_{\m}\slashed{\mathcal{D}} \omega^\dagger_{\m} + i \omega^\dagger_{\m}\big(\omega_{\m}\sigma_{\m}-{\sigma}_{\p}\omega_{\m}\big)& \nn\\&- W^\dagger_{\m} W_{\m}\sigma^2_{\m} +  2 W^\dagger_{\m}{\sigma}_{\p} W_{\m}{\sigma}_{\m} -W^\dagger_{\m} {\sigma}^2_{\p} W_{\m} \Big]\;.&
\end{align}
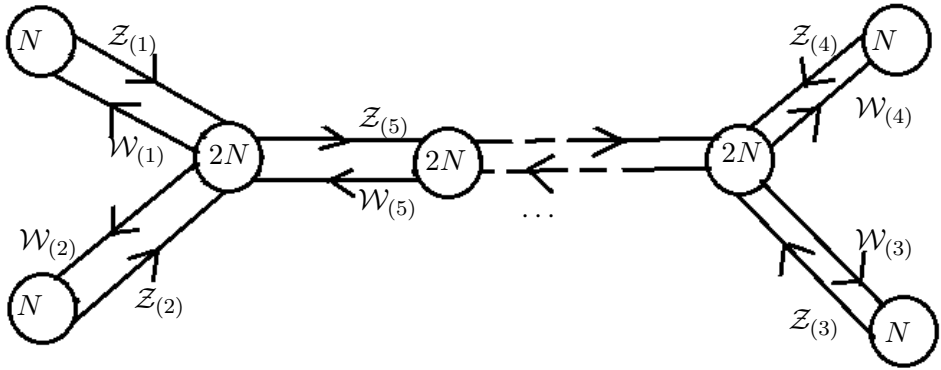
\begin{figure}[h]
	\centering
	\includegraphics[width=5in]{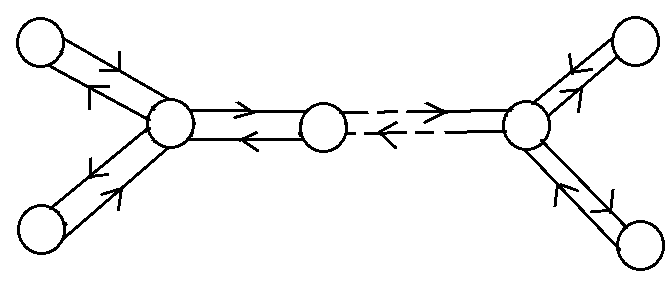}
	\put (-24,15) {$N$} \put (-29,125) {$N$}\put (-349,125) {$N$} \put (-315,127) {$\mathcal{Z}_\1$}\put (-315,85) {$\mathcal{W}_\1$} 
	\put (-305,28) {$\mathcal{Z}_\2$}\put (-348,50) {$\mathcal{W}_\2$}
	\put (-221,96) {$\mathcal{Z}_\5$}\put (-277,82) {$2N$}
	\put (-60,127) {$\mathcal{Z}_\4$}\put (-35,100) {$\mathcal{W}_\4$}
	\put (-60,20) {$\mathcal{Z}_\3$}\put (-35,50) {$\mathcal{W}_\3$}
	\put (-349,25) {$N$} \put (-221,65) {$\WW_\5$} \put (-196,80) {$2N$} \put (-160,60) {$\cdots$} \put (-85,83) {$2N$ } 
	\caption{$\widehat D_n$ quiver diagram.}
	\label{fig:Dn}
\end{figure}
As shown in the figure \ref{fig:Dn} the internal nodes, i.e $\m=5,\cdots, r+1$ is same as the $\hat{A}$ case. The saddle point solution in this case is computed similarly  as before and is given below,
\be
\label{eqn:gauge_dn}
A_\m &=&\frac{E_\m}{2}\,\,(\pm1-\cos\theta)\,d\varphi\quad for\,\,\, \m=1,2,3,4\,,\nn\\
A_\m &=&\frac{I_\m}{2}\,\,(\pm1-\cos\theta)\,d\varphi\quad for\,\,\, \m=5, \cdots , r+1\,
\ee
where, $E_\m=\mathrm{diag}(t^{\scaleto{1}{4pt}}_\m,t^{\scaleto{2}{4pt}}_\m, \cdots, t^{\scaleto{N}{4pt}}_\m)$ and $I_\m=\mathrm{diag}(q^{\scaleto{1}{4pt}}_\m,q^{\scaleto{2}{4pt}}_\m, \cdots, q^{\scaleto{2N}{4pt}}_\m)$.
\subsection{Computation of the index for $\widehat{D}$-type quiver}
\subsubsection{Matter sector}
Plugging in the background solution and using the mode expansion \eqref{mode_expansion_scalar}   we obtain from \eqref{action_dn},
\begin{align}
	&\mathcal{S}^E(Z,W, Z^\g, W^\g)=\int d\tau\, d\Omega\,\sum_{\hat{\mathfrak{m}},\mathfrak{n}=1}^{2N} \sum_{\m=5}^r  \Bigg(
	Z^{\g jm}_{\m \hat{\mathfrak{m}}\mathfrak{n}}\Bigg( -\mathcal{D}_\tau^2 +( j+\frac{1}{2})^2\Bigg)\, Z^{jm}_{\m \mathfrak{n}\hat{\mathfrak{m}}}&\nn\\
	& +W^{\g jm}_{\m \mathfrak{n}\hat{\mathfrak{m}}} \Bigg( -\mathcal{D}_\tau^2 +( j+\frac{1}{2})^2\Bigg)\, W^{jm}_{\m \hat{\mathfrak{m}}\mathfrak{m}}&\nn\\
	&+\sum_{\hat{\mathfrak{m}},\mathfrak{n}=1}^N \sum_{\m=1}^2  \Bigg(
	Z^{\g jm}_{\m \hat{\mathfrak{m}}\mathfrak{n}}\Bigg( -\mathcal{D}_{\tau\5}^2 +( j+\frac{1}{2})^2\Bigg)\, Z^{jm}_{\m \mathfrak{n}\hat{\mathfrak{m}}}
	+W^{\g jm}_{\m \mathfrak{n}\hat{\mathfrak{m}}} \Bigg( -\mathcal{D}_{\tau\5}^2 +( j+\frac{1}{2})^2\Bigg)\, W^{jm}_{\m \hat{\mathfrak{m}}\mathfrak{m}}&\nn\\
	&+\sum_{\hat{\mathfrak{m}},\mathfrak{n}=1}^N \sum_{\m=3}^4  \Bigg(
	Z^{\g jm}_{\m \hat{\mathfrak{m}}\mathfrak{n}}\Bigg( -\mathcal{D}_{\tau(n+1)}^2 +( j+\frac{1}{2})^2\Bigg)\, Z^{jm}_{\m \mathfrak{n}\hat{\mathfrak{m}}}
	+W^{\g jm}_{\m \mathfrak{n}\hat{\mathfrak{m}}} \Bigg( -\mathcal{D}_{\tau(n+1)}^2 +( j+\frac{1}{2})^2\Bigg)\, W^{jm}_{\m \hat{\mathfrak{m}}\mathfrak{m}}&
\end{align}
where the expression of $\mathcal{D}_\tau$ is, 
\be
\label{eqn:derivative_on_sphere}
\hspace*{-1cm} \mathcal{D}_{\tau\5}=\partial_\tau+ i\eta\frac{\alpha_\m^\mathfrak{n}-\alpha_\5^{\mathfrak{\hat{m}}}}{\beta+\beta^\prime} -\frac{\beta-\beta^\prime}{\beta+\beta^\prime}\, j_3+ \frac{\beta^\prime}{\beta+\beta^\prime}\, h\;.
\ee
One has to replace $\alpha_\5^{\mathfrak{\hat{m}}}$ by $\alpha_{(n+1)}^{\mathfrak{\hat{m}}}$ while writing $\mathcal{D}_{\tau(n+1)}$ and for the internal nodes its same as $\hat{A}$ case.
The one loop effective action is calculated following the same steps as before. Therefore we directly write the result below,
\be
\hspace*{-2.5cm}\Gamma(Z,W, Z^\g, W^\g)
&=&-
\sum_{\hat{\mathfrak{m}},\mathfrak{n}=1}^{2N}\sum_{\m=5}^n\sum_{j,j_3} \Bigg[\frac{i\eta}{2}(\alpha_\m^\mathfrak{n}-\alpha_\p^{\mathfrak{\hat{m}}})+\frac{\beta}{2} ( \epsilon_j+ j_3)+\frac{1}{2}\beta^\prime(\epsilon_j- j_3-h) \Bigg]+\cdots\nn\\
&+&\sum_{\hat{\mathfrak{m}},\mathfrak{n}=1}^{2N}\sum_{\m=5}^n\sum_{r=1}^\infty \frac{1}{r}\Big[ e^{-in(\alpha_\m^\mathfrak{n}-\alpha_\p^{\mathfrak{\hat{m}}})}f^{+B}_{\mathfrak{n}\hat{\mathfrak{m}}\m}(x^r, (x^\prime)^r) +e^{in(\alpha_\m^\mathfrak{n}-\alpha_\p^{\mathfrak{\hat{m}}})}f^{-B}_{\mathfrak{n}\hat{\mathfrak{m}}\m}(x^r, (x^\prime)^r)\Big]\nn\\
&+&\sum_{\hat{\mathfrak{m}},\mathfrak{n}=1}^N\sum_{\m=1}^2\sum_{r=1}^\infty \frac{1}{r}\Big[ e^{-in(\alpha_\m^\mathfrak{n}-\alpha_\5^{\mathfrak{\hat{m}}})}f^{+B}_{\mathfrak{n}\hat{\mathfrak{m}}\m}(x^r, (x^\prime)^r) +e^{in(\alpha_\m^\mathfrak{n}-\alpha_\5^{\mathfrak{\hat{m}}})}f^{-B}_{\mathfrak{n}\hat{\mathfrak{m}}\m}(x^r, (x^\prime)^r)\Big]\nn\\
&+&\sum_{\hat{\mathfrak{m}},\mathfrak{n}=1}^N\sum_{\m=3}^4\sum_{r=1}^\infty \frac{1}{r}\Big[ e^{-in(\alpha_\m^\mathfrak{n}-\alpha_{(n+1)}^{\mathfrak{\hat{m}}})}f^{+B}_{\mathfrak{n}\hat{\mathfrak{m}}\m}(x^r, (x^\prime)^r) +e^{in(\alpha_\m^\mathfrak{n}-\alpha_{(n+1)}^{\mathfrak{\hat{m}}})}f^{-B}_{\mathfrak{n}\hat{\mathfrak{m}}\m}(x^r, (x^\prime)^r)\Big]\nn\\
\ee
where, $\cdots$ denotes the counterparts for the external nodes and
\be 
\textit{for $\m=5,\cdots n$}
&&f^{+B}_{\m\mathfrak{n}\hat{\mathfrak{m}}}(x^r, (x^\prime)^r)=\sum_{j=\frac{|q_\m^{\mathfrak{n}}-q_\p^{\hat{\mathfrak{m}}}|}{2}}^\infty\Big(x^{ \frac{1}{2}}(x^\prime)^{2j+1}+x^{ \frac{3}{2}}(x^\prime)^{2j}+...+x^{ 2j+\frac{1}{2}}(x^\prime)\Big)\nn\\
&&+\sum_{j=\frac{|q_\m^{\mathfrak{n}}-q_\p^{\hat{\mathfrak{m}}}|}{2}}^\infty\Big(x^{ \frac{1}{2}}(x^\prime)^{2j}+x^{ \frac{3}{2}}(x^\prime)^{2j- 1}+...+x^{ 2j+\frac{1}{2}}\Big)=f_{\m\mathfrak{n}\hat{\mathfrak{m}}}^{-B}(x^r, (x^\prime)^r)\nn\\
\ee
\be 
\textit{for $\m=1,2$}
&&f^{+B}_{\mathfrak{n}\hat{\mathfrak{m}}\m}(x^r, (x^\prime)^r)=\sum_{j=\frac{|t_\m^{\mathfrak{n}}-q_\5^{\hat{\mathfrak{m}}}|}{2}}^\infty\Big(x^{ \frac{1}{2}}(x^\prime)^{2j+1}+x^{ \frac{3}{2}}(x^\prime)^{2j}+...+x^{ 2j+\frac{1}{2}}(x^\prime)\Big)\nn\\
&&+\sum_{j=\frac{|t_\m^{\mathfrak{n}}-q_\5^{\hat{\mathfrak{m}}}|}{2}}^\infty\Big(x^{ \frac{1}{2}}(x^\prime)^{2j}+x^{ \frac{3}{2}}(x^\prime)^{2j- 1}+...+x^{ 2j+\frac{1}{2}}\Big)=f_{\mathfrak{n}\hat{\mathfrak{m}}\m}^{-B}(x^r, (x^\prime)^r)\nn\\
\ee
\be 
\textit{for $\m=3,4$}
&&f^{+B}_{\mathfrak{n}\hat{\mathfrak{m}}\m}(x^r, (x^\prime)^r)=\sum_{j=\frac{|t_\m^{\mathfrak{n}}-q_{n+1}^{\hat{\mathfrak{m}}}|}{2}}^\infty\Big(x^{ \frac{1}{2}}(x^\prime)^{2j+1}+x^{ \frac{3}{2}}(x^\prime)^{2j}+...+x^{ 2j+\frac{1}{2}}(x^\prime)\Big)\nn\\
&&+\sum_{j=\frac{|t_\m^{\mathfrak{n}}-q_{n+1}^{\hat{\mathfrak{m}}}|}{2}}^\infty\Big(x^{ \frac{1}{2}}(x^\prime)^{2j}+x^{ \frac{3}{2}}(x^\prime)^{2j- 1}+...+x^{ 2j+\frac{1}{2}}\Big)=f_{\mathfrak{n}\hat{\mathfrak{m}}\m}^{-B}(x^r, (x^\prime)^r)\nn\\
\ee
\par The fermionic effective action follows simply from the previous case and we write down the result below,
\begin{align}
	&\Gamma(\xi,\xi^\g)
	= \sum_{\hat{\mathfrak{m}},\mathfrak{n}=1}^{2N}\sum_{\m=1}^n\sum_{j,j_3}\frac{1}{2}\Big( i\eta(\alpha_\m^\mathfrak{n}-\alpha_\p^{\mathfrak{\hat{m}}}) +\beta(\epsilon_j+j_3) +\beta^\prime(\epsilon_j- j_3-h) \Big)+\cdots&\nn\\
	&+\sum_{\hat{\mathfrak{m}},\mathfrak{n}=1}^{2N}\sum_{\m=1}^n\sum_{r=1}^\infty \frac{1}{r}\Big[ e^{-in(\alpha_\m^\mathfrak{n}-\alpha_\p^{\mathfrak{\hat{m}}})}f^{+F}_{\m\mathfrak{n}\hat{\mathfrak{m}}}(x^r, (x^\prime)^r) +e^{in(\alpha_\m^\mathfrak{n}-\alpha_\p^{\mathfrak{\hat{m}}})}f^{-F}_{\mathfrak{n}\hat{\mathfrak{m}}\m}(x^r, (x^\prime)^r)\Big]&\nn\\
&+\sum_{\hat{\mathfrak{m}},\mathfrak{n}=1}^N\sum_{\m=1}^2\sum_{r=1}^\infty \frac{1}{r}\Big[ e^{-in(\alpha_\m^\mathfrak{n}-\alpha_\5^{\mathfrak{\hat{m}}})}f^{+F}_{\mathfrak{n}\hat{\mathfrak{m}}\m}(x^r, (x^\prime)^r) +e^{in(\alpha_\m^\mathfrak{n}-\alpha_\5^{\mathfrak{\hat{m}}})}f^{-F}_{\mathfrak{n}\hat{\mathfrak{m}}\m}(x^r, (x^\prime)^r)\Big]&\nn\\
&+\sum_{\hat{\mathfrak{m}},\mathfrak{n}=1}^N\sum_{\m=3}^4\sum_{r=1}^\infty \frac{1}{r}\Big[ e^{-in(\alpha_\m^\mathfrak{n}-\alpha_{(n+1)}^{\mathfrak{\hat{m}}})}f^{+F}_{\mathfrak{n}\hat{\mathfrak{m}}\m}(x^r, (x^\prime)^r) +e^{in(\alpha_\m^\mathfrak{n}-\alpha_{(n+1)}^{\mathfrak{\hat{m}}})}f^{-F}_{\mathfrak{n}\hat{\mathfrak{m}}\m}(x^r, (x^\prime)^r)\Big]&
\end{align} 
where,
\be 
\textit{for $\m=5,\cdots, r$}
&&f^{+F}_{\mathfrak{n}\hat{\mathfrak{m}}\m}=\,\sum_{j=\frac{|q_\m^{\mathfrak{n}}-q_\p^{\hat{\mathfrak{m}}}|+1}{2}}^\infty\Big(x^{ \frac{1}{2}}(x^\prime)^{2j}+x^{ \frac{3}{2}}(x^\prime)^{2j- 1}+...+x^{ 2j+\frac{1}{2}}\Big)\nn\\ 
&&+\sum_{j=\frac{|q_\m^{\mathfrak{n}}-q_\p^{\hat{\mathfrak{m}}}|-1}{2}}^\infty\Big(x^{ \frac{1}{2}}(x^\prime)^{2j+1}+x^{ \frac{3}{2}}(x^\prime)^{2j}+...+x^{ 2j+\frac{1}{2}}(x^\prime)\Big)=f^{-F}_{\mathfrak{n}\hat{\mathfrak{m}}\m}\nn\\
\ee
\be 
\textit{for $\m=1,2$}
&&f^{+F}_{\mathfrak{n}\hat{\mathfrak{m}}\m}=\sum_{j=\frac{|t_\m^{\mathfrak{n}}-q_\5^{\hat{\mathfrak{m}}}|+1}{2}}^\infty\Big(x^{ \frac{1}{2}}(x^\prime)^{2j}+x^{ \frac{3}{2}}(x^\prime)^{2j- 1}+...+x^{ 2j+\frac{1}{2}}\Big)\nn\\ [2mm]
&&+\sum_{j=\frac{|t_\m^{\mathfrak{n}}-q_\5^{\hat{\mathfrak{m}}}|-1}{2}}^\infty\Big(x^{ \frac{1}{2}}(x^\prime)^{2j+1}+x^{ \frac{3}{2}}(x^\prime)^{2j}+...+x^{ 2j+\frac{1}{2}}(x^\prime)\Big)=f^{-F}_{\mathfrak{n}\hat{\mathfrak{m}}\m}\nn\\
\ee 
\be 
\textit{for $\m=3,4$}
&&f^{+F}_{\mathfrak{n}\hat{\mathfrak{m}}\m}=\sum_{j=\frac{|t_\m^{\mathfrak{n}}-q_{n+1}^{\hat{\mathfrak{m}}}|+1}{2}}^\infty\Big(x^{ \frac{1}{2}}(x^\prime)^{2j}+x^{ \frac{3}{2}}(x^\prime)^{2j- 1}+...+x^{ 2j+\frac{1}{2}}\Big)\nn\\ [2mm]
&&+\sum_{j=\frac{|t_\m^{\mathfrak{n}}-q_{n+1}^{\hat{\mathfrak{m}}}|-1}{2}}^\infty\Big(x^{ \frac{1}{2}}(x^\prime)^{2j+1}+x^{ \frac{3}{2}}(x^\prime)^{2j}+...+x^{ 2j+\frac{1}{2}}(x^\prime)\Big)=f^{-F}_{\mathfrak{n}\hat{\mathfrak{m}}\m}\;.\nn\\
\ee 
Now, adding contributions from matter scalars and fermions we find,
\begin{align}
	&\Gamma(\mathrm{ matter})
	=- \sum_{\hat{\mathfrak{m}},\mathfrak{n}=1}^{2N}\sum_{\m=1}^n\sum_{j,j_3}\frac{1}{2}(-1)^F\Big( i\eta(\alpha_\m^\mathfrak{n}-\alpha_\p^{\mathfrak{\hat{m}}}) +\beta(\epsilon_j+j_3) +\beta^\prime(\epsilon_j- j_3-h) \Big)+\cdots&\nn\\
	&+\sum_{\hat{\mathfrak{m}},\mathfrak{n}=1}^{2N}\sum_{\m=1}^n\sum_{r=1}^\infty \frac{1}{r}\Big[ e^{-ir(\alpha_\m^\mathfrak{n}-\alpha_\p^{\mathfrak{\hat{m}}})}f^{+}_{\mathfrak{n}\hat{\mathfrak{m}}\m}(x^r) +e^{ir(\alpha_\m^\mathfrak{n}-\alpha_\p^{\mathfrak{\hat{m}}})}f^{-}_{\mathfrak{n}\hat{\mathfrak{m}}\m}(x^r)\Big]&\nn\\
	&+\sum_{\hat{\mathfrak{m}},\mathfrak{n}=1}^N\sum_{\m=1}^2\sum_{r=1}^\infty \frac{1}{r}\Big[ e^{-ir(\alpha_\m^\mathfrak{n}-\alpha_\5^{\mathfrak{\hat{m}}})}f^{+}_{\mathfrak{n}\hat{\mathfrak{m}}\m}(x^r) +e^{ir(\alpha_\m^\mathfrak{n}-\alpha_\5^{\mathfrak{\hat{m}}})}f^{-}_{\mathfrak{n}\hat{\mathfrak{m}}\m}(x^r)\Big]&\nn\\
	&+\sum_{\hat{\mathfrak{m}},\mathfrak{n}=1}^N\sum_{\m=3}^4\sum_{r=1}^\infty \frac{1}{r}\Big[ e^{-ir(\alpha_\m^\mathfrak{n}-\alpha_{(n+1)}^{\mathfrak{\hat{m}}})}f^{+}_{\mathfrak{n}\hat{\mathfrak{m}}\m}(x^r) +e^{ir(\alpha_\m^\mathfrak{n}-\alpha_{(n+1)}^{\mathfrak{\hat{m}}})}f^{-}_{\mathfrak{n}\hat{\mathfrak{m}}\m}(x^r)\Big]&
\end{align} 
Adding the series one obtains,
\be 
f^{+}_{\mathfrak{n}\hat{\mathfrak{m}}\m}(x^r)=x^{|q_\m^{\mathfrak{n}}-q_\p^{\hat{\mathfrak{m}}}|}.\frac{x^{\frac{1}{2}}}{1-x^2}-x^{|q_\m^{\mathfrak{n}}-q_\p^{\hat{\mathfrak{m}}}|}.\frac{x^{\frac{3}{2}}}{1-x^2}=f^{-}_{\mathfrak{n}\hat{\mathfrak{m}}\m}(x^r)
\ee 
for $\m=5,\cdots, n$. The external nodes are straight forward from the above.
\subsubsection{Adjoint sector}
This computation is fairly straight forward from the $\widehat{A}$ case and we directly write the result below,
\begin{align}
	&\Gamma(\mathrm{ adjoint})
	=\sum_{{\mathfrak{m}},\mathfrak{n}=1}^{2N}\sum_{\m=5}^{r+1}\sum_{r=1}^\infty \frac{1}{r}\Big[ e^{-ir(\alpha_\m^\mathfrak{n}-\alpha_\m^{\mathfrak{{m}}})}f^{\mathrm{adj}}_{\mathfrak{n}{\mathfrak{m}}\m}(x^r)\Big]+\sum_{{\mathfrak{m}},\mathfrak{n}=1}^{N}\sum_{\m=1}^4\sum_{r=1}^\infty \frac{1}{r}\Big[ e^{-ir(\alpha_\m^\mathfrak{n}-\alpha_\m^{\mathfrak{{m}}})}f^{\mathrm{adj}}_{\mathfrak{n}{\mathfrak{m}}\m}(x^r)\Big]&
\end{align} 
where,
\be 
\textit{for $\m=5,\cdots , r+1$}
&&f^{\mathrm{adj}}_{\mathfrak{n}{\mathfrak{m}}\m }(x)=(-1+\delta_{q_{\mathfrak{m}}q_{\mathfrak{n}}})\,x^{|q^\mathfrak{m}_\m-q^\mathfrak{n}_\m|}\nn\\
\textit{for $\m=1,\cdots , 4$}
&&f^{\mathrm{adj}}_{\mathfrak{n}{\mathfrak{m}}\m }(x)=(-1+\delta_{q_{\mathfrak{m}}q_{\mathfrak{n}}})\,x^{|t^\mathfrak{m}_\m-t^\mathfrak{n}_\m|}
\ee

\section{$\widehat{E}_6$ quiver}\label{sec_e6_index}
\begin{figure}[h]
	\centering
	\includegraphics[width=5in]{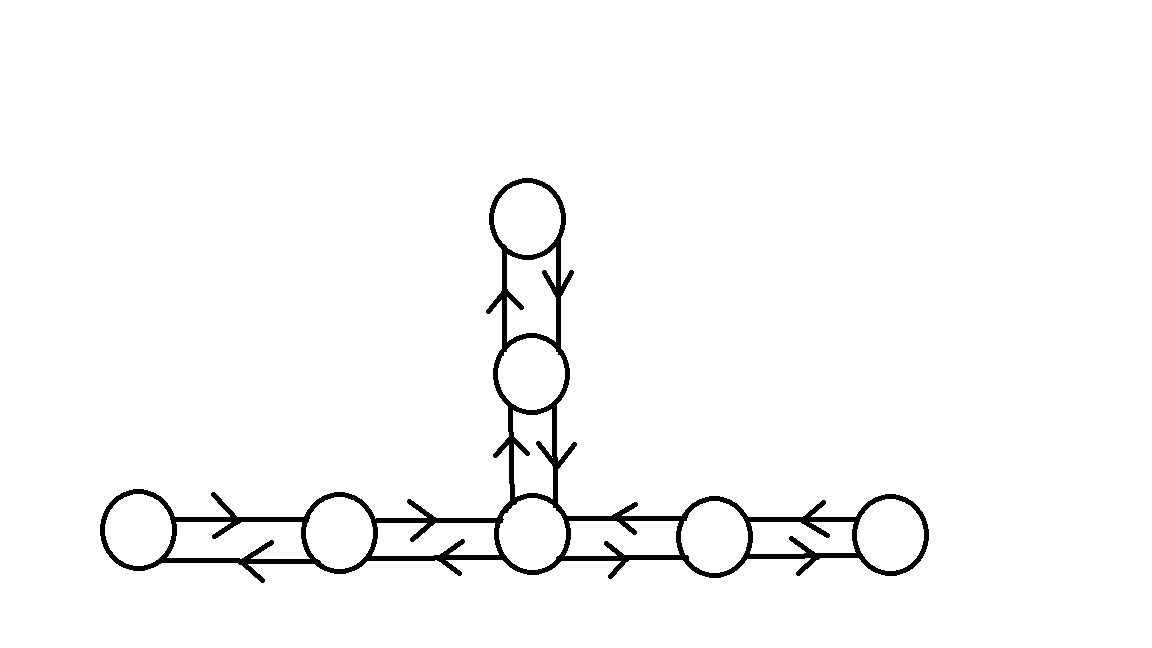}
	\put (-324,32) {$N$} \put (-322,55) {$1$} \put (-261,32) {$2N$} \put (-261,55) {$2$} \put (-200,32) {$3N$}\put (-198,10) {$7$}\put (-199,130) {$N$} \put (-180,132) {$5$} \put (-200,82) {$2N$} \put (-178,82) {$6$} 
	\put (-144,30) {$2N$} \put (-142,50) {$4$} \put (-84,30) {$N$} \put (-84,50) {$3$}
	\caption{$\widehat{E}_6$ quiver diagram with gauge group $U(2N)^3\times U(N)^3\times U(3N)$. The numbers beside the nodes are labelling of their $\m$ values. The CS levels satisfy $ 
		\kappa_\1+\kappa_\3+\kappa_\5 +2\big(\kappa_\2+\kappa_\4+ \kappa_\6\big)+ 3\kappa_\7=0 $.}
	\label{e6_rank}
\end{figure}
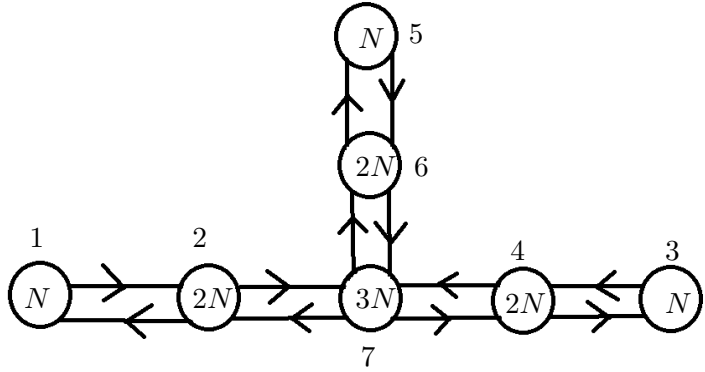

The index for $\widehat{E}$ quiver is  straight forward from the $\widehat{D}$ case. Hence  we do not present the calculations in detail. The quiver diagram is shown in \ref{e6_rank}. 
The matter part of the action is,
\begin{align}
	&\mathcal{L}_{\mathrm{mat}}=
	\sum_{\m=2,4,6} \Tr\Big[-\mathcal{D}_m Z_{\m}\mathcal{D}^m Z^\dagger_{\m}+i\zeta^\dagger_{\m}\slashed{\mathcal{D}} \zeta_{\m}  - i \zeta^\dagger_{\m}\big(\sigma_{\m}\zeta_{\m}-\zeta_{\m}{\sigma}_\7\big)- Z^\dagger_{\m} Z_{\m}\sigma_\7^2  &\nn\\&+  2 Z^\dagger_{\m}\sigma_{\m} Z_{\m}{\sigma}_\7-Z^\dagger_{\m} \sigma^2_{\m} Z_{\m}
	-(\mathcal{D}_m W_{\m})(\mathcal{D}^m W^\g_{\m})+i\omega_{\m}\slashed{\mathcal{D}} \omega^\dagger_{\m} 
	+ i \omega^\dagger_{\m}\big(\omega_{\m}\sigma_{\m}-{\sigma}_\7\omega_{\m}\big) &\nn\\-& W^\dagger_{\m} W_{\m}\sigma^2_{\m} +  2 W^\dagger_{\m}{\sigma}_\7 W_{\m}{\sigma}_{\m} -W^\dagger_{\m} {\sigma}^2_\5 W_{\m} \Big] + \sum_{\m=1,3,5} \Tr\Big[-\mathcal{D}_m Z_{\m}\mathcal{D}^m Z^\dagger_{\m}+i\zeta^\dagger_{\m}\slashed{\mathcal{D}} \zeta_{\m} &\nn\\ &- i \zeta^\dagger_{\m}\big(\sigma_{\m}\zeta_{\m}-\zeta_{\m}{\sigma}_{\p}\big)- Z^\dagger_{\m} Z_{\p}{\sigma^2}_{\p} +  2 Z^\dagger_{\m}\sigma_{\m} Z_{\m}{\sigma}_{\p} -Z^\dagger_{\m} \sigma^2_{\m} Z_{\m}\Big]&\nn\\
	& -(\mathcal{D}_m W_{\m})(\mathcal{D}^m W^\dagger_{\m})+i\omega_{\m}\slashed{\mathcal{D}} \omega^\dagger_{\m}+ i \omega^\dagger_{\m}\big(\omega_{\m}\sigma_{\m}-{\sigma}_{\p}\omega_{\m}\big)& \nn\\&- W^\dagger_{\m} W_{\m}\sigma^2_{\m} +  2 W^\dagger_{\m}{\sigma}_{\p} W_{\m}{\sigma}_{\m} -W^\dagger_{\m} {\sigma}^2_{\p} W_{\m} \Big] &
\end{align}
and the rest of the action is similar to the previous cases. Localization leads us to the following saddle point equations for gauge fields.
\be
\label{eqn:gauge_e6}
A_\m &=&\frac{E_\m}{2}\,\,(\pm1-\cos\theta)\,d\varphi\quad for\,\,\, \m=1,3,5\,,\nn\\
A_\m &=&\frac{I_\m}{2}\,\,(\pm1-\cos\theta)\,d\varphi\quad for\,\,\, \m=2,4,6\,,\nn\\
A_\7 &=&\frac{C_\7}{2}\,\,(\pm1-\cos\theta)\,d\varphi\quad for\,\,\, \m=7\,,
\ee
where, $E_\m=\mathrm{diag}(t^1_\m,t^2_\m,...t_\m^N), I_\m=\mathrm{diag}(q^1_\m,q^2_\m,...q^{2N}_\m), C_\7=\mathrm{diag}(c^1_\7,c^2_\7,...c^{3N}_\7)$. We also have similar expressions for the $\sigma_\m$'s.
The final expression of the index is,
\be\label{final_index_e6}
&&\hspace*{-2.5cm}I(x)=
\mathcal{C}\int\frac{1}{\rm (symmetry)}
	\prod_{\m=1}^7 	\left[\frac{d\alpha_\m^{\mathfrak{m}}}{2\pi}\right]	\prod_{\mathfrak{{m}}<\mathfrak{{n}}}
	\left[2\sin\left(\frac{\alpha_\m^\mathfrak{m}\!-\!\alpha_\m^\mathfrak{n}}{2}\right)\right]^2\times e^{i\sum_{\m=2,4,6} k_\m\sum_{\mathfrak{m}=1}^{2N} q_\m^\mathfrak{m}\alpha_\m^\mathfrak{{m}}\!}\nn\\
	&&\hspace*{-2.5cm}\times e^{i\sum_{\m=1,3,5} k_\m\sum_{\mathfrak{m}=1}^N t_\m^\mathfrak{m}\alpha_\m^\mathfrak{{m}}\!}
	\times e^{i k_\7\sum_{\mathfrak{m}=1}^{3N} c_\7^\mathfrak{m}\alpha_\7^\mathfrak{{m}}\!}\times \prod_{\hat{\mathfrak{m}},\mathfrak{n}=1}^{N}\exp\Bigg[\sum_{\m=1,3,5}\sum_{r=1}^\infty \frac{1}{r}\Big[ e^{-ir(\alpha_\m^\mathfrak{n}-\alpha_\p^{\mathfrak{\hat{m}}})}f^{+}_{\mathfrak{n}\hat{\mathfrak{m}}\m}(x^r)\nn\\ &&\hspace*{-2.5cm}+e^{ir(\alpha_\m^\mathfrak{n}-\alpha_\p^{\mathfrak{\hat{m}}})}f^{-}_{\mathfrak{n}\hat{\mathfrak{m}}\m}(x^r)\Bigg]\times \prod_{\hat{\mathfrak{m}},\mathfrak{n}=1}^{2N}\exp\Bigg[\sum_{\m=2,4,6}\sum_{r=1}^\infty \frac{1}{r}\Big[ e^{-ir(\alpha_\m^\mathfrak{n}-\alpha_\7^{\mathfrak{\hat{m}}})}f^{+}_{\mathfrak{n}\hat{\mathfrak{m}}\m}(x^r) +e^{ir(\alpha_\m^\mathfrak{n}-\alpha_\7^{\mathfrak{\hat{m}}})}f^{-}_{\mathfrak{n}\hat{\mathfrak{m}}\m}(x^r)\Bigg]\nn\\
	&&\hspace*{-2.5cm}\times\!\prod_{\mathfrak{{m}},\mathfrak{{n}}=1}^{2N}\!\exp\left[\sum_{r=1}^\infty\frac{1}{r}
	\sum_{\m=2,4,6} f^{{\rm adj}}_{\mathfrak{{m}}\mathfrak{{n}}\m}(x^r)e^{-ir(\alpha^\mathfrak{{m}}_\m\!-\!\alpha^\mathfrak{n}_\m)}
	\right]\times\!\prod_{\mathfrak{{m}},\mathfrak{{n}}=1}^{N}\!\exp\left[\sum_{r=1}^\infty\frac{1}{r}
	\sum_{\m=1,3,5} f^{{\rm adj}}_{\mathfrak{{m}}\mathfrak{{n}}\m}(x^r)e^{-ir(\alpha^\mathfrak{{m}}_\m\!-\!\alpha^\mathfrak{n}_\m)}
	\right]\nn\\
	&& \hspace*{-2.5cm} \times\!\prod_{\mathfrak{{m}},\mathfrak{{n}}=1}^{3N}\!\exp\left[\sum_{r=1}^\infty\frac{1}{r}
	f^{{\rm adj}}_{\mathfrak{{m}}\mathfrak{{n}}\7}(x^r)e^{-ir(\alpha^\mathfrak{{m}}_\7\!-\!\alpha^\mathfrak{n}_\7)}
	\right]
\ee
The expressions of the letter indices is very similar to the $\widehat{D}$ case. One has to write the summation limit of $\m$ by keeping in mind the appropriate representation of the associated fields. 
\section{Large $N$ limit of  the index }\label{sec:large_N}
In this section we compute the large $N$, i.e $N\to \infty$ limit of the index $I(x)$. This limit is necessary to match the index with that of the dual gravity side.
\subsection{$\widehat{A}$-type quiver}\label{large_N_A}
We want to do a $\frac{1}{N}$ expansion of the integral in eqn.\eqref{final_index_an} and then take $N\to \infty$ limit. To do so one has to know the magnetic flux configuration $\{q^\mm_\m\}$'s, i.e the diagonal entries of the $H_\m$'s. The magnetic fluxes are constrained by the 
 Gauss law constraint,
\be 
\label{eqn:constraint_magcharges_an}
 \sum_{\m=1}^n k_\m m_\m=0,\quad where\quad  m_\m=\frac{1}{2\pi}\int \Tr F_\m,\quad  \Tr F_\m=\sum_{\mathfrak{m}=1}^N q^{\mathfrak{{m}}}_\m\;. 
\ee 
We restrict ourselves to the case of diagonal monopole operators which have,
\be 
m_\1=m_\2=\cdots= m_{(n)}=m_{\rm diag}\;.
\ee 
Therefore there are $n$ magnetic charges constrained by eqn.\eqref{eqn:constraint_magcharges_an} which makes $n-1$  magnetic fluxes to be independent. Out of this $n-1$ fluxes one corresponds to the M-direction with momentum $m_{\rm diag}$.
\par For the  $1/N$ expansion we can choose the flux distribution to be - all zero or some zero and rest non zero. These two cases are analyzed below one by one. For each node $\m$ the flux configuration is described by a set of $N$ integers $\{q_\m^1, q_\m^2,\cdots q_\m^N  \}$. Let, $N_{\m}$ is the number of $U(1)$'s with zero flux and $M_\m$ be the number of $U(1)$ with non zero flux, i.e 
\be 
N=N_{\m}+M_\m\;.
\ee 
Now, to carry out $\frac{1}{N}$ expansion we define,
\be 
\rho_{r\m}= \frac{1}{N_{\m}}\sum_{\mm=1}^{N_{\m}} e^{-ir\alpha^\mm_\m}=\frac{1}{N-M_{\m}}\sum_{\mm=1}^{N_{\m}} e^{-ir\alpha^\mm_\m},\quad r\in \mathbbm{Z}, r\neq 0\;.
\ee 
\underline{\textbf{Zero flux sector:}}\\
In this case the index integral becomes,
\begin{align}
&			I_0(x)=
\int\frac{1}{\rm (symmetry)}
\left[\frac{d\alpha_\m^{\mathfrak{m}}}{(2\pi)^n}\right]
\prod_{\m=1}^n	\prod_{\mathfrak{{m}}<\mathfrak{{n}}}
\left[2\sin\left(\frac{\alpha_\m^\mathfrak{m}\!-\!\alpha_\m^\mathfrak{n}}{2}\right)\right]^2\nn\\
&\prod_{\hat{\mathfrak{m}},\mathfrak{n}=1}^N\exp\Bigg[\sum_{\m=1}^n\sum_{r=1}^\infty \frac{1}{r}\Big[ e^{-in(\alpha_\m^\mathfrak{n}-\alpha_\p^{\mathfrak{\hat{m}}})}f^{+}_{\mathfrak{n}\hat{\mathfrak{m}}\m}(x^r) +e^{in(\alpha_\m^\mathfrak{n}-\alpha_\p^{\mathfrak{\hat{m}}})}f^{-}_{\mathfrak{n}\hat{\mathfrak{m}}\m}(x^r)\Bigg]&
\end{align}
where,
\be 
f^{+}_{\mathfrak{n}\hat{\mathfrak{m}}\m}(x)=x^{\frac{1}{2}+|q_\m^{\mathfrak{n}}-q_\p^{\hat{\mathfrak{m}}}|}.\frac{1}{1-x^2}-x^{\frac{3}{2}+|q_\m^{\mathfrak{n}}-q_\p^{\hat{\mathfrak{m}}}|}.\frac{1}{1-x^2}=\frac{x^{\frac{1}{2}}}{1-x^2}-\,\frac{x^{\frac{3}{2}}}{1-x^2}=f^+=f^-\nn\\
\ee 
and the contribution from the adjoint sector vanishes
\be 
f^{\mathrm{adj}}_{\mathfrak{n}{\mathfrak{m}}\m }(x)=(-1+\delta_{q_{\mathfrak{m}}q_{\mathfrak{n}}})\,x^{|q^\mathfrak{m}_\m-q^\mathfrak{n}_\m|}=0
\ee 
Since the Fadeev-Popov determinant occurs for the zero flux sector, we write it in the following way 
\be 
\prod_{\mm<\N}\left[2\sin\left(\frac{\alpha_\m^\mm-\!\alpha_\m^\N}{2}\right)\right]^2 =\prod_{\mm\neq \N}\exp\Big[-\sum_{r=1}^\infty \frac{1}{r} e^{-ir(\alpha_\m^\mm-\alpha_\m^\N)}\Big]= \prod\exp\Big[-N_{\m}^2\sum_{r=1}^\infty \frac{\rho_{r\m}\, \rho_{-r\m}}{r} \Big]\nn\\
\ee 
The final expression of the index  in terms of  $\rho_{r\m}$'s is,
\be
&&	I_0(x)=
\int\frac{1}{\rm (symmetry)}
\prod_{\m=1}^n\left[d\rho_{r\m}\right]	\prod_{\hat{\mathfrak{m}},\mathfrak{n}=1}^N
\exp\Big[-N_{\m}^2\sum_{r=1}^\infty \frac{\rho_{r\m}\, \rho_{-r\m}}{r} \Big]\nn\\
&&\exp\Bigg[\sum_{\m=1}^n\sum_{r=1}^\infty \frac{1}{r}\Big[ e^{-ir(\alpha_\m^\mathfrak{n}-\alpha_\p^{\mathfrak{\hat{m}}})}f^+(x^r) +e^{ir(\alpha_\m^\mathfrak{n}-\alpha_\p^{\mathfrak{\hat{m}}})}f(x^r)\Bigg]\nn\\
&&=
\int\frac{1}{\rm (symmetry)}
\left[\frac{d\alpha_\m^{\mathfrak{m}}}{(2\pi)^n}\right]
\exp\Bigg[\sum_{\m=1}^n\sum_{r=1}^\infty \frac{1}{r}\Bigg(-N_{\m}^2 \rho_{r\m}\, \rho_{-r\m}\nn\\
&& +N_{\m} N_{\p}\, f^+(x^r)\, \rho_{r\m} \rho_{-r\p} +N_{\m} N_{\p}f^-(x^r)\,\rho_{r\p} \rho_{-r\m} \Bigg)\Bigg]\;.
\ee
Writing the above as a matrix equation and taking $N$ to be very large, i.e $N-M_\m\sim N $ we find,
\be \label{eqn:largeN_zero_an}
&&	I_0(x)=
\int\frac{1}{\rm (symmetry)}
\left[\frac{d\alpha_\m^{\mathfrak{m}}}{(2\pi)^n}\right]
\exp\Bigg[N^2\sum_{\m,\s=1}^r\sum_{r=1}^\infty \frac{1}{r}\Bigg( \rho_{r\m}\,\mathcal{N}_{\m\s}(x^r) \rho_{-r\s} \Bigg)\Bigg]\nn\\
\ee
where,
\be 
\mathcal{N}(x)=\begin{pmatrix}
	1 & f^+&0 &\cdots & &f^-\\
	f^-&1&f^+& 0&\cdots\\
	0&f^-&1& f^+&0\\
	&&&\ddots&&\\
	&&&&&\\
	f^+&&&&&1
\end{pmatrix}\;.
\ee 
It is easy to see that the zero point energy vanishes in the zero flux sector. Observe that the index in this case is of $\mathcal{O}(\frac{1}{N^2})$ since $\rho_{r\m}$ is of $\mathcal{O}(\frac{1}{N})$ in the large $N$ limit.
In the large $N$ limit various $\rho_r$'s are treated as independent variables and one can perform a Gaussian integral to obtain $I_0(x)$ which is,
\be 
I_0(x)=\frac{1}{\det \mathcal{N}(x^r) }\;.
\ee 
To evaluate the determinant we write $\NN(x)$ as
\be 
\NN(x)=\frac{1}{1+x}(1-A)(1-B)
\ee 
where,
\be 
A=\begin{pmatrix}
	0& \sqrt{x}&0&\cdots\\
	0&0&\sqrt{x}&0\cdots\\
	\vdots&&\ddots\\
	\sqrt{x}&&&\cdots0
\end{pmatrix},\qquad B=\begin{pmatrix}
	0&0&\cdots&\sqrt{x}\\
	\sqrt{x}&0&0&0\cdots\\
	\vdots&&\ddots\\
	0&&&\cdots0
\end{pmatrix}\;.
\ee 
Using the relation,
\be 
\det (1+tA)=1+t\Tr A+\frac{t^2}{2!}\Big((\Tr A)^2- \Tr A^2\Big)++\frac{t^3}{3!}\Big((\Tr A)^3-3(\Tr A)(\Tr A^2)+ \Tr A^3\Big)+\cdots+t^n \det A\nn
\ee 
we obtain,
\be 
\det \mathcal{N}(x)=\frac{(1-x^{\frac{n}{2}})(1-x^{\frac{n}{2}})}{(1+x)^n}\;.
\ee 
\underline{\textbf{Non-zero flux sector:}}\\
In this  case  we turn on some non zero magnetic fluxes,
\be
q_\m^\mathfrak{{m}}=0 \quad \mm\in N_\m,\quad q_\m^\mathfrak{{m}}\neq0\quad \mm \in M_\m 
\ee 
i.e modes connecting zero and non zero flux. 
Substituting the above in eqn.\eqref{final_index_an}  we find,
\begin{align}\label{large_an_nonzero}
&			I(x)=
x^{\epsilon_0}\int\frac{1}{\rm (symmetry)}
\left[\frac{d\alpha_\m^{\mathfrak{m}}}{(2\pi)^2}\right]
\prod_{\m=1}^n	\prod_{\mathfrak{{m}}<\mathfrak{{n}}}
\left[2\sin\left(\frac{\alpha_\m^\mathfrak{m}\!-\!\alpha_\m^\mathfrak{n}}{2}\right)\right]^2\times e^{i\sum_{\m=1}^n k_\m\sum_{\mathfrak{m}=1}^{M_\m} q_\m^\mathfrak{m}\alpha_\m^\mathfrak{{m}}\!}&\nn\\
&\prod_{\hat{\mathfrak{m}},\mathfrak{n}=1}^{M_\m}\exp\Bigg[\sum_{\m=1}^n\sum_{r=1}^\infty \frac{1}{r}\Big[ e^{-ir(\alpha_\m^\mathfrak{n}-\alpha_\p^{\mathfrak{\hat{m}}})}F^{+}_{\mathfrak{n}\hat{\mathfrak{m}}\m}(x^r) +e^{ir(\alpha_\m^\mathfrak{n}-\alpha_\p^{\mathfrak{\hat{m}}})}F^{-}_{\mathfrak{n}\hat{\mathfrak{m}}\m}(x^r)\Bigg]&\nn \\
&			\times\!\prod_{\mathfrak{{m}},\mathfrak{{n}}=1}^{M_\m}\!\exp\left[\sum_{r=1}^\infty\frac{1}{r}
\sum_{\m=1}^n			F^{{\rm adj}}_{\mathfrak{{m}}\mathfrak{{n}}\m}(x^r)e^{-ir(\alpha^\mathfrak{{m}}_\m\!-\!\alpha^\mathfrak{n}_\m)}
\right]&
\end{align}
where,
\begin{align}
&F^{+}_{\mathfrak{n}\hat{\mathfrak{m}}\m}(x)=\Big(x^{|q_\m^{\mathfrak{n}}-q_\p^{\hat{\mathfrak{m}}}|}-x^{|q_\m^{\mathfrak{n}}|+|q_\p^{\hat{\mathfrak{m}}}|}\Big)f^+&\nn\\
&F^{-}_{\mathfrak{n}\hat{\mathfrak{m}}\m}(x)=\Big(x^{|q_\m^{\mathfrak{n}}-q_\p^{\hat{\mathfrak{m}}}|}-x^{|q_\m^{\mathfrak{n}}|+|q_\p^{\hat{\mathfrak{m}}}|}\Big)f^-&\nn\\
&F^{\mathrm{adj}}_{\mathfrak{n}{\mathfrak{m}}\m }(x)=(-1+\delta_{q_{\mathfrak{m}}q_{\mathfrak{n}}})\,x^{|q^\mathfrak{m}_\m-q^\mathfrak{n}_\m|}+x^{|q^\mathfrak{m}_\m|+|q^\mathfrak{n}_\m|}&\nn\\
&\epsilon_0
=\frac{1}{2}\sum_{\m=1}^n\sum_{\mathfrak{m}\in M_\m}\Big(N_\p|q^\mathfrak{m}_\m|+N_\m|q^\mathfrak{m}_\p|\Big) -\sum_{\m=1}^n\sum_{\mathfrak{m}\in M_\m} N_\m|q^\mathfrak{m}_\m|&
\end{align}
Furthermore, using the fact that $x^{|q_\m^{\mathfrak{n}}-q_\p^{\hat{\mathfrak{m}}}|}=x^{|q_\m^{\mathfrak{n}}|+|q_\p^{\hat{\mathfrak{m}}}|}
$ for fluxes with opposite sign, one can show that the index is completely factorized in zero, positive and negative flux sectors, i.e $I=I_0\, I^+\, I^-$ \cite{Imamura:2009hc}. 
\subsection{$\widehat{D}$-type quiver}\label{sec:large_N_D}
In this case the magnetic fluxes are constrained by the following relation,
\be 
\label{eqn:constraint_magcharges_dn}
 \sum_\m  m_\m k_\m =0,\quad m_\m=\frac{1}{2\pi}\int \Tr F_\m,\quad \m=1,...,n+1\;.
\ee 
\underline{\textbf{Zero flux sector}}\\
The index in eqn.\eqref{final_index_dn} for $\{q^\mm_\m\}=0$ reduces to,
\begin{align}
&			I_0(x)=
\int\frac{1}{\rm (symmetry)}
\prod_{\m=1}^{n+1}\left[\frac{d\alpha_\m^{\mathfrak{m}}}{2\pi}\right]	\prod_{\mathfrak{{m}}<\mathfrak{{n}}}
\left[2\sin\left(\frac{\alpha_\m^\mathfrak{m}\!-\!\alpha_\m^\mathfrak{n}}{2}\right)\right]^2\times &\nn\\
&\times \prod_{\hat{\mathfrak{m}},\mathfrak{n}=1}^{2N}\exp\Bigg[\sum_{\m=5}^n\sum_{r=1}^\infty \frac{1}{r}\Big[ e^{-ir(\alpha_\m^\mathfrak{n}-\alpha_\p^{\mathfrak{\hat{m}}})}f^{+}(x^r) +e^{ir(\alpha_\m^\mathfrak{n}-\alpha_\p^{\mathfrak{\hat{m}}})}f^{-}(x^r)\Bigg]&\nn \\
&\prod_{\hat{\mathfrak{m}},\mathfrak{n}=1}^N\exp\Bigg[\sum_{\m=1}^2\sum_{r=1}^\infty \frac{1}{r}\Big[ e^{-ir(\alpha_\m^\mathfrak{n}-\alpha_\5^{\mathfrak{\hat{m}}})}f^{+}(x^r) +e^{ir(\alpha_\m^\mathfrak{n}-\alpha_\5^{\mathfrak{\hat{m}}})}f^{-}(x^r)\Bigg]&\nn\\
&\times\prod_{\hat{\mathfrak{m}},\mathfrak{n}=1}^N\exp\Bigg[\sum_{\m=3}^4\sum_{r=1}^\infty \frac{1}{r}\Big[ e^{-ir(\alpha_\m^\mathfrak{n}-\alpha_{(n+1)}^{\mathfrak{\hat{m}}})}f^{+}(x^r) +e^{ir(\alpha_\m^\mathfrak{n}-\alpha_{(n+1)}^{\mathfrak{\hat{m}}})}f^{-}(x^r)\Bigg]&
\end{align}
Like the previous case we define,
\be 
\rho_{r\m}= \frac{1}{N_{\m}}\sum_{\mm=1}^{N_{\m}} e^{-ir\alpha^\mm_\m}\nn
\ee 
in terms of which the index  is,
\begin{align}
&			I_0(x)=
\int\frac{1}{\rm (symmetry)}
\left[\frac{d\alpha_\m^{\mathfrak{m}}}{(2\pi)^{n+1}}\right]
\prod_{\m=1}^{n+1}	\exp\Big[-N^2\sum_{r=1}^\infty \frac{\rho_{r\m}\, \rho_{-r\m}}{r} \Big]\times &\nn\\
&\times \prod_{\hat{\mathfrak{m}},\mathfrak{n}=1}^{2N}\exp \Bigg[N^2\sum_{\m=5}^n\sum_{r=1}^\infty \frac{1}{r}\Big[  f^+(x^n)\, \rho_{r\m} \rho_{-r\p} +f^-(x^n)\,\rho_{r\p} \rho_{-r\m}\Bigg]&\nn \\
&\prod_{\hat{\mathfrak{m}},\mathfrak{n}=1}^N\exp \Bigg[N^2\sum_{\m=1}^2\sum_{n=1}^\infty \frac{1}{r}\Big[ \rho_{r\m} \rho_{-r\5} f^{+}(x^r) +\rho_{-r\m} \rho_{r\5}f^{-}(x^r)\Bigg]&\nn\\
&\times\prod_{\hat{\mathfrak{m}},\mathfrak{n}=1}^N\exp\Bigg[N^2\sum_{\m=3}^4\sum_{r=1}^\infty \frac{1}{r}\Big[ \rho_{r\m} \rho_{-r(n+1)}f^{+}(x^r) +\rho_{-r\m} \rho_{r(n+1)}f^{-}(x^r)\Bigg]\;.&
\end{align}
Writing the above as a matrix equation we obtain,
\be
&&	I_0(x)=
\int\frac{1}{\rm (symmetry)}
\left[\frac{d\alpha_\m^{\mathfrak{m}}}{(2\pi)^{n+1}}\right]
\exp\Bigg[\sum_{\m,\s=1}^{n+1}\sum_{r=1}^\infty \frac{1}{r}\Bigg( \rho_{r\m}\,\mathcal{N}_{\m\s}(x^r) \rho_{-r\s} \Bigg)\Bigg]\nn\\
\ee
where,
\be 
\mathcal{N}(x)=\begin{pmatrix}
	1 &0&0&0& f^+&0 &\cdots  &0\\
	0&1&0& 0&f^+&\cdots\\
	0&0&1& 0&0&\cdots & &f^+\\
	0&0&0& 1&0&\cdots & &f^+\\
	f^-&f^-&0& 0&1&f^+ &\cdots &0\\
	0&0&0& 0&f^-&1 &\cdots &0\\
	\vdots&&&\ddots&&\\
	&&&&&\\
	0&0&f^-&f^-&&&f^-&1
\end{pmatrix}\;.
\ee 
\underline{\textbf{Non-zero flux sector}}\\
The index integral in eqn.\eqref{final_index_dn} for non zero fluxes reduces to the following,
\begin{align}\label{index_dn_large}
&			I(x)=
x^{\epsilon_0}\int\frac{1}{\rm (symmetry)}
\left[\frac{d\alpha_\m^{\mathfrak{m}}}{(2\pi)^{n+1}}\right]
\prod_{\m=1}^{n+1}	\prod_{\mathfrak{{m}}<\mathfrak{{n}}}
\left[2\sin\left(\frac{\alpha_\m^\mathfrak{m}\!-\!\alpha_\m^\mathfrak{n}}{2}\right)\right]^2\times e^{i\sum_{\m=5}^{n+1} k_\m\sum_{\mathfrak{m}=1}^{ M_\m} q_\m^\mathfrak{m}\alpha_\m^\mathfrak{{m}}\!}&\nn\\
&\times e^{i\sum_{\m=1}^4 k_\m\sum_{\mathfrak{m}=1}^{ M_\m} t_\m^\mathfrak{m}\alpha_\m^\mathfrak{{m}}\!}\times \prod_{\hat{\mathfrak{m}},\mathfrak{n}=1}^{M_\m}\exp\Bigg[\sum_{\m=5}^n\sum_{r=1}^\infty \frac{1}{r}\Big[ e^{-ir(\alpha_\m^\mathfrak{n}-\alpha_\p^{\mathfrak{\hat{m}}})}F^{+}_{\mathfrak{n}\hat{\mathfrak{m}}\m}(x^r) +e^{ir(\alpha_\m^\mathfrak{n}-\alpha_\p^{\mathfrak{\hat{m}}})}F^{-}_{\mathfrak{n}\hat{\mathfrak{m}}\m}(x^r)\Bigg]&\nn \\
&\prod_{\hat{\mathfrak{m}},\mathfrak{n}=1}^{M_\m}\exp\Bigg[\sum_{\m=1}^2\sum_{r=1}^\infty \frac{1}{r}\Big[ e^{-ir(\alpha_\m^\mathfrak{n}-\alpha_\5^{\mathfrak{\hat{m}}})}F^{+}_{\mathfrak{n}\hat{\mathfrak{m}}\m}(x^r) +e^{ir(\alpha_\m^\mathfrak{n}-\alpha_\5^{\mathfrak{\hat{m}}})}F^{-}_{\mathfrak{n}\hat{\mathfrak{m}}\m}(x^r)\Bigg]&\nn\\
&\times\prod_{\hat{\mathfrak{m}},\mathfrak{n}=1}^{M_\m}\exp\Bigg[\sum_{\m=3}^4\sum_{r=1}^\infty \frac{1}{r}\Big[ e^{-ir(\alpha_\m^\mathfrak{n}-\alpha_{(n+1)}^{\mathfrak{\hat{m}}})}F^{+}_{\mathfrak{n}\hat{\mathfrak{m}}\m}(x^r) +e^{ir(\alpha_\m^\mathfrak{n}-\alpha_{(n+1)}^{\mathfrak{\hat{m}}})}F^{-}_{\mathfrak{n}\hat{\mathfrak{m}}\m}(x^r)\Bigg]&\nn\\
&			\times\!\prod_{\mathfrak{{m}},\mathfrak{{n}}=1}^{M_\m}\!\exp\left[\sum_{r=1}^\infty\frac{1}{r}
\sum_{\m=5}^{n+1} F^{{\rm adj}}_{\mathfrak{{m}}\mathfrak{{n}}\m}(x^r)e^{-ir(\alpha^\mathfrak{{m}}_\m\!-\!\alpha^\mathfrak{n}_\m)}
\right]\times\!\prod_{\mathfrak{{m}},\mathfrak{{n}}=1}^{M_\m}\!\exp\left[\sum_{r=1}^\infty\frac{1}{r}
\sum_{\m=1}^4 F^{{\rm adj}}_{\mathfrak{{m}}\mathfrak{{n}}\m}(x^r)e^{-ir(\alpha^\mathfrak{{m}}_\m\!-\!\alpha^\mathfrak{n}_\m)}
\right]&
\end{align}
where,
\be
\epsilon_0	
&=&\frac{1}{2}\sum_{\m=5}^n\sum_{\mathfrak{m}\in M_\m}\Big(N_\p|q^\mathfrak{m}_\m|+N_\m|q^\mathfrak{m}_\p|\Big)+\frac{1}{2}\sum_{\m=1}^2\sum_{\mathfrak{m}\in M_\m}\Big(N_\5|q^\mathfrak{m}_\m|+N_\m|q^\mathfrak{m}_\5|\Big)\nn\\
& +&\frac{1}{2}\sum_{\m=3}^4\sum_{\mathfrak{m}\in M_\m}\Big(N_{n+1}|q^\mathfrak{m}_\m|+N_\m|q^\mathfrak{m}_{n+1}|\Big)-\sum_{\m=1}^{n+1}\sum_{\mathfrak{m}\in M_\m} N_\m|q^\mathfrak{m}_\m|\nn
\ee
\begin{align}
\textit{for $\m=5,\cdots, n+1$}\quad	&F^{+}_{\mathfrak{n}\hat{\mathfrak{m}}\m}(x)=\Big(x^{|q_\m^{\mathfrak{n}}-q_\p^{\hat{\mathfrak{m}}}|}-x^{|q_\m^{\mathfrak{n}}|+|q_\p^{\hat{\mathfrak{m}}}|}\Big)f^+&\nn\\
&F^{-}_{\mathfrak{n}\hat{\mathfrak{m}}\m}(x)=\Big(x^{|q_\m^{\mathfrak{n}}-q_\p^{\hat{\mathfrak{m}}}|}-x^{|q_\m^{\mathfrak{n}}|+|q_\p^{\hat{\mathfrak{m}}}|}\Big)f^-&\nn\\
&F^{\mathrm{adj}}_{\mathfrak{n}{\mathfrak{m}}\m }(x)=(-1+\delta_{q_{\mathfrak{m}}q_{\mathfrak{n}}})\,x^{|q^\mathfrak{m}_\m-q^\mathfrak{n}_\m|}+x^{|q^\mathfrak{m}_\m|+|q^\mathfrak{n}_\m|}&\nn\\ \nn\\
\textit{for $\m=1,2$}\quad	&F^{+}_{\mathfrak{n}\hat{\mathfrak{m}}\m}(x)=\Big(x^{|t_\m^{\mathfrak{n}}-q_\5^{\hat{\mathfrak{m}}}|}-x^{|t_\m^{\mathfrak{n}}|+|q_\5^{\hat{\mathfrak{m}}}|}\Big)f^+&\nn\\
&F^{-}_{\mathfrak{n}\hat{\mathfrak{m}}\m}(x)=\Big(x^{|t_\m^{\mathfrak{n}}-q_\5^{\hat{\mathfrak{m}}}|}-x^{|t_\m^{\mathfrak{n}}|+|q_\5^{\hat{\mathfrak{m}}}|}\Big)f^-&\nn\\
&F^{\mathrm{adj}}_{\mathfrak{n}{\mathfrak{m}}\m }(x)=(-1+\delta_{t_{\mathfrak{m}}t_{\mathfrak{n}}})\,x^{|t^\mathfrak{m}_\m-t^\mathfrak{n}_\m|}+x^{|t^\mathfrak{m}_\m|+|t^\mathfrak{n}_\m|}&\nn\\ \nn\\
\textit{for $\m=3,4$}\quad	&F^{+}_{\mathfrak{n}\hat{\mathfrak{m}}\m}(x)=\Big(x^{|t_\m^{\mathfrak{n}}-q_{(n+1)}^{\hat{\mathfrak{m}}}|}-x^{|t_\m^{\mathfrak{n}}|+|q_{(n+1)}^{\hat{\mathfrak{m}}}|}\Big)f^+&\nn\\
&F^{-}_{\mathfrak{n}\hat{\mathfrak{m}}\m}(x)=\Big(x^{|t_\m^{\mathfrak{n}}-q_{(n+1)}^{\hat{\mathfrak{m}}}|}-x^{|t_\m^{\mathfrak{n}}|+|q_{(n+1)}^{\hat{\mathfrak{m}}}|}\Big)f^-&\nn\\
&F^{\mathrm{adj}}_{\mathfrak{n}{\mathfrak{m}}\m }(x)=(-1+\delta_{t_{\mathfrak{m}}t_{\mathfrak{n}}})\,x^{|t^\mathfrak{m}_\m-t^\mathfrak{n}_\m|}+x^{|t^\mathfrak{m}_\m|+|t^\mathfrak{n}_\m|}\;.&
\end{align}
Now it is fairly straight forward to find the large $N$ limit of the index for $\widehat{E}_6$ quiver.

\section{Conclusion and outlook}\label{sec:conclusion}
To summarize our results-
\begin{itemize}
	\item In this note we computed the superconformal  index for $\widehat{ADE}$ quiver gauge theories with $\NN=3$ supersymmetry. The index is function of  magnetic fluxes and the holonomy variables. We have used method of suppersymmetric localization to compute the index. The index of ABJM theory using the same technique has been evaluated for $\NN=6$ circular quiver gauge theory(ABJM) in \cite{Kim:2009wb}. We apply the same technique in $\widehat{ADE}$ class of theories. Therefore extending the applicability of the localization procedure. 
	\item We have  performed a   large $N$ analysis of the index which is needed to match with the index over gravitons in dual geometry.  We have explicitly computed the index for $\widehat{A}_3$ and $\widehat{D}_4$ case for one unit of magnetic flux by performing integral over holonomy variables $\alpha_\m^\mm$'s. The final result is a series in $x$ which need to match with the gravity index order by order in $x$.
	
	\item The deformation used in \cite{Kim:2009wb} is different from ours \cite{bkk}. The main differences are - (i) we have turned on three dynamical adjoint  scalars through the $Q$-exact deformation $\mathcal{S}_{\rm YM}+\mathcal{S}_{\rm adj}$, which was needed to preserve the supersymmtry to be $\NN=3$, while in \cite{Kim:2009wb} only one dynamical adjoint scalar  is turned on through the deformation $\mathcal{S}_{\rm YM}$, (ii)  we have a  Euclidean time dependent deformation parameter in the radially quantized theory while \cite{Kim:2009wb} uses time independent deformation.  Therefore we have shown that the deformation used in \cite{bkk} can also be used to calculate the superconformal index.
	
	\item We have performed a consistency check of the  obtained results by comparing them to the superconformal index of ABJM theory obtained in \cite{Kim:2009wb}.  Let us consider $\widehat{A}$ quiver for example whose final result is given in eqn.\eqref{final_index_an}.	The $\widehat{A}$ quiver represents   $\NN=3$ CS matter theory with gauge group $\prod_{\m=1}^n U(N)_{\m}$ which becomes $\NN=6$  $U(N)\times U(N)$ ABJM theory if one takes $n=2$. The supersymmetry enhancement is a consequence of the bigger symmetry group of the superpotential for two gauge groups. Therefore one should reproduce the index of ABJM theory obtained in \cite{Kim:2009wb} starting from eqn.\eqref{final_index_an} or vice versa. Here, we will show how to reproduce the $\NN=3$ superconformal index starting from the index integral of ABJM theory. To see this first note that R-symmetry group in  case of $\NN=3$ supersymmetric case is $SO(3)$ which is a subgroup of $SO(6)$ R-symmetry group of ABJM theory. Therefore, roughly one has to switch off two of the Cartans of $SO(6)$ to land on  $SO(3)$. This is essentially making $y_1=y_2=1$ in the definition of superconformal index  in \cite{Kim:2009wb}.  After carrying out all necessary steps we have  reproduced eqn\eqref{final_index_an}   from  \cite{Kim:2009wb}. The details of these checks are given in appendix \ref{app:consistency_check}.

\end{itemize} 
\par Finally, we  comment on our results by comparing them to some previously obtained results in the literature as in  \cite{Kim:2010vwa}. This paper evaluates superconformal index of a variant of $\NN=6$ ABJM theory which preserves $\NN=3$ supersymmetry and has gauge group $U(N)\times U(N)$. This theory is called dual ABJM theory whose filed content is little different from the usual ABJM, i.e one has additional adjoint fields $\phi_1, \phi_2$ of the first gauge group factor. The superpotential is also different from the $\NN=3$ quiver theories. Therefore, though the quiver  theories and the dual ABJM  preserve $\NN=3$ supersymmetry, they have different field content, superpotential term and moduli spaces. Also, the index evaluated in \cite{Kim:2010vwa} (eqn 3.1) is different from ours \eqref{eqn:definition_index}. Therefore to compare it to our result one has to set $y_1=y_2=1$ in eqn(3.1)of that paper. By doing so we find the contribution from the matter sector $f_\pm$ in eqn(3.2) of \cite{Kim:2010vwa} exactly matches with our $f^\pm$. In addition to this, \cite{Kim:2010vwa} has contribution from the additional adjoint fields $\phi_1, \phi_2$ which is denoted by $g$ in eqn(3.2) of that paper. This factor of $g$ is not present in our case since we do not have these adjoint fields. Setting $g=0$ in (3.3) of \cite{Kim:2010vwa} we recover the contribution from the adjoint sector in our case 
(eqn. \eqref{eqn:matter_contribution})  exactly. In addition to these one has to keep in mind that in case of $\NN=3$ quiver theories we have $n>2$ gauge groups which brings one extra index in all entities which is denoted by subscript $\m$ in our paper. Although the final answer in  \cite{Kim:2010vwa} (eqn. (3.4)) looks similar to our result schematically, as it should due to similarity in the field content, the values of the one loop determinants ($f_m, f_{adj}$) are different from ours. Moreover, the deformation used in \cite{Kim:2010vwa} to carry out supersymmetric localization is the $Q$ exact chiral part of the vector multiplet fields in $U(N)\times U(N)$ which is exactly same as our deformation (in addition to this \cite{Kim:2010vwa} has extra deformations arising from extra fields $\phi_1, \phi_2$ present in there). 

\par 
To the best of our  knowledge the multiparticle index of the corresponding gravity dual of $\NN=3$ $\widehat{ADE}$ theories is not yet computed. It would be interesting to  compare the large $N $ index on the gauge theory to the corresponding multi-particle index for M-theory in the dual geometry $\ads\times M_7$. For example in the case of $\widehat{A}_3$ $M_7$ has the isometry group $U(2)\times U(1)$. Therefore we need to find the graviton spectrum of $Osp(3|4)\times U(2)\times U(1)$ which   corresponds to the global symmetry group on the dual CFT side. 

\begin{center}
	\textbf{Acknowledgements}
\end{center}
I would like to thank Ashoke Sen  for many valuable discussions.
I would  also like to thank Tarun Sharma, Sudhakar Panda, Palash Dubey, Sayantani Bhattacharyya  for useful discussions and comments on the manuscript. This work is partly supported by J.C Bose Fellowship under Prof. Sudhakar Panda.


\appendix

\section{Consistency check}\label{app:consistency_check}
Here we present the details of the consistency check of the results obtained in this paper. In order to perform the consistency check we match our results to the previously obtained results in the literature viz. \cite{Kim:2009wb} and \cite{Jain:2019lqb}.\\
\underline{\bf{Matching with \cite{Kim:2009wb}}}:
We have performed a consistency check of the final result by comparing it to the results of \cite{Kim:2009wb}. This check seems most straight forward and has been explained in  the Conclusion section of the manuscript. Here we write it in more detail. 

One can perform a consistency check of the  obtained results by comparing them to the superconformal index of ABJM theory obtained in eqn(2.38) of \cite{Kim:2009wb}.  Let us consider $\widehat{A}$ quiver for example whose final result is given in \eqref{final_index_an}.
The $\widehat{A}$ quiver represents   $\NN=3$ CS matter theory with gauge group $\prod_{\m=1}^n U(N)_{\m}$ which becomes $\NN=6$  $U(N)\times U(N)$ ABJM theory if one takes $n=2$. The supersymmetry enhancement is a consequence of the bigger symmetry group of the superpotential for two gauge groups. Therefore one should reproduce the index obtained in \cite{Kim:2009wb} starting from \eqref{final_index_an} or vice versa. 

Here, we will show how to reproduce the $\NN=3$ superconformal index in \eqref{final_index_an} of our manuscript starting from the index integral of ABJM theory. As explained in section \ref{sec:conclusion} keeping in mind the the R-symmetry group  we have to  switch off two of the Cartans of $SO(6)$ to land on  $SO(3)$. This is essentially making $y_1=y_2=1$ in the definition of superconformal index eqn.(2.13) in  \cite{Kim:2009wb}. Following this, the contribution from the matter multiplet $f^{\pm}$ in eqn. (2.30), (2.31) in \cite{Kim:2009wb} will reduce to our results exactly, keeping in mind that instead of two matter multiplets we have one matter multiplet in the $\NN=3$ case. The contribution from the adjoint multiplet in eqn.(2.35) of \cite{Kim:2009wb}  is straight forward to generalise to the $\NN=3$ case for $n$ gauge groups. The contribution from the CS term (second line in (2.38) of \cite{Kim:2009wb}) is obtained after generalising  the constraint $k_\1+k_\2=0$ (in ABJM $k_\1=-k_\2=k$) to $\sum _{\m=1}^n k_\m=0$.
After carrying out all these steps we have shown to reproduce eqn \eqref{final_index_an} of our paper from  eqn.(2.38) of \cite{Kim:2009wb}. 

Eqn \eqref{final_index_dn} and \eqref{final_index_e6}  which is for $\widehat{DE}$ type quivers is a generalisation of \eqref{final_index_an} with a different field content.  The final result of the index is always of the form  of eqn.\eqref{eqn:schematic_index}. Therefore taking the appropriate field content for  $\widehat{DE}$ quivers one can generalize the final result of the index from the $\widehat{A}$ case.

\underline{\bf{Matching with \cite{Jain:2019lqb} }}: Our final result can also be matched to that of \cite{Jain:2019lqb}  but since the index computed there is not exactly the same as ours matching will not be straight forward. The index in   \cite{Jain:2019lqb} is a twisted index on $\Sigma_\mathfrak{g}\times S^1$, which implies turning on a background for the R-symmetry which is quantized magnetic flux on $\Sigma_\mathfrak{g}$. The following steps should be carried out in order to match our result to that of \cite{Jain:2019lqb}.
\begin{itemize}
	\item  In our paper we take $\Sigma_\mathfrak{g}=S^2$. $S^2$ is a genus $0$ surface which means putting $\mathfrak{g}=0$ in eqn. (3.1) of \cite{Jain:2019lqb}.
	\item  In addition to this we need to switch off the twist along $S^2$, which is done by turning off $\nu$. The flux of background vector multiplet has to be turned off as well by making $\mathfrak{n}=0$.
	\item As we are dealing with $\NN=3$ theory the R-charges for the hypermutiplets are fixed to their canonical value. Hence  we need to make  $\Delta=\frac{1}{2}$ in eqn.(1.2) of \cite{Jain:2019lqb}.
	\item
	After these changes one has to take  appropriate representation of the fields in eqn.(3.1) of \cite{Jain:2019lqb} to substitute the weights and roots $\rho(\mathfrak{m}), \alpha(\mathfrak{m})$ respectively.  For example in the $U(N)$ gauge group case the root vectors are $\alpha(\mathfrak{m})=\mathfrak{m}_i-\mathfrak{m}_j$ where $i,j$ are the gauge indices (in our notation $\mathfrak{m}$ is $q$, $u$ is $\alpha$). 
	\item Carrying out all these steps our  results can be reproduced. This is possible as $\NN=2$ and $\NN=3$ theory has similar field content, hence same schematic structure of the superconformal index	viz. (i) contribution from the vector multiplet (ii) contribution from the hypermultiplets (iii) Classical contribution coming from the CS term. 
\end{itemize}

\section{$\widehat{A}_3$ quiver}\label{app:a3}
In this case the gauge group of the theory is $ U(N)\times U(N)\times U(N)$. In the zero flux sector from eqn.\eqref{eqn:largeN_zero_an} we have,
\be 
I_0(x)=\frac{1}{\det \mathcal{N}(x^r) },\quad\qquad
\mathcal{N}(x)=\begin{pmatrix}
	1 & f^+ &f^-\\
	f^-&1&f^+\\
	f^+&f^-&1
\end{pmatrix},\nn\\
\det \mathcal{N}=(1-f^+f^-)-f^+(f^--{f^+}^2)-f^-({f^-}^2-f^+)
\ee 
The non-zero flux sector of the index has the form,
\be 
I(x)=I^{(\pm 1)}(x)+I^{(\pm 2)}(x)+ I^{(\pm 3)}(x)+\cdots
\ee 
Now, we evaluate the index by fixing the diagonal magnetic flux below.\\
\underline{$\boldsymbol{m_{\rm diag}=1}$}\\
The flux configuration for each gauge group can be expressed by a Young diagram. Hence we need  three Young diagrams to encode the flux configuration. The number of boxes in the $ii$-th row is the $ii$-th entry of $H$. Hence, $m_{\rm diag}=1$ can be expressed as follows,
\be 
I^{(+1)}(x)=I_{\Yboxdim4pt\yng(1)~\yng(1) ~\yng(1)}\ee 
Substituting the above in eqn.\eqref{large_an_nonzero}  we find,
\begin{align}
&			I(x)=
x^{\epsilon_0}\int\frac{1}{\rm (symmetry)}
\left[\frac{d\alpha_\m^{\mathfrak{m}}}{(2\pi)^3}\right]
e^{i( k_\1 \alpha_\1+k_\2 \alpha_\2+k_\3 \alpha_\3)}&\nn\\
&\times \exp\Bigg[\sum_{r=1}^\infty \frac{1}{r}(1-x^{2r})\Big( e^{-ir(\alpha_\1-\alpha_\2)}f^+(x^r) +e^{ir(\alpha_\1-\alpha_\2)}f^{-}(x^r)&\nn\\
&+e^{-ir(\alpha_\2-\alpha_\3)}f^{+}(x^r) +e^{ir(\alpha_\2-\alpha_\3)}f^{-}(x^r)+e^{-ir(\alpha_\3-\alpha_\1)}f^{+}(x^r) +e^{ir(\alpha_\3-\alpha_\1)}f^{-}(x^r)\Big)\Bigg]&\nn \\
&	\times		\exp \sum_{r=1}^\infty\frac{1}{r}
\left[3x^{2r}
\right]&
\end{align}
where we have substituted,
\be 
&&F^{+}_{1\hat{1}\m}(x)=(1-x^2)f^+=\sqrt{x}-x\sqrt{x}\quad
F^{-}_{1\hat{1}\m}(x)=(1-x^2)f^-=\sqrt{x}-x\sqrt{x}\quad
F^{\mathrm{adj}}_{11\m }(x)=x^2\nn\\
\ee 
Now let $k_\1,k_\2>0$ and using $k_\3=-(k_\1+k_\2)$ we can write the CS term as follows,
\be 
&&  e^{i( k_\1 \alpha_\1+k_\2 \alpha_\2+k_\3 \alpha_\3)}
= e^{ik_\1(\alpha_\1-\alpha_\3)}e^{ik_\2(\alpha_\2-\alpha_\3)}
\ee 
Observing that the index integral depends only on the variables $\alpha_\m-\alpha_\p$ we define the following,
\be 
&& u=e^{-i(\alpha_\1-\alpha_\2)},\quad v=e^{-i(\alpha_\2-\alpha_\3)},\quad w=e^{-i(\alpha_\3-\alpha_\1)}
\ee 
In terms of these variable the index is,
\be
I(x)&=&
x^{\epsilon_0}\int_0^{2\pi}
\frac{du}{2\pi i u}
\exp\Bigg[\sum_{r=1}^\infty \frac{1}{r}(x^{r/2}-x^{3r/2})\Big( u^{r}  +u^{-r}\Big)\Bigg]\nn\\
&&			\int_0^{2\pi}
\frac{dv}{2\pi i v}  v^{-k_\2}
\times \exp\Bigg[\sum_{r=1}^\infty \frac{1}{r}(x^{r/2}-x^{3r/2})\Big( v^{r} +v^{-r}\Big) \Bigg]\nn\\
&&	\int_0^{2\pi}
\frac{dw}{2\pi i w}   w^{k_\1}
\times \exp\Bigg[\sum_{r=1}^\infty \frac{1}{r}(x^{r/2}-x^{3r/2})\Big( w^{r} +w^{-r}\Big) \Bigg]\exp\sum_{r=1}^\infty \frac{1}{r} 3x^{2r}
\ee
Using the following relations,
\begin{eqnarray}
\hspace*{-2cm}f(x,p,r,z)
=(\sqrt{x}-x\sqrt{x})z+(\sqrt{x}-x\sqrt{x})z^{-1}\nn\\
\exp\left(\sum_{n=1}^\infty\frac{1}{n}f(x^n,p^n,r^n,z^n)\right)
=\frac{(1-x\sqrt{x}z^{-1})(1-x\sqrt{x}z)}
{(1-\sqrt{x}z)(1-\sqrt{x}z^{-1})}
\end{eqnarray}
we find,

\be
\int_0^{2\pi}
\frac{du}{2\pi i u}
\exp\Bigg[\sum_{r=1}^\infty \frac{1}{r}(x^{r/2}-x^{3r/2})\Big( u^{r}  +u^{-r}\Big)\Bigg]=\int_0^{2\pi}
\frac{du}{2\pi i u}\;
\frac{(1-x\sqrt{x}u^{-1})(1-x\sqrt{x}u)}
{(1-\sqrt{x}u)(1-\sqrt{x}u^{-1})}
\ee
This integral can be evaluated using Cauchy's integral formula which implies we need to pick up $\mathcal{O}(u^0)$ terms from the above integrand. 
Now, expanding the denominator we find,
\be
\frac{1}{(1-\sqrt{x}u)}.\frac{1}{(1-\sqrt{x}u^{-1})}=(1-\sqrt{x}u)^{-1}(1-\sqrt{x}u^{-1})^{-1}\nn\\
=\Big(1+x^{\frac{1}{2}}u+xu^2+\cdots\Big)\Big(1+x^{\frac{1}{2}}u^{-1}+xu^{-2}+\cdots\Big)\nn\\
=1+x^{\frac{1}{2}}u^{-1}+xu^{-2}+\cdots+x^{\frac{1}{2}}u+x+x^{\frac{3}{2}}u^{-1}+\cdots+ xu^2+x^{\frac{3}{2}}u+x^2+\cdots
\ee
And the numerator is,
\be 
(1-x\sqrt{x}u^{-1})(1-x\sqrt{x}u)
=1-x\sqrt{x}u-x\sqrt{x}u^{-1}+x^3\;.
\ee 
Therefore we have,
\be
&&	\int_0^{2\pi}
\frac{du}{2\pi i u}
\exp\Bigg[\sum_{r=1}^\infty \frac{1}{r}(x^{r/2}-x^{3r/2})\Big( u^{r}  +u^{-r}\Big)\Bigg]=1+x-2x^2-x^3+\cdots
\ee
Now, moving on to the $v$ integral we find
observe that, since $k_\2$ is positive we need to pick all positive powers of $v$ from  the infinite series $\frac{(1-x\sqrt{x}v^{-1})(1-x\sqrt{x}v)}
{(1-\sqrt{x}v)(1-\sqrt{x}v^{-1})}$ to find $v^0$ terms (after fixing the value of $k_\2$) which is,
\be 
&&\hspace*{-1cm}	\frac{(1-x\sqrt{x}v^{-1})(1-x\sqrt{x}v)}
{(1-\sqrt{x}v)(1-\sqrt{x}v^{-1})}= \Big(1-x\sqrt{x}v-x\sqrt{x}v^{-1}+x^3\Big)\Big(1+x^{\frac{1}{2}}v+xv^2+\cdots\Big)\Big(1+x^{\frac{1}{2}}v^{-1}+xv^{-2}+\cdots\Big)\nn\\
&&\hspace*{-1cm}=\Big(x^{\frac{1}{2}}v+xv^2+x^{\frac{3}{2}}v^3+\cdots\Big)-x\sqrt{x}v\Big(1+x^{\frac{1}{2}}v+xv^2+\cdots\Big)+\cdots
\ee 
Now the remaining integral is,
\be 
\int_0^{2\pi}
\frac{dw}{2\pi i w}   w^{k_\1}
\times \exp\Bigg[\sum_{r=1}^\infty \frac{1}{r}(x^{r/2}-x^{3r/2})\Big( w^{r} +w^{-r}\Big) \Bigg]=\int_0^{2\pi}
\frac{dw}{2\pi i w}   w^{k_\1}
\times \frac{(1-x\sqrt{x}w^{-1})(1-x\sqrt{x}w)}
{(1-\sqrt{x}w)(1-\sqrt{x}w^{-1})}.\nn\\
\ee
In this case, since $k_\1$ is positive we need to pick all negative powers of $w$ from  the infinite series in order to find $w^0$ terms, which is,
\be 
&&\hspace*{-1cm}\frac{(1-x\sqrt{x}v^{-1})(1-x\sqrt{x}w)}
{(1-\sqrt{x}w)(1-\sqrt{x}w^{-1})}= \Big(1-x\sqrt{x}w-x\sqrt{x}w^{-1}+x^3\Big)\Big(1+x^{\frac{1}{2}}w+xw^2+\cdots\Big)\Big(1+x^{\frac{1}{2}}w^{-1}+xw^{-2}+\cdots\Big)\nn\\
&&\hspace*{-1cm}=\Big(x^{\frac{1}{2}}w^{-1}+xw^{-2}+\cdots\Big)-x\sqrt{x}w^{-1}+\cdots
\ee 
Now one can fix different values of the CS level and obtain the index  an infinite series in $x$.

\section{$\widehat{D}_4$ quiver}\label{app:d4}
The gauge group associated with $\widehat{D}_4$ quiver is $U(N)^4\times U(2N)$.
In this case the large N limit of the index in zero flux sector is,
\be
&&	I_0(x)=
x^{\epsilon_0}\int\frac{1}{\rm (symmetry)}
\left[\frac{d\alpha_\m^{\mathfrak{m}}}{(2\pi)^{5}}\right]
\exp\Bigg[\sum_{\m,\s=1}^{5}\sum_{r=1}^\infty \frac{1}{r}\Bigg( \rho_{r\m}\,\mathcal{N}_{\m\s}(x^r) \rho_{-r\s} \Bigg)\Bigg]\nn\\
\ee
where,
\be 
\mathcal{N}(x)=\begin{pmatrix}
	1 &0&0&0& f^+\\
	0&1&0& 0&f^+\\
	0&0&1& 0&f^+\\
	0&0&0& 1&f^+\\
	f^-&f^-&f^-& f^-&1
\end{pmatrix}
\ee  
The constraint on the CS level in this case is,
\be 
\label{eqn:const_cs_d4}
k_\1+k_\2+k_\3+k_\4+2k_\5=0\;.
\ee 
Now, choosing,
\be 
&&For\quad \m=1,...4 \quad m_\m= m_{\rm diag}\nn\\
&&For\quad \m=5 \quad m_\5= 2m_{\rm diag}  \qquad (\textit{Consistent with \eqref{eqn:const_cs_d4}})
\ee 
the index is,
\begin{align}
&			I(x)=
x^{\epsilon_0}\int\frac{1}{\rm (symmetry)}
\left[\frac{d\alpha_\m^{\mathfrak{m}}}{(2\pi)^{5}}\right]
\prod_{\m=1}^5	\prod_{\mathfrak{{m}}<\mathfrak{{n}}}
\left[2\sin\left(\frac{\alpha_\m^\mathfrak{m}\!-\!\alpha_\m^\mathfrak{n}}{2}\right)\right]^2\times e^{i k_\5\sum_{\mathfrak{m}=1}^{ M_\5} q_\5^\mathfrak{m}\alpha_\5^\mathfrak{{m}}\!}&\nn\\
&\times e^{i\sum_{\m=1}^4 k_\m\sum_{\mathfrak{m}=1}^{ M_\m} t_\m^\mathfrak{m}\alpha_\m^\mathfrak{{m}}\!}&\nn \\
&\prod_{\hat{\mathfrak{m}},\mathfrak{n}=1}^{M_\m}\exp\Bigg[\sum_{\m=1}^2\sum_{r=1}^\infty \frac{1}{r}\Big[ e^{-ir(\alpha_\m^\mathfrak{n}-\alpha_\5^{\mathfrak{\hat{m}}})}F^{+}_{\mathfrak{n}\hat{\mathfrak{m}}\m}(x^r) +e^{ir(\alpha_\m^\mathfrak{n}-\alpha_\5^{\mathfrak{\hat{m}}})}F^{-}_{\mathfrak{n}\hat{\mathfrak{m}}\m}(x^r)\Bigg]&\nn\\
&\times\prod_{\hat{\mathfrak{m}},\mathfrak{n}=1}^{M_\m}\exp\Bigg[\sum_{\m=3}^4\sum_{r=1}^\infty \frac{1}{r}\Big[ e^{-ir(\alpha_\m^\mathfrak{n}-\alpha_{\5}^{\mathfrak{\hat{m}}})}F^{+}_{\mathfrak{n}\hat{\mathfrak{m}}\m}(x^r) +e^{ir(\alpha_\m^\mathfrak{n}-\alpha_{\5}^{\mathfrak{\hat{m}}})}F^{-}_{\mathfrak{n}\hat{\mathfrak{m}}\m}(x^r)\Bigg]&\nn\\
&			\times\!\prod_{\mathfrak{{m}},\mathfrak{{n}}=1}^{M_\5}\!\exp\left[\sum_{r=1}^\infty\frac{1}{r}
F^{{\rm adj}}_{\mathfrak{{m}}\mathfrak{{n}}\5}(x^r)e^{-ir(\alpha^\mathfrak{{m}}_\5\!-\!\alpha^\mathfrak{n}_\5)}
\right]\times\!\prod_{\mathfrak{{m}},\mathfrak{{n}}=1}^{M_\m}\!\exp\left[\sum_{r=1}^\infty\frac{1}{r}
\sum_{\m=1}^4 F^{{\rm adj}}_{\mathfrak{{m}}\mathfrak{{n}}\m}(x^r)e^{-ir(\alpha^\mathfrak{{m}}_\m\!-\!\alpha^\mathfrak{n}_\m)}
\right]&
\end{align}
\underline{$\boldsymbol{m_{\rm diag}=1}$}
The index with one unit of magnetic flux containts contribution from two flux configurations as follows.
\be 
I^{(+2,+1)}(x)=I_{\Yboxdim4pt \{\yng(2)\}~\{\yng(1) ~\yng(1)~\yng(1) ~\yng(1)\}}+I_{\Yboxdim4pt \{\yng(1,1)\}~\{\yng(1) ~\yng(1)~\yng(1) ~\yng(1)\}}
\ee
The first entry in the superscript of $I$ is the flux of the internal nodes and second entry is the magnetic flux associated with the external nodes. Since we can have $2$ units of flux for the internal node, we have two choices for the corresponding flux configuration, viz. $\{1,1, 0, \cdots\}$ and $\{2,0, \cdots\}$. We determine the index integral for these two cases below. \\ \\
\underline{ $H_\m=\{1,0,\cdots\}, H_\5=\{2,0,\cdots\}$:}\\ \\
Plugging in the flux configuration in eqn.\eqref{index_dn_large}  we find,
\be
&&\hspace*{-2cm}		I(x)=
x^{\epsilon_0}\int
\left[\frac{d\alpha_\m^{\mathfrak{m}}}{(2\pi)^5}\right]
e^{i( k_\1 \alpha_\1+k_\2 \alpha_\2+k_\3 \alpha_\3+k_\4\alpha_\4+2k_\5\alpha_\5)}\nn\\
&&\hspace*{-2cm}\exp\Bigg[\sum_{r=1}^\infty \frac{1}{r}\Big[ e^{-ir(\alpha_\1-\alpha_\5)}F^{+}_{\1}(x^r) +e^{ir(\alpha_\1-\alpha_\5)}F^{-}_{\1}(x^r)+e^{-ir(\alpha_\2-\alpha_\5)}F^{+}_{\2}(x^r) +e^{ir(\alpha_\2-\alpha_\5)}F^{-}_{\2}(x^r)\Bigg]\nn\\
&&\hspace*{-2cm}\exp\Bigg[\sum_{r=1}^\infty \frac{1}{r}\Big[ e^{-ir(\alpha_\3-\alpha_\5)}F^{+}_{\3}(x^r) +e^{ir(\alpha_\3-\alpha_\5)}F^{-}_{\3}(x^r)+e^{-ir(\alpha_\4-\alpha_\5)}F^{+}_{\4}(x^r) +e^{ir(\alpha_\4-\alpha_\5)}F^{-}_{\4}(x^r)\Bigg]\nn\\			&&\hspace*{-2cm}\times\!\prod_{\mathfrak{{m}},\mathfrak{{n}}=1}^{M_\5}\!\exp\left[\sum_{r=1}^\infty\frac{1}{r}
F^{{\rm adj}}_{\mathfrak{{m}}\mathfrak{{n}}\5}(x^r)e^{-ir(\alpha^\mathfrak{{m}}_\5\!-\!\alpha^\mathfrak{n}_\5)}
\right]\times\!\prod_{\mathfrak{{m}},\mathfrak{{n}}=1}^{M_\m}\!\exp\left[\sum_{r=1}^\infty\frac{1}{r}
\sum_{\m=1}^4 F^{{\rm adj}}_{\mathfrak{{m}}\mathfrak{{n}}\m}(x^r)e^{-ir(\alpha^\mathfrak{{m}}_\m\!-\!\alpha^\mathfrak{n}_\m)}
\right]
\ee
where,
\be 
For \quad \m=1,...,4\qquad
F^{+}_{1\hat{1}\m}(x)&=&\Big(x^{|q_\1^1-q_\5^{\hat 1}|}-x^{|q_\1^1|+|q_\5^{\hat 1}|}\Big)f^+= \Big(x-x^3\Big)f^+=x^{\frac{3}{2}}-x^{\frac{5}{2}}\nn\\
F^{-}_{1\hat{1}\m}(x)&=&\Big(x^{|q_\m^{\mathfrak{n}}-q_\5^{\hat{\mathfrak{m}}}|}-x^{|q_\m^{\mathfrak{n}}|+|q_\5^{\hat{\mathfrak{m}}}|}\Big)f^-=(x-x^3)f^-=x^{\frac{3}{2}}-x^{\frac{5}{2}}\nn\\
For \quad \m=1,...,4\qquad F^{\mathrm{adj}}_{11\m }(x)&=&x^2,\nn\\
For \quad \m=5\qquad F^{\mathrm{adj}}_{11\5}(x)&=&x^4
,\quad \rm symmetry=1
\ee 
Further plugging the letter indices in the index we obtain,
\begin{align}
&			I(x)=
x^{\epsilon_0}\int
\left[\frac{d\alpha_\m^{\mathfrak{m}}}{(2\pi)^5}\right]
e^{i( k_\1 \alpha_\1+k_\2 \alpha_\2+k_\3 \alpha_\3+k_\4\alpha_\4+2k_\5\alpha_\5)}&\nn\\
&\exp\Bigg[\sum_{r=1}^\infty \frac{1}{r}\Big(x^{\frac{3r}{2}}-x^{\frac{5r}{2}}\Big)\Big[ e^{-ir(\alpha_\1-\alpha_\5)} +e^{ir(\alpha_\1-\alpha_\5)}+e^{-ir(\alpha_\2-\alpha_\5)} +e^{ir(\alpha_\2-\alpha_\5)}&\nn\\
&+ e^{-ir(\alpha_\3-\alpha_\5)} +e^{ir(\alpha_\3-\alpha_\5)}+e^{-ir(\alpha_\4-\alpha_\5)} +e^{ir(\alpha_\4-\alpha_\5)}\Bigg]&\nn\\
&			\exp\left[\sum_{r=1}^\infty\frac{1}{r}
\Big(4x^{2r}+x^{4r}\Big)\right]&
\end{align}
Now, similarly as the previous case we define 
\be 
u_\m= e^{-i  (\alpha_\m-\alpha_\5)},\quad k_\m>0\quad for \quad \m=1,...4
\ee 
In terms of the $u_\m$'s the index takes the following form,
\begin{align}
&			I(x)=
x^{\epsilon_0}\int
\left[\frac{du_\m}{2\pi u_\m}\right]
u_\1^{-k_\1}u_\2^{-k_\2}u_\3^{-k_\3}u_\4^{-k_\4}&\nn\\
&\exp\Bigg[\sum_{r=1}^\infty \frac{1}{r}\Big(x^{\frac{3r}{2}}-x^{\frac{5r}{2}}\Big)\Big[ u_\1^r +u_\1^{-r}+u_\2^r +u_\2^{-r}
+ u_\3^r+u_\3^{-r}+ u_\4^r +u_\4^{-r}\Bigg]&\nn\\
&			\exp\left[\sum_{r=1}^\infty\frac{1}{r}
\Big(4x^{2r}+x^{4r}\Big)\right]&
\end{align}
We make use of  the following result\cite{Kim:2009wb},
\begin{eqnarray}
\hspace*{-2cm}f_1(x,p,r,z)
=(x\sqrt{x}-x^2\sqrt{x})z+(x\sqrt{x}-x^2\sqrt{x})z^{-1}\nn\\
\exp\left(\sum_{n=1}^\infty\frac{1}{n}f_1(x^n,p^n,r^n,z^n)\right)
=\frac{(1-x^2\sqrt{x}z^{-1})(1-x^2\sqrt{x}z)}
{(1-x\sqrt{x}z)(1-x\sqrt{x}z^{-1})}
\end{eqnarray}
to evaluate the  integrals. The $u_\1$ integral,
\be
&&\hspace*{-2cm}\int_0^{2\pi}
\frac{du_\1}{2\pi i u_\1}\, u_\1^{-k_\1}
\exp\Bigg[\sum_{r=1}^\infty \frac{1}{r}\Big(x^{\frac{3r}{2}}-x^{\frac{5r}{2}}\Big)\Big( u_\1^{r} +u_\1^{-r}\Big)\Bigg]=\int_0^{2\pi}
\frac{du_\1}{2\pi i u_\1}\;u_\1^{-k_\1}
\frac{(1-x^2\sqrt{x}u_\1^{-1})(1-x^2\sqrt{x}u_\1)}
{(1-x\sqrt{x}u_\1)(1-x\sqrt{x}u_\1^{-1})}\nn\\
\ee
is solved exactly as before, i.e expanding the numerator and the denominator and picking up appropriate powers of $u_\1$.
Now expanding the denominator we obtain,
\be
&&\hspace*{-2.5cm}(1-x\sqrt{x}u_\1)^{-1}(1-x\sqrt{x}u_\1^{-1})^{-1}
=\Big(1+x\sqrt{x}u_\1+x^3 u_\1^2+\cdots\Big) \Big(1+x\sqrt{x}u_\1^{-1}+x^3 u_\1^{-2}+\cdots\Big)
\ee
while the numerator is,
\be 
(1-x^2\sqrt{x}u^{-1}_\1)(1-x^2\sqrt{x}u_\1)=1-x^2\sqrt{x}u_\1-x^2\sqrt{x}u^{-1}_\1+x^5
\ee
which gives, 
\be
&&	
\int_0^{2\pi}
\frac{du_\1}{2\pi i u_\1}\, u_\1^{-k_\1}
\exp\Bigg[\sum_{r=1}^\infty \frac{1}{r}\Big(x^{\frac{3r}{2}}-x^{\frac{5r}{2}}\Big)\Big( u_\1^{r} +u_\1^{-r}\Big)\Bigg]\nn\\
&&=	\int_0^{2\pi}
\frac{du_\1}{2\pi i u_\1}\, u_\1^{-k_\1}
\Big(1+x\sqrt{x}u_\1+x^3 u_\1^2+\cdots\Big) \Big(1+x\sqrt{x}u_\1^{-1}+x^3 u_\1^{-2}+\cdots\Big)\nn\\
&&\Big(1-x^2\sqrt{x}u_\1-x^2\sqrt{x}u^{-1}_\1+x^5\Big)\nn
\ee
Since $k_\1$ is positive we need to pick all positive powers of $u_\1$ from  the infinite series  to find $u^0$ terms. Some of such terms are given below,
\be 
&&\Big(1+x\sqrt{x}u_\1+x^3 u_\1^2+\cdots\Big) \Big(1+x\sqrt{x}u_\1^{-1}+x^3 u_\1^{-2}+\cdots\Big)\Big(1-x^2\sqrt{x}u_\1-x^2\sqrt{x}u^{-1}_\1+x^5\Big)\nn\\
&&=x^3-x^4+x^5+\cdots
\ee 
Now, it is straight forward to evaluate the $u_\2, u_\3, u_\4$ integrals.\\ \\
\underline{ $H_\m=\{1,0,\cdots\},\, H_\5=\{1,1,0,\cdots\}$:}\\  \\
In this case the letter indices will be different since we have a different flux configuration. All the non zero letter indices are,
\be 
For \quad \m=1,...,4\qquad
F^{+}_{1\hat{1}\m}(x)&=&\Big(x^{|q_\1^1-q_\5^{\hat 1}|}-x^{|q_\1^1|+|q_\5^{\hat 1}|}\Big)f^+= \Big(1-x^2\Big)f^+=x^{\frac{1}{2}}-x^{\frac{3}{2}}\nn\\
F^{-}_{1\hat{1}\m}(x)&=&\Big(x^{|q_\m^1-q_\5^{\hat{1}}|}-x^{|q_\m^1|+|q_\5^{\hat{1}}|}\Big)f^-=(1-x^2)f^-=x^{\frac{1}{2}}-x^{\frac{3}{2}}\nn\\
F^{+}_{1\hat{2}\m}(x)&=&x^{\frac{1}{2}}-x^{\frac{3}{2}},\quad F^{-}_{1\hat{2}\m}(x)=x^{\frac{1}{2}}-x^{\frac{3}{2}}\nn\\
For \quad \m=1,...,4\qquad F^{\mathrm{adj}}_{11\m }(x)&=&x^2,\nn\\
For \quad \m=5 \qquad F^{\mathrm{adj}}_{11\5 }(x)&=&F^{\mathrm{adj}}_{22\5 }(x)=F^{\mathrm{adj}}_{12\5 }(x)=F^{\mathrm{adj}}_{21\5 }(x)=x^2,\quad \rm symmetry=2\nn\\
\ee 
The explicit index integral  is,
\begin{align}
&			I(x)=
x^{\epsilon_0}\int\frac{1}{2}
\left[\frac{d\alpha_\m^{\mathfrak{m}}}{(2\pi)^6}\right]\left[2\sin\left(\frac{\alpha_\5^1-\alpha_\5^2}{2}\right)\right]^2
e^{i( k_\1 \alpha_\1+k_\2 \alpha_\2+k_\3 \alpha_\3+k_\4\alpha_\4+k_\5\alpha^1_\5+k_\5\alpha^2_\5)}&\nn\\
&\exp\Bigg[\sum_{r=1}^\infty \frac{1}{r}\Big[ e^{-ir(\alpha_\1-\alpha_\5^1)}F^{+}_{\1}(x^r) +e^{ir(\alpha_\1-\alpha_\5^1)}F^{-}_{\1}(x^r)+e^{-ir(\alpha_\1-\alpha_\5^2)}F^{+}_{\1}(x^r) +e^{ir(\alpha_\1-\alpha_\5^2)}F^{-}_{\1}(x^r)&\nn\\
&+e^{-ir(\alpha_\2-\alpha_\5^1)}F^{+}_{\2}(x^r) +e^{ir(\alpha_\2-\alpha_\5^1)}F^{-}_{\2}(x^r)+e^{-ir(\alpha_\2-\alpha_\5^2)}F^{+}_{\2}(x^r) +e^{ir(\alpha_\2-\alpha_\5^2)}F^{-}_{\2}(x^r)&\nn\\
& + e^{-ir(\alpha_\3-\alpha_\5^1)}F^{+}_{\3}(x^r) +e^{ir(\alpha_\3-\alpha_\5^1)}F^{-}_{\3}(x^r)+ e^{-ir(\alpha_\3-\alpha_\5^2)}F^{+}_{\3}(x^r) +e^{ir(\alpha_\3-\alpha_\5^2)}F^{-}_{\3}(x^r)&\nn\\
&+e^{-ir(\alpha_\4-\alpha_\5^1)}F^{+}_{\4}(x^r) +e^{ir(\alpha_\4-\alpha_\5^1)}F^{-}_{\4}(x^r)+e^{-ir(\alpha_\4-\alpha_\5^2)}F^{+}_{\4}(x^r) +e^{ir(\alpha_\4-\alpha_\5^2)}F^{-}_{\4}(x^r)\Big]\Bigg]&\nn\\
&			\times\exp\Bigg[\sum_{r=1}^\infty\frac{1}{r}\Big[
x^{2r}e^{-ir(\alpha^1_\5\!-\!\alpha^2_\5)}+x^{2r}e^{-ir(\alpha^2_\5\!-\!\alpha^1_\5)}+2x^{2r}+
4x^{2r}\Big]\Bigg]&
\end{align}
Observe that we have a non zero Faddeev-Popov term in this case.
Using \eqref{eqn:const_cs_d4} the CS term reduces to the following,
\ben 
&&\exp{\Big[i\Big(  k_\1 \alpha_\1+k_\2 \alpha_\2+k_\3 \alpha_\3+k_\4\alpha_\4+k_\5(\alpha^1_\5+\alpha^2_\5)\Big)\Big]}\nn\\
&&=\exp{\Big[i\Big(  k_\1 \alpha_\1+k_\2 \alpha_\2+k_\3 \alpha_\3+k_\4\alpha_\4-\frac{1}{2}(k_\1+k_\2+k_\3+k_\4)(\alpha^1_\5+\alpha^2_\5)\Big)\Big]}\\
&&= \exp\Big[i\Big(k_\1( \alpha_\1-\frac{1}{2}\alpha^1_\5-\frac{1}{2}\alpha^2_\5)+k_\2( \alpha_\2-\frac{1}{2}\alpha^1_\5-\frac{1}{2}\alpha^2_\5)
+k_\3(\alpha_\3-\frac{1}{2}\alpha^1_\5-\frac{1}{2}\alpha^2_\5)\nn\\
&&+k_\4(\alpha_\4-\frac{1}{2}\alpha^1_\5-\frac{1}{2}\alpha^2_\5)\Big)\Big]\;.
\en 
Now defining,
\be 
u_\m= e^{-i  (\alpha_\m-\alpha^1_\5)},\quad v_\m= e^{-i  (\alpha_\m-\alpha^2_\5)}
\ee 
the CS term can be written as,
\be 
\prod_{\m=1}^4(u_\m v_\m )^{-\frac{k_\m}{2}}\;.
\ee 
The Faddeev-Popov term can also be expressed as follows,
\be 
\left[2\sin\left(\frac{\alpha_\5^1-\!\alpha_\5^2}{2}\right)\right]^2 =\exp\Big[-\sum_{r=1}^\infty \frac{1}{r} \frac{v^r_\1}{u^r_\1}\Big]\;.
\ee 
In terms of the new integration variables we write the index as,
\begin{align}
&			I(x)=\frac{1}{2}
x^{\epsilon_0}\int 
\prod_{\m=1}^4\left[\frac{du_\m}{2\pi u_\m}\right]\left[\frac{dv_\m}{2\pi v_\m}\right]
(u_\m v_\m )^{-\frac{k_\m}{2}}&\nn\\
&\exp\Bigg[\sum_{r=1}^\infty \frac{1}{r}\Big[ u_\1^r (x^{\frac{r}{2}}-x^{\frac{3r}{2}}) +u_\1^{-r}(x^{\frac{r}{2}}-x^{\frac{3r}{2}})+v_\1^r (x^{\frac{r}{2}}-x^{\frac{3r}{2}})+v_\1^{-r} (x^{\frac{r}{2}}-x^{\frac{3r}{2}})&\nn\\
&+u_\2^r (x^{\frac{r}{2}}-x^{\frac{3r}{2}}) +u_\2^{-r}F^{-}_{\2}(x^r)+v_\2^r (x^{\frac{r}{2}}-x^{\frac{3r}{2}}) +v_\2^{-r}F^{-}_{\2}(x^r)&\nn\\
& + u_\3^r (x^{\frac{r}{2}}-x^{\frac{3r}{2}})+u_\3^{-r}F^{-}_{\3}(x^r)+ v_\3^3 (x^{\frac{r}{2}}-x^{\frac{3r}{2}}) +v_\3^{-r}(x^{\frac{r}{2}}-x^{\frac{3r}{2}})&\nn\\
&+u_\4^r (x^{\frac{r}{2}}-x^{\frac{3r}{2}}) +u_\4^{-r}F^{-}_{\4}(x^r)+v_\4^r (x^{\frac{r}{2}}-x^{\frac{3r}{2}}) +v_\4^{-r}(x^{\frac{r}{2}}-x^{\frac{3r}{2}})-\frac{v^r_\1}{u^r_\1}+
x^{2r}\frac{v^r_\1}{u^r_\1}+x^{2r}\frac{u^r_\1}{v^r_\1}+2x^{2r}+
4x^{2r}\Big]\Bigg]\;.&
\end{align}
Now one can use the  previous results to evaluate the above integral.


\begin{thebibliography}{9}
\bibitem{Kim:2009wb}
S.~Kim,
``The Complete superconformal index for N=6 Chern-Simons theory,''
Nucl.\ Phys.\ B { 821} (2009) 241
\texttt{\arxivref{0903.4172}}.

\bibitem{charge_ade}
M. ~Patra, ``Charges of Monopole Operators in $\widehat {ADE} $ Chern-Simons Quiver Gauge Theories." JHEP 06, 008 (2020) \texttt{\arxivref{2002.03685}} .

\bibitem{Kinney:2005ej}
J.~Kinney, J.~M.~Maldacena, S.~Minwalla and S.~Raju,
``An index for 4 dimensional super conformal theories,''
Commun.\ Math.\ Phys.\  {\bf 275}, 209 (2007)
\texttt{\arxivref{hep-th/0510251}}.

\bibitem{Bhattacharya:2008zy}
J.~Bhattacharya, S.~Bhattacharyya, S.~Minwalla and S.~Raju,
``Indices for superconformal field theories in 3,5 and 6 dimensions,''
JHEP {\bf 0802}, 064 (2008)
\texttt{\arxivref{0801.1435}}.

\bibitem{Aharony:2005bq}
O.~Aharony, J.~Marsano, S.~Minwalla, K.~Papadodimas and M.~Van Raamsdonk,
``A first order deconfinement transition in large N Yang-Mills theory on
a small 3-sphere,''
Phys.\ Rev.\  D {\bf 71}, 125018 (2005)
\texttt{\arxivref{hep-th/0502149}}.


\bibitem{abjm}
O. Aharony, O. Bergman, D. L. Jafferis and J. Maldacena, “N=6 superconformal
Chern-Simons-matter theories, M2-branes and their gravity duals”, \texttt{\arxivref{0806.1218}}.

\bibitem{bkk}
M. K. Benna, I. R. Klebanov, and T. Klose, ``Charges of monopole operators in Chern-Simons Yang-Mills theory."  JHEP {1001} (2010) 110 \texttt{\arxivref{0906.3008}}.

\bibitem{Aharony:2003sx}
O.~Aharony, J.~Marsano, S.~Minwalla, K.~Papadodimas and M.~Van Raamsdonk,
``The Hagedorn/deconfinement phase transition in weakly coupled large
N gauge theories,''
Adv.\ Theor.\ Math.\ Phys.\  {\bf 8}, 603 (2004)
{\tt [arXiv:hep-th/0310285]}.

\bibitem{Weinberg:1993sg}
E.~J.~Weinberg,
``Monopole vector spherical harmonics,''
Phys.\ Rev.\  D {\bf 49}, 1086 (1994) \texttt{\arxivref{hep-th/9308054}}.

\bibitem{nishioka}
D. R. Gulotta,  C. P. Herzog and T. Nishioka, `` The ABCDEF’s of matrix models for supersymmetric Chern-Simons theories" JHEP {{2012.4}} (2012): 138  \texttt{\arxivref{1201.6360}}.
\bibitem{gulotta}
D. R. Gulotta,  J. P. Ang and C. P. Herzog,``Matrix models for supersymmetric Chern-Simons theories with an ADE classification." JHEP  2012.1 (2012): 132 \texttt{\arxivref{1111.1744}}.
\bibitem{non-toric}
P. Marcos Crichigno, and D. Jain, ``Non-toric Cones and Chern-Simons Quivers", JHEP 2017.05:046 \texttt{\arxivref{1702.05486}}.

	\bibitem{jaff}
D. L. Jafferis  and A. Tomasiello, ``A simple class of $\superN=3$ gauge/gravity duals." JHEP {{0810}} (2008) 101   \texttt{\arxivref{0808.0864}}.

\bibitem{Maldacena:1997re}
J.~M.~Maldacena,
``The Large N limit of superconformal field theories and supergravity,''
Int.\ J.\ Theor.\ Phys.\  { 38} (1999) 1113
[Adv.\ Theor.\ Math.\ Phys.\  { 2} (1998) 231]
\bibitem{Szabo:2014zua}
R.~J.~Szabo and O.~Valdivia,
``Covariant Quiver Gauge Theories,''
JHEP \textbf{06} (2014), 144
\texttt{\arxivref{1404.4319}}.

\bibitem{Imamura:2009hc}
Y.~Imamura and S.~Yokoyama,
Nucl. Phys. B \textbf{827} (2010), 183-216
\texttt{\arxivref{0908.0988}}.



\bibitem{Kim:2010vwa}
S.~Kim and J.~Park,
JHEP \textbf{08} (2010), 069
 \texttt{\arxivref{1003.4343}}.

\bibitem{Jain:2019lqb}
D.~Jain and A.~Ray,
Phys. Rev. D \textbf{100} (2019) no.4, 046007
\texttt{\arxivref{1902.10498}}.

\end{thebibliography}
\end{document}